\documentclass[11pt]{article}
\usepackage[a4paper,margin=1in]{geometry}
\usepackage[T1]{fontenc}
\usepackage{lmodern}
\usepackage{amsmath,amssymb,amsthm}
\usepackage{graphicx}
\usepackage{pgfplots}
\pgfplotsset{compat=1.17}
\usepgfplotslibrary{groupplots}
\usetikzlibrary{arrows.meta,positioning}
\usepackage[dvipsnames]{xcolor}
\usepackage[authoryear,round]{natbib}
\usepackage{xurl}
\usepackage[colorlinks=true,linkcolor=blue!50!black,citecolor=blue!50!black,urlcolor=blue!50!black]{hyperref}
\usepackage{enumitem}
\usepackage{booktabs}
\usepackage{setspace}
\setlist{itemsep=2pt,topsep=4pt}

\newif\ifblind
\blindfalse

\newtheorem{theorem}{Theorem}
\newtheorem{proposition}{Proposition}
\newtheorem{lemma}{Lemma}
\newtheorem{corollary}{Corollary}
\newtheorem*{proposition*}{Proposition}
\theoremstyle{definition}
\newtheorem{assumption}{Assumption}

\theoremstyle{remark}
\newtheorem*{remark*}{Remark}
\newtheorem{remark}{Remark}

\newcommand{\chiu}{\underline{\chi}}

\title{Certification under Cheap Generation:\\ What Journals Certify When GenAI Makes Papers Cheap}
\ifblind
\author{Anonymous authors}
\else
\author{Martin Bichler\thanks{School of Computation, Information and Technology, Technical University of Munich.} \and Ram D.\ Gopal\thanks{Warwick Business School, University of Warwick.}}
\fi
\date{November 2026}

\begin{document}
\maketitle

\begin{abstract}
\noindent Generative AI makes submissions to certifiers cheap to produce and to polish, while expert attention remains scarce. We model a digital certifier, with academic journals as the leading application, as an information intermediary that bundles AI-scalable technical validation with accountable human endorsement of significance, and ask when this bundle strengthens incentives for substantive work relative to an AI validation platform. Three theorems organise the analysis. First, the certifier raises net substantive output if and only if its marginal reward to presentation, relative to substance, lies below a cost-dependent threshold; the condition holds for any smooth single-index endorsement rule, and with competition for resources held fixed, cheaper presentation lowers the threshold. Second, committing to disregard presentation becomes more valuable as presentation becomes cheaper: in a genuine-only pool, the least self-consistent linear weight on presentation weakly rises as its cost falls, so best linear prediction need not restore the condition. Third, endorsed junk makes credibility a coordination outcome with a stable collapse equilibrium and discontinuous tipping; when readers also value the substance behind the certificate, a review norm that rewards presentation beyond a cost threshold enlarges the basin of collapse relative to a certificate valued by its genuine share alone. AI validity screening complements human endorsement exactly when the threshold condition holds, and the relative value of significance triage is the ratio of marginal to flawed submissions, adjusted for deterrence and endorsement leakage. Adaptive review weights amplify leakage when low-value submissions are more polished than genuine work and polish receives positive weight. With endogenous participation, the policy inducing more substantive work and a more credible endorsement can draw fewer genuine submissions.

\medskip
\noindent\textbf{Keywords:} peer review; certification; generative AI; human--AI complementarity; information intermediaries
\end{abstract}

\section{Introduction}

Generative AI lowers the cost of producing research and of presenting it well. Writing, coding, data analysis and routine checks become cheaper, and repositories already provide rapid dissemination. What remains scarce is expert and accountable attention. Journals therefore compete increasingly on the credibility of their endorsement, and they must decide how to divide evaluation between AI tools and human reviewers. The same design problem arises for any information intermediary that certifies outputs whose observable presentation is becoming cheap to improve: app stores, freelance and code-sharing platforms, grant panels and marketplaces with quality badges. We develop the model for academic journals, the leading application, and state it in primitives that apply to any such digital certifier (Table~\ref{tab:mapping}). We ask when such a certificate rewards substantive work, which AI capability relieves its human bottleneck, and when its credibility survives a flood of cheap submissions.

The shock is under way. Submissions to \emph{Organization Science} rose by about 42\% after the release of ChatGPT \citep{gartenberg2026}, and large language models have measurably increased scientific production \citep{kusumegi2025}. Weak manuscripts now arrive fluent and with citations that survive automated checks \citep{atlantic2026}. Cheap submissions come in two kinds: papers reporting research that was never performed, which integrity checks can catch, and papers produced by systems that actually ran their experiments, whose merit turns on significance \citep{skuse2026}. Autonomous research agents already produce the second kind at scale: their studies score well on rigor but below human work on novelty and significance, and text-only review struggles to separate flawed from corrected drafts.\footnote{In a twelve-week experiment, 936 autonomous AI agents deployed by 35 human-run laboratories produced 936 studies with code and data. In blinded comparisons the agents' studies scored higher than recent human preprints on experimental rigor and on how well claims were supported by measurements, while the human preprints scored higher on novelty and significance. In a journal founded by the agents, text-only referees flagged 11 of 16 flawed drafts but also 10 of 16 corrected versions \citep{howarth2026,ahmad2026}.} Certifiers are beginning to unbundle what a certificate says: a pilot of the Discovery Stack model scored manuscripts separately on Quality and Impact and found Quality scores more consistent across reviewers \citep{mcgargill2026,sciencenews2026}.

These observations motivate the distinction on which our model rests. A journal's certificate bundles two services that cheap generation affects very differently. \emph{Technical validation} asks whether a paper meets a stated standard of correctness, consistency or reproducibility; AI increasingly scales this service. \emph{Editorial endorsement} asks whether a paper that reaches assessment is original and significant. It draws on scarce, accountable expert attention and, because technical screening is imperfect, can still endorse flawed work. The distinction parallels the separation of prediction from judgment in the economics of AI \citep{agrawal2019}: checking a stated standard is increasingly a prediction task, whereas the criteria of significance remain subject to deliberation and revision by a scholarly community. We do not assume that human reviewers judge significance more accurately than AI. What distinguishes a journal is that its significance judgments are made under community-governed criteria by people accountable for them.

\paragraph{Model and results.} Authors allocate effort between substance and presentation. Flawed and valid-but-marginal submissions enter through separate low-cost production technologies. AI screening and human review determine delay and the credibility of endorsement, defined as the genuine share of the endorsed pool. Three theorems organise the analysis.
\begin{enumerate}
\item \emph{When certification rewards substance} (Theorem~\ref{thm:look}). Relative to a validation platform with the same screen, the journal raises net epistemic output if and only if reviewers' marginal reward to presentation, relative to substance, lies below a threshold $L$ that depends on production costs. The condition is not an artefact of a linear decision rule: it holds for any smooth endorsement rule that depends on a single index of what reviewers read and how the paper is presented, including threshold rules and fixed acceptance quotas. With competition for research resources held fixed, cheaper presentation lowers $L$, so a review process that was harmless when presentation was costly can become harmful without any change in reviewer behaviour.
\item \emph{Why commitment gains value} (Theorem~\ref{thm:rational}). A contribution-first rule that disregards presentation becomes more valuable as presentation becomes cheaper, at a quantifiable cost in selection accuracy. The mechanism is that rational reviewers need not restore the condition: when presentation scales with capability, the least self-consistent linear weight on presentation weakly rises as its cost falls. Whether this weight exceeds the threshold depends on parameters; we give a necessary condition, which requires noisy reading, a strong endorsement reward and presentation that is cheap relative to substance.
\item \emph{When credibility survives} (Theorem~\ref{thm:trap}). With endorsed junk, collapse is a stable equilibrium whose basin weakly expands with the junk load, and positive participation, where it exists at low junk loads, disappears discontinuously at a finite threshold. At given review weights and leakage rates, the mechanism survives when effort, both credibilities, delay and junk generation adjust jointly (Proposition~\ref{prop:integrated}). When readers value the certificate also by the substance of the genuine papers behind it, the incentive and credibility channels are coupled: at a given review norm that rewards presentation beyond $\kappa/\rho$, the threshold for substantive effort, this valuation lowers the certificate's worth relative to a certificate valued by its genuine share alone, enlarges the basin of collapse and lowers the tipping threshold (Proposition~\ref{prop:reputation}). The comparison holds the norm fixed and does not rank norms, because rewarding presentation also raises the private value of submitting (Remark~\ref{rem:norms}).
\end{enumerate}
Further results connect these mechanisms to the design of the review process. In the linear benchmark, AI validity screening and human endorsement are complements exactly when the threshold condition holds and screening raises credibility (Proposition~\ref{prop:complement}). Valid but marginal papers pass validity screens, so the marginal value of significance triage relative to validity screening equals the ratio of marginal to flawed submissions, adjusted for deterrence and, for credibility, for endorsement leakage (Proposition~\ref{prop:triage}). Inference links the credibility channel to the incentive channel: an adaptive review rule raises endorsement leakage when low-value submissions are more polished than genuine work and polish receives positive weight, and weakly lowers it otherwise. Whether adaptive review or commitment to disregard presentation yields more substantive output and higher credibility thus depends on the polish of the junk a journal receives (Proposition~\ref{prop:adaptive}). With endogenous participation, fewer genuine submissions can coexist with more aggregate substantive work and a more credible endorsement; Section~\ref{sec:policy-equilibrium} characterises this comparison and illustrates a reversal of the policy ranking as the cost of presentation changes.
The design implications are concrete: separate the basis for endorsement from presentation through contribution-first review forms when presentation is cheap; allocate AI budgets by the composition of the submission stream; and monitor endorsed junk and participation rather than average publication quality, which can rise while aggregate output falls (Proposition~\ref{prop:aggregate}).

The paper contributes to research on information intermediaries and human--AI evaluation systems in three ways. First, certification badges and reputation systems resolve information asymmetries on digital platforms, and their value depends on how informative and how rare certification is \citep{dellarocas2003,elfenbein2015,hui2016}. We show how the credibility of a certificate becomes endogenous when generative AI makes low-value submissions cheap, and why a certifier can tip into a low-credibility equilibrium. Second, research on human--AI complementarity studies how AI advice changes human decisions and what unique knowledge humans contribute \citep{fugener2021}. We characterise complementarity at the level of the evaluation system: when AI screening and human endorsement reinforce each other in their effect on incentives, and which AI capability, validity screening or significance triage, relieves the human bottleneck. Third, generative AI changes participation and content production in online knowledge communities \citep{burtch2024}. Our participation results distinguish congestion of a certifier from dilution of its certificate and identify when the latter becomes self-reinforcing. Review rules, screening capabilities and monitoring metrics are design variables under the control of editors and platform operators, and our results are stated in terms of them.

\paragraph{Scope.} By substance we mean the work that determines what a paper establishes: execution (coding, data work, derivations) and judgment about which question to ask, how to answer it and how to interpret the results. We study a shock that lowers execution and presentation costs while holding the cost of this judgment fixed. This is a comparative-static benchmark, not a claim that AI cannot assist scientific judgment. Our outcome measure captures incentives for substantive work; it is not a complete measure of scientific significance or welfare. Reward slopes and credibility shares are reduced forms, the inference results concern best linear prediction, and the participation analysis holds the review architecture fixed. Each result states the quantities it holds fixed. The ICLR evidence in Section~\ref{sec:iclr} describes evaluation practices; it does not test the technology shock or the welfare comparisons.

The paper proceeds as follows. Section~\ref{sec:model} introduces the regimes, review rule and production costs. Section~\ref{sec:effort} derives the incentive condition and its extension to general endorsement rules. Section~\ref{sec:cheap} studies cheaper execution and presentation, best linear prediction and the value of commitment. Section~\ref{sec:review} analyses screening, attention, delay and adaptive review. Section~\ref{sec:participation} endogenises participation and credibility and compares the two review rules in equilibrium. Section~\ref{sec:empirics} discusses empirical implications and Section~\ref{sec:conclusion} implications for review design. Appendix~\ref{app:proofs} contains the proofs; Appendix~\ref{app:ext} contains three extensions: joint reductions in presentation cost and resource competition, saturated presentation, and significance triage with false exclusions. Appendices~\ref{app:policy-computation} and~\ref{app:iclr} report the numerical comparison of review rules and the ICLR estimates.

A companion Lean~4 formalisation covers selected algebraic steps of the linear benchmark: Lemmas~\ref{lem:effort}, \ref{prop:regimes}, \ref{prop:cheap}, \ref{lem:frontier} and~\ref{lem:capacity} and Theorem~\ref{thm:look}(a) and~(c), together with order-theoretic steps of Propositions~\ref{prop:complement}, \ref{prop:triage} and~\ref{prop:fixed} and Theorem~\ref{thm:trap}(a)--(d), under stated analytical hypotheses. Lemma~\ref{lem:projection}, Theorem~\ref{thm:rational}(a)--(b), Proposition~\ref{prop:junk}, Corollary~\ref{cor:invest}, Theorem~\ref{thm:trap}(e) and Propositions~\ref{prop:integrated}, \ref{prop:reputation} and~\ref{prop:aggregate} rely on the written proofs. Theorem~\ref{thm:look}(b), Theorem~\ref{thm:rational}(c)--(f), Proposition~\ref{prop:adaptive}, Lemma~\ref{lem:contested}, the refined tipping statement, the policy comparison in Section~\ref{sec:policy-equilibrium} and the extensions in Appendix~\ref{app:ext} follow from the written arguments and, for the policy illustration, the numerical computation. \ifblind\else The companion repository is \url{https://github.com/bichler/lean-genai-journal-certification}.\fi

\paragraph{Related literature.} The closest antecedent is \citet{ellison2002}, whose $q$--$r$ model separates the main contribution from other dimensions of paper quality and studies the evolution of evaluation standards. Our substance--presentation split uses this distinction. We hold review weights fixed or determine their ratio by best linear prediction, and examine a technological cost change. The additional questions concern the moving incentive threshold, its relation to an endogenous review rule, and the credibility of the pool that receives the certificate.

The effort-allocation mechanism follows \citet{holmstrom1991}; the value of a credible career signal relates to \citet{holmstrom1999}. Strategic classification and manipulable-signal models study how evaluation rules alter the behaviour that generates their data \citep{hardt2016,kleinberg2020,frankel2019,frankel2022,ball2025}. In particular, \citet{frankel2022} explain why a decision maker may benefit from committing to underuse manipulable data. We apply this incentive--information tension to a certificate with separate validity and endorsement components. Our additions are the response of the self-consistent weight to a fall in presentation cost, the resulting increase in the value of commitment, and the interaction between the review weight and the leakage of junk into the endorsed pool. We do not solve the certifier's general optimal information policy.

Certification and grading research examines disclosure, investment and participation \citep{lizzeri1999,ostrovsky2010,boleslavsky2015,harbaugh2018}. Our credibility channel shares these concerns. Its coordination structure relates to network effects and intermediation \citep{katz1985,caillaud2003}, and collapse resembles market unravelling \citep{akerlof1970}. The existence of a zero-participation equilibrium is not by itself our contribution. We characterise how endorsed junk changes its basin and the viability of positive participation when congestion is also present. Global-games methods \citep{morris1998} offer a possible route to equilibrium selection, which we do not pursue. Delay as a screening device and its comparison with fees follow \citet{leslie2005,cotton2013}.

Recent work addresses the AI submission flood through capacity, fees and quotas \citep{lopezlira2026,lawford2026,fourie2026}; \citet{fourie2026b} compares rejection without review with cheaper AI checking. \citet{achim2026} models a verification bottleneck with optimal editorial scrutiny and endogenous reviewer supply, and \citet{bergstrom2026} show how submission volume and review accuracy reinforce each other. Our contribution is complementary: we separate invalid from valid-but-marginal submissions, distinguish congestion from dilution of endorsement, and connect the participation problem to the incentive condition for substantive effort. Empirical work documents changes in production and review \citep{liang2024,thakkar2025,kusumegi2025,gartenberg2026,ahmad2026} and experiments with unbundled assessment \citep{mcgargill2026}.

\section{Model}\label{sec:model}

The model has three building blocks: institutional regimes that differ in how much technical screening and how much expert assessment they provide; a review rule that translates an author's substance and polish into a probability of endorsement; and authors who choose how to split effort between the two dimensions. Section~\ref{sec:review} later adds the review process and the composition of submissions.

\subsection{Institutional regimes}

We compare three regimes $\ell \in \{R, P, J\}$, described by an intensity of AI technical screening $r_\ell \ge 0$ and an intensity of expert, accountable editorial assessment $e_\ell \ge 0$. The baseline comparison sets AI-supported significance triage to zero. Section~\ref{sec:review} later adds a triage capability $\zeta \in [0,1)$ for a regime with a human review stage; thus $\zeta_R=\zeta_P=0$ and, when triage is considered, $\zeta_J=\zeta$:
\begin{itemize}
\item a repository ($R$): $r_R = e_R = 0$. It makes a paper visible.
\item an AI-validated platform ($P$): $r_P = r > 0$, $e_P = 0$. It adds AI-supported technical validation and may support novelty assessment, but it provides no accountable endorsement of significance. The agents' preprint server of \citet{ahmad2026}, where every reported number is traced to an output file, is an example.
\item a journal ($J$): $r_J = r \ge 0$, $e_J = e > 0$. It adds an expert and accountable assessment of significance. This assessment may be supported by AI, but its criteria are governed by the relevant scholarly community. We do not assume that human reviewers judge significance more accurately than AI. Human reviewers observe substance with noise and may respond to polish. What distinguishes a journal is that its significance judgments are made under criteria that a scholarly community governs, and by people accountable for them.
\end{itemize}

\paragraph{Other digital certifiers.} The three regimes have counterparts wherever a platform certifies submissions whose presentation is becoming cheap to improve. A repository corresponds to open listing, a validation platform to automated checks that a submission meets a stated standard, and a journal to accountable human curation on top of such checks. Table~\ref{tab:mapping} maps the model's primitives for app stores and marketplace quality badges, settings that the literature on certification and reputation on digital platforms studies \citep{dellarocas2003,elfenbein2015,hui2016}. What the model requires of the endorsing stage is that its criteria are not set by the submitter and that it is accountable for what it endorses. We use journal terminology throughout because journals are our leading application; the results are stated in terms of the primitives and apply to other certifiers to the extent that the mapping holds.

\begin{table}[htbp]
\centering
\caption{Model primitives for three digital certifiers}
\label{tab:mapping}
\small
\renewcommand{\arraystretch}{1.1}
\begin{tabular}{@{}p{0.19\textwidth}p{0.245\textwidth}p{0.245\textwidth}p{0.245\textwidth}@{}}
\toprule
Primitive & Journal & App store & Marketplace badge \\
\midrule
Validation $r$ & code execution, consistency and reproducibility checks & automated security and policy scans & automated identity and transaction checks \\
Endorsement $e$ & accountable judgment of significance & human curation, featured placement & human-verified quality badge \\
Substance $x$ & what the paper establishes & functionality and reliability & product or service quality \\
Presentation $y$ & writing and exposition & screenshots, descriptions, store optimisation & photos, descriptions, listing optimisation \\
Flawed stream $N_f$ & fabricated or erroneous results & malware, non-functional apps & counterfeit or fraudulent listings \\
Marginal stream $N_m$ & valid but insignificant papers & functional clone or template apps & valid but undifferentiated generated listings \\
Leakage $\psi$ & endorsing flawed or marginal work & featuring clones & badges for low-quality sellers \\
\bottomrule
\end{tabular}
\end{table}

\subsection{Endorsement and reward slopes}

A genuine author chooses substantive effort $x \ge 0$ and presentation effort $y \ge 0$. Over the relevant range of this production technology, regime $\ell$ endorses the paper with probability
\begin{equation}\label{eq:endorse}
\pi_\ell(x,y) = e_\ell\,(\pi_0 + s_x x + s_y y),
\end{equation}
where $\pi_0 \ge 0$, $s_x > 0$ is reviewers' sensitivity to substance and $s_y \in \mathbb{R}$ their sensitivity to polish. If $s_y < 0$, reviewers penalise polish; if $s_y=0$, they ignore it; if $s_y > 0$, they reward it. We use ``look through polish'' for $s_y\leq0$, while the condition for beneficial certification derived below can also accommodate a modest positive weight. A language-editing experiment shows that presentation can affect evaluation: economists gave edited versions of working papers higher ratings and were more willing to recommend them for a conference \citep{feld2024}. Language editing can also make the contribution easier to assess, so the experiment does not isolate the effect of polish in our narrower sense. Sections~\ref{sec:effort}--\ref{sec:cheap} first treat $(s_x,s_y)$ as primitives; Section~\ref{sec:bayes} derives their ratio from a best linear predictor. A regime without editorial assessment ($e_\ell = 0$) endorses nothing. We maintain that, for every regime and genuine author type considered, the optimal choices satisfy $0 \le \pi_\ell(x_\theta^{\ell*},y_\theta^{\ell*}) \le 1$. Equation~\eqref{eq:endorse} describes incentives within the genuine production technology; it is not extrapolated to a marginal or fabricated paper with true substance set to zero. Section~\ref{sec:review} specifies the erroneous endorsements of those papers separately.

A regime's certificate has two components, each with its own credibility, derived in Section~\ref{sec:review}. An endorsement is worth $\gamma_\ell V$ to its recipient, where $V > 0$ is the career value of a credible endorsement and the \emph{endorsement credibility} $\gamma_\ell \in [0,1]$ is the share of genuine contributions among the papers the regime endorses: endorsing the significance of an invalid or worthless result is worth nothing, so the market discounts endorsements accordingly. If readers observe only whether a paper is endorsed, $\gamma_\ell$ is exactly the posterior probability that an endorsed paper is genuine, and $\gamma_\ell V$ is the expected value of an endorsement when endorsed junk is worth nothing. Technical validation contributes $\gamma^T_\ell \alpha_T r_\ell$ per unit of substance, where $\alpha_T>0$ and the \emph{validation credibility} $\gamma^T_\ell \in [0,1]$ is the share of technically valid papers among those that pass the regime's screen. The checks that AI performs (code execution, consistency, reproducibility, proof checking) verify the substantive components of a paper, so validation is worth more to papers with more substance. Because a valid but marginal paper is valid, it dilutes $\gamma_\ell$ but not $\gamma^T_\ell$. The career value of a paper is
\begin{equation}\label{eq:W}
W_\ell(x,y) = v_\ell + A_\ell x + B_\ell y,
\end{equation}
where $v_\ell$ collects the value of visibility and the effort-independent part of endorsement. The reward slopes are
\begin{equation}\label{eq:slopes}
A_\ell = a_0 + \gamma^T_\ell\alpha_T r_\ell + \gamma_\ell\,\alpha_E e_\ell, \qquad B_\ell = b_0 - \gamma_\ell\, \beta_E\, e_\ell, \qquad\text{with } \alpha_E := V s_x,\ \ \beta_E := -V s_y,
\end{equation}
with $a_0, b_0 > 0$ the rewards readers attach to substance and polish without certification; the effort-independent term $\gamma_\ell V e_\ell\pi_0$ enters $v_\ell$. The sign of $\beta_E$ is unrestricted: $\beta_E \ge 0$ if reviewers look through polish ($s_y\le0$) and $\beta_E < 0$ if they reward it ($s_y>0$). Since the platform does not endorse, its endorsement credibility plays no role. When $P$ and $J$ apply the same screen to pools with the same share of flawed papers ($\gamma^T_P = \gamma^T_J$), $A_J - A_P = \gamma\alpha_E e$ and $B_J - B_P = -\gamma\beta_E e$, with $\gamma := \gamma_J$. Whether the journal rewards polish less than the platform is governed by $\beta_E$.

\paragraph{What the reduced forms do and do not drive.} Three reduced forms carry the analysis. First, endorsement credibility is the posterior of a reader who observes only the certificate; readers who observe more than the certificate would discount it differently. Section~\ref{sec:reputation} lets readers also value the substance of the genuine papers behind the certificate. Second, the endorsement probability~\eqref{eq:endorse} is affine over the relevant range of genuine papers, and the restriction $\pi_\ell\in[0,1]$ prevents extrapolation beyond that range. By Theorem~\ref{thm:look}(b), the sign of the comparison between journal and platform depends on the endorsement rule only through the ratio of the marginal endorsement values of substance and presentation, which equals the relative weight $t$ under any single-index rule, including thresholds and quotas. The affine form, and the slope $s_x$ held fixed in Section~\ref{sec:bayes}, therefore determine magnitudes, not the direction of the comparison at a given weight; they do affect effort levels and, through them, the self-consistent weight. Third, the relative rates at which junk is endorsed, $\psi_f$ and $\psi_m$ in Section~\ref{sec:review}, are held fixed except in Section~\ref{sec:adaptive}, where $\psi_m$ is derived jointly with the review weight. The editorial intensity $e$ scales the return to endorsement and, below, human time per manuscript; it does not parameterise screening accuracy or leakage, which are represented separately. Each result states which of these quantities it holds fixed.

\subsection{Authors, costs and cheap generation}

Authors differ in research capability $\theta > 0$. The cost of producing a paper is
\begin{equation}\label{eq:cost}
C_\theta(x,y) = \frac{1}{2\theta}\left(\chi x^2 + 2\rho xy + \kappa y^2\right), \qquad \rho \ge 0,
\end{equation}
where $\chi$ is the cost of substance, $\kappa$ the cost of presentation, and $\rho$ captures competition between the two for an author's time. The quadratic form gives convex effort costs and therefore diminishing marginal returns, while preserving a closed-form solution for effort. Its cross term makes time spent improving one dimension more costly when the author is also investing in the other. We decompose $\chi = \chiu + \chi_X$, where $\chi_X \ge 0$ is the cost of execution (coding, calculation, data work, drafting) and $\chiu > 0$ is the cost of producing and interpreting an original idea that remains after execution costs are removed. The technology shock lowers $\chi_X$ (cheap execution) and $\kappa$ (cheap polish) while holding $\chiu$ fixed. Call a cost of polish \emph{admissible} if $\kappa > \rho^2/\chiu$. We restrict attention to admissible $\kappa$, so $D := \chi\kappa - \rho^2 > 0$ for every $\chi \ge \chiu$. Authors maximise $U_\theta^\ell(x,y) = W_\ell(x,y) - C_\theta(x,y)$.

To assess scientific value, define net epistemic output
\begin{equation}\label{eq:Q}
Q_\theta^\ell = x_\theta^{\ell*} - \lambda\, y_\theta^{\ell*}, \qquad \lambda \ge 0,
\end{equation}
where $\lambda$ values the residual social burden of presentation rewarded beyond its contribution to communicating the research. The measure is an effort-based accounting criterion, not a complete welfare function or a direct measure of scientific significance. It excludes research and review costs other than those represented by the presentation term, and it does not value clear communication negatively. At $\lambda=0$, the criterion is substantive effort alone, $Q=x$; results that hold in that case do not depend on a social-waste penalty. A positive $\lambda$ is a maintained valuation of the residual burden, not an estimate from the review-score data. Because a reduction in production cost can also change that burden, comparisons with $\lambda$ fixed should be read as local comparative statics. We report separately below when lower presentation cost reduces $x$ and when it reduces only this net-output measure.

\begin{assumption}[Interior effort]\label{ass:interior}
In every regime considered, $\rho B_\ell < \kappa A_\ell$ and $\rho A_\ell < \chi B_\ell$.
\end{assumption}

Since $\rho \ge 0$ and $D > 0$, Assumption~\ref{ass:interior} implies $A_\ell > 0$ and $B_\ell > 0$: in an interior regime the market still rewards polish at the margin, however strongly reviewers look through it.

\paragraph{Analytical strategy.} We first solve the authors' quadratic effort-allocation problem in closed form (Lemma~\ref{lem:effort}). We then compare the reward slopes induced by a journal and a validation platform to derive the look-through condition, study how cheap polish and cheap execution change effort and the value of certification, extend the condition to general endorsement rules, and ask whether self-consistent linear review rules would undo the result and what commitment to a different rule is worth. Next, we differentiate the review-load, capacity, and credibility expressions to compare validity screening with significance triage, and link review weights to the leakage of junk. Finally, we characterise participation and credibility as a fixed-point problem, first with exogenous junk and then in an integrated model. The functional forms are intended to identify these mechanisms transparently, not to claim that every journal literally operates according to them.

\paragraph{Modelling choices.} Three choices deserve comment. First,
genuine contributions and junk come from different technologies.
Genuine authors choose substance and presentation, and even a
low-capability author's paper carries positive substance. Flawed and
marginal papers are produced by mass generation at low cost, with
substance close to zero; they enter as separate streams in
Section~\ref{sec:review}. A continuum of paper types would blur this
boundary, but the mechanisms below require only that the endorsed pool
contains papers whose substance falls well short of genuine
contributions. The separation excludes a margin that cheap generation
makes salient: genuine authors can themselves produce marginal papers,
alongside their main work or instead of it. Theorems~\ref{thm:look}
and~\ref{thm:rational} concern the effort of authors who submit genuine
work and do not depend on who produces the marginal stream. The
participation results depend on it only through the response of the
junk load to participation; Remark~\ref{rem:genuine-marginal} states
which cases the integrated model covers. Second, we treat the reviewers' weights as primitives or
derive them from inference rather than letting the journal choose them.
A journal that could commit to a review rule could rank papers by the
reading signal alone ($t = 0$), which satisfies the look-through
condition (Theorem~\ref{thm:look}(c)). By Lemma~\ref{lem:projection},
however, such a rule discards information about substance whenever
polish and substance covary, so the journal trades selection accuracy
for incentives, as in the literature on manipulable signals
\citep{frankel2019,frankel2022}. In practice the rule is implemented by
reviewers, whose weights the journal shapes only indirectly, for
instance through separate scoring or review forms; this is why we
model the weights as behaviour rather than as a choice, and treat
commitment as a design option whose value we quantify
(Theorem~\ref{thm:rational}(c)--(e)). Third, we
consider a single journal and one outside option. The outside option
can be read as the next-best certifier available to an author. With
competing journals, a journal that loses credibility would partly
redistribute authors rather than lose them; the aggregate effects in
Section~\ref{sec:participation} then depend on how much weaker the
incentives of the next-best certifier are ($q_J > q_o$ in
Proposition~\ref{prop:aggregate}).

\section{Effort allocation and certification}\label{sec:effort}

Because career rewards are linear and effort costs are quadratic, optimal effort has a closed form. This makes transparent how each regime's reward slopes translate into substance and polish, and hence into net epistemic output.

\begin{lemma}[Optimal effort]\label{lem:effort}
Under Assumption~\ref{ass:interior}, the unique maximiser of $U_\theta^\ell$ is
\begin{equation}\label{eq:effort}
x_\theta^{\ell*} = \frac{\theta(\kappa A_\ell - \rho B_\ell)}{D}, \qquad y_\theta^{\ell*} = \frac{\theta(\chi B_\ell - \rho A_\ell)}{D}.
\end{equation}
Net epistemic output and the author's indirect utility are
\begin{equation}\label{eq:qPhi}
\begin{gathered}
Q_\theta^\ell = \theta q_\ell, \qquad q_\ell := \frac{(\kappa + \lambda\rho)A_\ell - (\rho + \lambda\chi)B_\ell}{D};\\
U_\theta^{\ell*} = v_\ell + \theta\Phi_\ell, \qquad \Phi_\ell := \frac{\kappa A_\ell^2 - 2\rho A_\ell B_\ell + \chi B_\ell^2}{2D} > 0.
\end{gathered}
\end{equation}
\end{lemma}

Effort on each dimension rises with its own reward and falls with the reward to the competing dimension, to the extent that the two compete for the author's time ($\rho > 0$). Net epistemic output is proportional to capability $\theta$ and linear in the two reward slopes. Comparing regimes therefore reduces to comparing how they shift these slopes.

\begin{lemma}[Regime comparison]\label{prop:regimes}
Under Assumption~\ref{ass:interior}, regime $m$ yields higher net epistemic output than regime $\ell$, for every $\theta$, if and only if
\begin{equation}\label{eq:compare}
(\kappa + \lambda\rho)(A_m - A_\ell) > (\rho + \lambda\chi)(B_m - B_\ell).
\end{equation}
If $\rho+\lambda\chi=0$, the condition reduces to $A_m>A_\ell$. If $\rho+\lambda\chi>0$, it holds whenever $A_m\ge A_\ell$ and $B_m\le B_\ell$, with at least one strict inequality. It fails whenever $A_m\le A_\ell$ and $B_m\ge B_\ell$. In the remaining cases the two changes have the same non-zero sign, and their magnitudes determine the outcome.
\end{lemma}

A regime is better for science if the additional reward it places on substance outweighs the additional reward it places on polish, where the weights reflect how strongly each reward moves effort toward or away from substance. Applied to the journal and the platform, condition~\eqref{eq:compare} becomes a condition on the reviewers alone. The affine endorsement probability~\eqref{eq:endorse} delivers closed forms, but journals typically accept a fixed share of submissions or apply a threshold to an overall assessment. The first theorem therefore states the condition both for the linear benchmark and for general endorsement rules. For the latter, let the journal endorse a genuine paper with effort $(x,y)$ with expected probability $\Pi(x,y)\in[0,1]$, differentiable in $(x,y)$, so that the career value at $J$ is $v_J + A_Px + B_Py + \gamma V\,\Pi(x,y)$, where $A_P$ and $B_P$ are the platform's reward slopes. At an author's optimum $(x_\theta^{J*},y_\theta^{J*})$ at $J$, write $g_x := \gamma V\,\partial\Pi/\partial x$ and $g_y := \gamma V\,\partial\Pi/\partial y$ for the marginal endorsement values of substance and presentation.

\begin{theorem}[When certification rewards substance]\label{thm:look}
Suppose $J$ and $P$ apply the same screen with the same validation credibility, $r_J = r_P$ and $\gamma^T_J = \gamma^T_P$, and let $\gamma := \gamma_J > 0$. Let
\[
L(\kappa,\chi) := \frac{\kappa + \lambda\rho}{\rho + \lambda\chi}\quad\text{if } \rho+\lambda\chi>0,\qquad L(\kappa,\chi):=+\infty\ \text{otherwise}.
\]
\begin{enumerate}[label=(\alph*)]
\item \emph{Linear benchmark.} Under Assumption~\ref{ass:interior}, $Q^J_\theta > Q^P_\theta$ if and only if
\begin{equation*}\tag{C}\label{eq:C}
(\kappa + \lambda\rho)\,\alpha_E + (\rho + \lambda\chi)\,\beta_E > 0,
\end{equation*}
that is, if and only if $s_y/s_x < L(\kappa,\chi)$. The sign of the comparison does not depend on credibility $\gamma$ or on the scale of human review $e$; these determine only its magnitude, $Q^J_\theta - Q^P_\theta = \theta\gamma e\,[(\kappa+\lambda\rho)\alpha_E + (\rho+\lambda\chi)\beta_E]/D$.
\item \emph{General endorsement rules.} Suppose an author of capability $\theta$ has an interior maximiser $(x_\theta^{J*},y_\theta^{J*})$ at $J$ and that Assumption~\ref{ass:interior} holds at $P$. Then
\begin{equation}\label{eq:link}
\begin{gathered}
x_\theta^{J*}-x_\theta^{P*}=\frac{\theta(\kappa g_x-\rho g_y)}{D},\qquad
y_\theta^{J*}-y_\theta^{P*}=\frac{\theta(\chi g_y-\rho g_x)}{D},\\
Q_\theta^J-Q_\theta^P=\frac{\theta\bigl[(\kappa+\lambda\rho)g_x-(\rho+\lambda\chi)g_y\bigr]}{D}.
\end{gathered}
\end{equation}
Hence, if $g_x>0$ and $\rho+\lambda\chi>0$, then $Q_\theta^J>Q_\theta^P$ if and only if $g_y/g_x<L(\kappa,\chi)$. In particular, suppose the journal applies an arbitrary non-decreasing rule to the index $s+t\tilde y$ of Section~\ref{sec:bayes}, for instance endorsement above a cutoff or within a fixed acceptance quota, and that $\varepsilon+tu$ has a bounded continuous density. Then $\Pi(x,y)=\Pi_1(x+ty)$, and wherever $\Pi_1'>0$ at the optimum, $g_y/g_x=t$: the journal improves on the platform for author $\theta$ if and only if $t<L(\kappa,\chi)$.
\item \emph{Cheap polish raises the bar.} Hold reward slopes, credibility terms and the review architecture fixed, and suppose $\rho + \lambda\chi > 0$. The threshold $L(\kappa,\chi)$ is strictly increasing in $\kappa$ and non-increasing in $\chi_X$, strictly if $\lambda > 0$. If reviewers look through polish ($s_y \le 0$), the journal improves on the platform at every cost of polish. If they are swayed by polish ($s_y > 0$), it does so if and only if
\begin{equation}\label{eq:kappahat}
\kappa > \hat\kappa := (\rho + \lambda\chi)\frac{s_y}{s_x} - \lambda\rho ,
\end{equation}
and cheaper execution weakly lowers $\hat\kappa$, strictly if $\lambda>0$. By part~(b), these statements hold with $s_y/s_x$ replaced by $t$ for any single-index endorsement rule, for every author at whose optimum $\Pi_1'>0$.
\end{enumerate}
\end{theorem}

Part~(a) makes precise the sense in which a journal is not automatically beneficial simply because it exists. We call~\eqref{eq:C} the look-through condition. It is weaker than looking through polish in the narrow sense $s_y\le0$: reviewers may be mildly swayed by polish ($s_y > 0$), provided their sensitivity to it, relative to substance, stays below $L$. Credibility amplifies whatever the reviewers reward. It cannot turn a harmful reward structure into a beneficial one.

Part~(b) shows that the condition is not an artefact of the affine rule. What matters is the ratio of the marginal endorsement values of substance and presentation at the author's optimum. Because these values are evaluated at the journal's own optimum, \eqref{eq:link} is an exact comparison, not a local approximation. Under a single-index rule the ratio $g_y/g_x$ equals the relative weight $t$ for every author, although the levels $g_x$ differ across authors; with a quota, the cutoff is an equilibrium object that each author takes as given. The threshold $L$ and the cost $\hat\kappa$ at which it binds therefore apply unchanged. What the affine benchmark adds are closed-form magnitudes, effort that is proportional to capability, which Section~\ref{sec:bayes} uses to derive the self-consistent weight, and the cross-partial in Proposition~\ref{prop:complement}(a). We retain the benchmark for these results. Part~(c) is discussed in Section~\ref{sec:bar}.

\begin{remark}[Corner solutions]\label{rem:corner}
If $\chi B_\ell \le \rho A_\ell$, the author sets $y = 0$ and $x = \theta A_\ell/\chi$, so $Q_\theta^\ell = \theta A_\ell/\chi$. Symmetrically, if $\kappa A_\ell \le \rho B_\ell$, then $x = 0$ and $Q_\theta^\ell = -\lambda\theta B_\ell/\kappa$. A journal that pushes $B_J$ below $\rho A_J/\chi$ eliminates presentation effort entirely; further reductions in $B_J$ have no effect on effort or output.
\end{remark}

\section{Cheap generation}\label{sec:cheap}

We now ask how generative AI, by lowering the costs of execution and of polish, changes research effort and the value of certification. Sections~\ref{sec:direct} and~\ref{sec:bar} hold the reward slopes fixed, so the results isolate the direct effect of cheaper production. Sections~\ref{sec:bayes} and~\ref{sec:commit} let the reviewers' weights respond.

\subsection{Direct effects}\label{sec:direct}

\begin{lemma}[Cheap generation, effort and the value of certification]\label{prop:cheap}\label{prop:value}
Hold reward slopes, credibility terms and the review architecture fixed, and let Assumption~\ref{ass:interior} hold.
\begin{enumerate}[label=(\alph*)]
\item \emph{Effort.} In any regime,
\begin{align}
\frac{\partial x^*}{\partial \kappa} &= \frac{\rho\, y^*}{D} \ge 0, &
\frac{\partial y^*}{\partial \kappa} &= -\frac{\chi\, y^*}{D} < 0, &
\frac{\partial Q}{\partial \kappa} &= \frac{(\rho + \lambda\chi)\, y^*}{D} \ge 0, \label{eq:dkappa}\\
\frac{\partial x^*}{\partial \chi_X} &= -\frac{\kappa\, x^*}{D} < 0, &
\frac{\partial y^*}{\partial \chi_X} &= \frac{\rho\, x^*}{D} \ge 0, &
\frac{\partial Q}{\partial \chi_X} &= -\frac{(\kappa + \lambda\rho)\, x^*}{D} < 0. \label{eq:dchi}
\end{align}
Cheaper polish (lower $\kappa$) lowers net epistemic output, strictly whenever $\rho + \lambda\chi > 0$. Cheaper execution (lower $\chi_X$) raises it, but only up to a bound, because $\chi \ge \chiu$.
\item \emph{Value of certification.} Let $\Delta^{m\ell} := Q_\theta^m - Q_\theta^\ell$ for two regimes. Then
\begin{equation}\label{eq:dDelta}
\frac{\partial \Delta^{m\ell}}{\partial \kappa} = \frac{\rho + \lambda\chi}{D}\left(y_\theta^{m*} - y_\theta^{\ell*}\right), \qquad
\frac{\partial \Delta^{m\ell}}{\partial \chi_X} = -\frac{\kappa + \lambda\rho}{D}\left(x_\theta^{m*} - x_\theta^{\ell*}\right).
\end{equation}
If $\rho + \lambda\chi > 0$, cheaper polish increases the advantage of regime $m$ if and only if $m$ induces less presentation effort than $\ell$. Cheaper execution increases the advantage of $m$ if and only if $m$ induces more substantive effort.
\end{enumerate}
\end{lemma}

Cheaper polish makes presentation effort more attractive, and because presentation competes with substance for the author's time, substance falls. Cheaper execution works in the opposite direction, but its benefit is bounded because AI can reduce the cost of substance only down to the irreducible cost $\chiu$ of producing and interpreting an original idea. A certificate gains value from cheap polish if it steers authors away from polish, and from cheap execution if it steers them toward substance. For the journal against the platform (under the conditions of Theorem~\ref{thm:look}(a)), $y_\theta^{J*} - y_\theta^{P*} = -\theta\gamma e\,(\chi\beta_E + \rho\alpha_E)/D$. A journal whose reviewers look through polish ($\beta_E \ge 0$) weakly gains from cheap polish, strictly whenever it induces strictly less presentation effort and $\rho + \lambda\chi > 0$.

The fixed-$\rho$ benchmark treats cheaper polish as cheaper but still time-consuming. Appendix~\ref{app:jointcost} lets presentation cost and resource competition fall together: along such a path, cheaper presentation lowers substantive effort if and only if the reduction in presentation cost is large relative to the reduction in competition for the author's time. Neither a decline in substantive effort nor a decline in the net-output measure is an unconditional consequence of cheaper writing.

\begin{remark}[Substantive effort and significance]\label{rem:significance}
Net epistemic output~\eqref{eq:Q} is an effort-based measure. The scientific significance of the resulting contribution is a separate consideration. Holding reward slopes and the review architecture fixed, a reduction in $\chiu$ would lower $\chi$ and, within the interior regime, raise $x^*$ by the same mechanism as cheaper execution. If the same technological change also lowers the cost of producing marginal papers, it can increase $N_m$ through~\eqref{eq:generation}, which raises the relative value of significance triage, other things equal (Proposition~\ref{prop:triage}).
\end{remark}

\subsection{Cheap polish raises the bar}\label{sec:bar}

Theorem~\ref{thm:look}(c) holds reward slopes, credibility terms and the review architecture fixed. Its content is that, if reviewers are swayed by polish, a fall in the cost of polish below $\hat\kappa$ turns a beneficial journal into a harmful one, although reviewers' behaviour has not changed. This statement has content when $\hat\kappa$ lies in the admissible range, $\hat\kappa > \rho^2/\chiu$, and Assumption~\ref{ass:interior} holds for both regimes on the range of $\kappa$ considered. If $\hat\kappa \le \rho^2/\chiu$, the journal improves on the platform at every admissible cost of polish.

When polish was costly, a review process that was moderately impressed by good writing did little harm, because authors could not cheaply supply what impressed it. When polish is nearly free, the same review process rewards a margin that every author can now exploit, and the tolerance for susceptibility to presentation shrinks. This does not imply that clear writing lacks value; the concern is presentation rewarded beyond its contribution to communicating the research.

\begin{remark}[What ``cheap polish'' means]\label{rem:polish}
Lemma~\ref{prop:cheap} lowers $\kappa$ holding $\rho$ fixed: polish becomes cheaper but still competes for an author's time. If generative AI makes polish effectively push-button, $\rho$ falls as well. At $\rho = 0$, $\partial x^*/\partial\kappa = 0$, and cheap polish no longer crowds out substance; it lowers output only through the social-waste term $\lambda$. If the market rewards polish positionally, the private return $b_0$ remains but the social return is nil, which is what $\lambda > 0$ captures. Presentation that improves communication beyond its private return would correspond to extending definition~\eqref{eq:Q} to $\lambda < 0$. Lemma~\ref{prop:regimes}, Theorem~\ref{thm:look}(a) and Lemma~\ref{prop:cheap} continue to hold under this extension as long as $\kappa + \lambda\rho > 0$ and $\rho + \lambda\chi > 0$; only the comparative statics of $L$ and $\hat\kappa$ in $\chi_X$ in Theorem~\ref{thm:look}(c) reverse sign, because $\partial L/\partial\chi$ has the sign of $-\lambda$.
\end{remark}

\subsection{Do rational reviewers look through polish?}\label{sec:bayes}

Theorem~\ref{thm:look}(c) holds the reviewers' weights fixed while the cost of polish falls. A natural objection is that reviewers should adapt: when every paper is polished, polish stops being informative. We examine this objection within the class of linear prediction rules. The best linear predictor minimises mean squared prediction error among affine functions of the observed signals; without additional distributional assumptions it need not equal the Bayesian conditional expectation. Thus the results below establish self-consistency within this class, rather than characterise all rational review mechanisms. Among genuine papers, the response depends on how presentation varies with capability. In a mixed pool, it also depends on the prevalence of polished marginal papers.

A reviewer reads a paper and obtains a noisy signal of its substance, $s = x + \varepsilon$, where $\varepsilon$ has mean zero and variance $\sigma_\varepsilon^2 > 0$ and is independent of everything else; $\sigma_\varepsilon^2$ falls with the care of the reading. The reviewer also observes presentation $\tilde y = y + u$, where $u$ is an endowment that the author does not choose (writing talent, language, access to editing), with mean zero and variance $\sigma_u^2 > 0$, independent of capability and, in a mixed pool, paper class. The reviewer predicts substance by the best linear predictor $\hat x$ given $(s,\tilde y)$, using the distribution in the pool under review. An institutional probability rule, specified below, translates its ranking into endorsements.

\begin{lemma}[Linear-prediction weight on polish]\label{lem:projection}
Let $\mathcal V_x := \mathrm{Var}(x) > 0$, $\mathcal V_y := \mathrm{Var}(\tilde y)$ and $C := \mathrm{Cov}(x,\tilde y)$ in the reviewed pool, with $\mathcal V_x\mathcal V_y > C^2$. The best linear predictor is $\hat x = c + w_s s + w_y \tilde y$ with
\begin{equation}\label{eq:projection}
w_s = \frac{\mathcal V_x\mathcal V_y - C^2}{\Delta_{\mathrm{pred}}} > 0, \qquad w_y = \frac{\sigma_\varepsilon^2\, C}{\Delta_{\mathrm{pred}}}, \qquad \Delta_{\mathrm{pred}} := (\mathcal V_x + \sigma_\varepsilon^2)\mathcal V_y - C^2 > 0 .
\end{equation}
A review rule that ranks papers by $\hat x$ ranks them by $s + t\,\tilde y$ with
\begin{equation}\label{eq:t}
t := \frac{w_y}{w_s} = \frac{\sigma_\varepsilon^2\, C}{\mathcal V_x\mathcal V_y - C^2},
\end{equation}
Consider an institutional rule that endorses with probability $e(\pi_0' + s_x(s + t\tilde y))$, where $s_x>0$ and $\pi_0'\ge0$ are specified independently of the prediction coefficients. Suppose this probability lies in $[0,1]$ on the relevant signal support. An author choosing $(x,y)$ then has expected endorsement probability~\eqref{eq:endorse} with $\pi_0=\pi_0'$ and $s_y=t\,s_x$. In particular, $s_y$ has the sign of $\mathrm{Cov}(x,\tilde y)$ in the reviewed pool.
\end{lemma}

Ranking by $\hat x$ is equivalent to ranking by $s+t\tilde y$, because $w_s>0$. This ordinal equivalence does not determine acceptance probabilities or marginal incentives. We therefore hold the institutional slope $s_x$ of the affine probability map fixed while inference determines $t=s_y/s_x$. Holding $s_x$ fixed restricts magnitudes: a fixed acceptance quota, a changing threshold, or a nonlinear probability map could induce different effort levels. For a given weight $t$, the comparison with the platform depends on the rule only through $t$ (Theorem~\ref{thm:look}(b)). Under this implementation, the rule penalises or ignores polish ($t\le0$) exactly when polish and substance have non-positive covariance in the reviewed pool.

Write $S := \gamma V e s_x > 0$ for the journal's reward to substance through endorsement, $A := a_0 + \gamma^T\alpha_T r + S$, and $B(t) := b_0 + S t$ for the reward slopes~\eqref{eq:slopes}. A \emph{rational review rule within the linear class} is a weight $t$ such that authors' best responses to $(A,B(t))$ generate a pool whose best linear predictor yields that same weight in~\eqref{eq:t}. We call these \emph{linear review rules} below. A rule is \emph{interior} if Assumption~\ref{ass:interior} holds at $(A,B(t))$. The fixed-point analysis concerns these effort and prediction conditions; an implementation as actual probabilities also requires the signal-support restriction in Lemma~\ref{lem:projection}.

A journal that could commit to a review rule could rank papers by the reading signal alone ($t=0$), which satisfies the look-through condition. Such a rule is not self-consistent when presentation is informative about substance, so commitment trades selection accuracy for incentives, as in the literature on manipulable signals \citep{frankel2019,frankel2022}. The second theorem characterises the self-consistent weight in a genuine pool and quantifies both sides of this trade-off.

\begin{theorem}[Self-consistent review weights and the value of commitment]\label{thm:rational}
Hold $\gamma$, $\gamma^T$, $r$, $e$ and $s_x$ fixed, and suppose the reviewed pool consists of genuine papers whose authors' capability has variance $\Sigma > 0$.
\begin{enumerate}[label=(\alph*)]
\item \emph{Positive self-consistent weight.} Every interior linear review rule satisfies
\begin{equation}\label{eq:tgenuine}
t = f(t;\kappa) := \frac{\sigma_\varepsilon^2}{\sigma_u^2}\cdot\frac{\chi B(t) - \rho A}{\kappa A - \rho B(t)} > 0 .
\end{equation}
The self-consistent linear predictor places positive weight on polish because capable authors produce more of both dimensions.
\item \emph{Cheaper polish, heavier weight.} Suppose $\chi b_0 > \rho A$. The set $\mathcal K$ of admissible costs of polish $\kappa$ at which an interior linear review rule exists is an interval unbounded above, and the least interior weight $t_*(\kappa)$ is non-increasing in $\kappa$. Thus cheaper polish weakly increases this relative weight. For $\kappa<\inf\mathcal K$ there is no interior fixed point within this class; this is not a non-existence result for corner solutions, capped effort, or other review mechanisms.
\item \emph{Incentive value of commitment.} Let $t_*$ be an interior linear review rule, and suppose the journal commits to a weight $t_c\in[0,t_*]$ such that Assumption~\ref{ass:interior} holds at $(A,B(t_c))$. Then, for every $\theta$,
\begin{equation}\label{eq:commit}
Q_\theta^J(t_c)-Q_\theta^J(t_*)=\frac{\theta(\rho+\lambda\chi)\,S\,(t_*-t_c)}{D}\ \ge\ 0,
\end{equation}
strictly if $t_c<t_*$ and $\rho+\lambda\chi>0$.
\item \emph{Commitment gains value as presentation becomes cheaper.} Under the conditions of part~(b), the value of committing to $t_c=0$ rather than following the least interior rule, $\mathcal V_c(\kappa):=\theta(\rho+\lambda\chi)S\,t_*(\kappa)/D(\kappa)$, is non-increasing in $\kappa$ on $\mathcal K$, and strictly decreasing if $\rho+\lambda\chi>0$.
\item \emph{Selection cost.} In the pool generated under $t_c=0$, with effort $\theta(a_c,b_c)$, ranking by the reading signal alone rather than by the best linear predictor raises the mean squared error of predicted substance by
\begin{equation}\label{eq:selcost}
\ell_c=\frac{a_c^2b_c^2\Sigma^2\sigma_\varepsilon^4}{(a_c^2\Sigma+\sigma_\varepsilon^2)\bigl(\sigma_u^2(a_c^2\Sigma+\sigma_\varepsilon^2)+b_c^2\Sigma\sigma_\varepsilon^2\bigr)}>0 .
\end{equation}
\item \emph{Where the threshold can bind.} Under the conditions of part~(b), let $\lambda=0$ and $\rho>0$, so $L=\kappa/\rho$. Write $A_P:=a_0+\gamma^T\alpha_T r$, and let $a_P:=(\kappa A_P-\rho b_0)/D$ and $b_P:=(\chi b_0-\rho A_P)/D$ be effort per unit of capability on a platform with the same technical reward, with $a_P>0$. If the least interior weight exceeds the threshold, $t_*(\kappa)>L$, then
\begin{equation}\label{eq:crossing}
\frac{\sigma_\varepsilon^2}{\sigma_u^2}\,\bigl(\rho\, b_P+S\bigr)>\kappa\, a_P .
\end{equation}
\end{enumerate}
\end{theorem}

The robust implication is part~(d): cheaper presentation makes commitment more valuable whether or not the self-consistent weight crosses the threshold. Parts~(a) and~(b) supply the mechanism, and part~(f) delimits when the threshold can bind. Condition~\eqref{eq:crossing} is necessary, not sufficient. It requires reading to be noisy relative to the presentation endowment, the endorsement reward $S$ to be large, and substantive effort on the platform, scaled by the cost of presentation, to be small. Section~\ref{sec:commit} discusses parts~(c)--(e).

\begin{figure}[t]
\centering
\begin{tikzpicture}
\begin{semilogyaxis}[width=0.72\textwidth,height=0.33\textwidth,xmin=0,xmax=2,ymin=0.006,ymax=25,
xlabel={Cost of polish $\kappa$},ylabel={Relative weight on polish $t$},
grid=major,grid style={gray!20},tick label style={font=\small},label style={font=\small},
legend style={at={(.98,.79)},anchor=north east,font=\scriptsize,draw=none,fill=white}]
\addplot[black,densely dashed] coordinates {(0.001000,0.010000) (0.045422,0.454222) (0.089844,0.898444) (0.134267,1.342667) (0.178689,1.786889) (0.223111,2.231111) (0.267533,2.675333) (0.311956,3.119556) (0.356378,3.563778) (0.400800,4.008000) (0.445222,4.452222) (0.489644,4.896444) (0.534067,5.340667) (0.578489,5.784889) (0.622911,6.229111) (0.667333,6.673333) (0.711756,7.117556) (0.756178,7.561778) (0.800600,8.006000) (0.845022,8.450222) (0.889444,8.894444) (0.933867,9.338667) (0.978289,9.782889) (1.022711,10.227111) (1.067133,10.671333) (1.111556,11.115556) (1.155978,11.559778) (1.200400,12.004000) (1.244822,12.448222) (1.289244,12.892444) (1.333667,13.336667) (1.378089,13.780889) (1.422511,14.225111) (1.466933,14.669333) (1.511356,15.113556) (1.555778,15.557778) (1.600200,16.002000) (1.644622,16.446222) (1.689044,16.890444) (1.733467,17.334667) (1.777889,17.778889) (1.822311,18.223111) (1.866733,18.667333) (1.911156,19.111556) (1.955578,19.555778) (2.000000,20.000000)};
\addlegendentry{$L(\kappa)=\kappa/\rho$}
\addplot[blue!70!black,thick] coordinates {(0.459972,1.274755) (0.460126,1.217474) (0.460588,1.162794) (0.461358,1.110625) (0.462436,1.060873) (0.463822,1.013446) (0.465516,0.968256) (0.467518,0.925211) (0.469828,0.884225) (0.472446,0.845210) (0.475372,0.808083) (0.478606,0.772761) (0.482148,0.739162) (0.485998,0.707209) (0.490157,0.676826) (0.494623,0.647937) (0.499397,0.620473) (0.504479,0.594364) (0.509869,0.569544) (0.515567,0.545948) (0.521573,0.523516) (0.527887,0.502188) (0.534509,0.481909) (0.541439,0.462625) (0.548678,0.444284) (0.556224,0.426836) (0.564078,0.410237) (0.572240,0.394440) (0.580710,0.379405) (0.589488,0.365090) (0.598575,0.351458) (0.607969,0.338473) (0.617671,0.326100) (0.627681,0.314307) (0.637999,0.303063) (0.648625,0.292340) (0.659560,0.282109) (0.670802,0.272345) (0.682352,0.263024) (0.694210,0.254122) (0.706376,0.245618) (0.718851,0.237490) (0.731633,0.229719) (0.744723,0.222287) (0.758121,0.215176) (0.771828,0.208370) (0.785842,0.201854) (0.800164,0.195612) (0.814794,0.189631) (0.829733,0.183898) (0.844979,0.178400) (0.860533,0.173126) (0.876396,0.168065) (0.892566,0.163207) (0.909044,0.158541) (0.925830,0.154058) (0.942925,0.149750) (0.960327,0.145608) (0.978037,0.141624) (0.996056,0.137791) (1.014382,0.134103) (1.033016,0.130551) (1.051959,0.127130) (1.071209,0.123835) (1.090767,0.120658) (1.110634,0.117596) (1.130808,0.114642) (1.151291,0.111793) (1.172081,0.109043) (1.193179,0.106388) (1.214586,0.103824) (1.236300,0.101348) (1.258323,0.098954) (1.280653,0.096641) (1.303291,0.094404) (1.326238,0.092241) (1.349492,0.090148) (1.373055,0.088122) (1.396925,0.086161) (1.421103,0.084262) (1.445590,0.082423) (1.470384,0.080641) (1.495487,0.078914) (1.520897,0.077239) (1.546616,0.075616) (1.572642,0.074041) (1.598977,0.072514) (1.625619,0.071031) (1.652570,0.069592) (1.679828,0.068195) (1.707395,0.066838) (1.735269,0.065520) (1.763452,0.064239) (1.791942,0.062994) (1.820741,0.061784) (1.849847,0.060608) (1.879262,0.059464) (1.908984,0.058351) (1.939015,0.057268) (1.969353,0.056214) (2.000000,0.055188)};
\addlegendentry{$\sigma_\varepsilon^2/\sigma_u^2=0.5$}
\addplot[teal!75!black,thick] coordinates {(0.274444,0.901388) (0.274617,0.850704) (0.275135,0.802908) (0.275997,0.757871) (0.277205,0.715463) (0.278758,0.675556) (0.280656,0.638025) (0.282900,0.602747) (0.285488,0.569601) (0.288421,0.538471) (0.291700,0.509243) (0.295324,0.481809) (0.299292,0.456063) (0.303606,0.431904) (0.308265,0.409236) (0.313269,0.387968) (0.318619,0.368011) (0.324313,0.349283) (0.330352,0.331706) (0.336737,0.315205) (0.343467,0.299710) (0.350541,0.285156) (0.357961,0.271481) (0.365726,0.258628) (0.373836,0.246541) (0.382292,0.235171) (0.391092,0.224470) (0.400237,0.214394) (0.409728,0.204901) (0.419564,0.195953) (0.429744,0.187514) (0.440270,0.179551) (0.451141,0.172033) (0.462357,0.164930) (0.473919,0.158217) (0.485825,0.151867) (0.498076,0.145858) (0.510673,0.140168) (0.523615,0.134777) (0.536901,0.129666) (0.550533,0.124817) (0.564510,0.120215) (0.578832,0.115844) (0.593500,0.111691) (0.608512,0.107741) (0.623869,0.103984) (0.639572,0.100407) (0.655620,0.096999) (0.672012,0.093752) (0.688750,0.090656) (0.705833,0.087701) (0.723261,0.084881) (0.741035,0.082188) (0.759153,0.079613) (0.777616,0.077152) (0.796425,0.074798) (0.815579,0.072544) (0.835077,0.070386) (0.854921,0.068319) (0.875110,0.066337) (0.895644,0.064436) (0.916524,0.062613) (0.937748,0.060863) (0.959317,0.059182) (0.981232,0.057567) (1.003492,0.056015) (1.026096,0.054522) (1.049046,0.053086) (1.072341,0.051705) (1.095981,0.050374) (1.119967,0.049093) (1.144297,0.047858) (1.168972,0.046668) (1.193993,0.045520) (1.219359,0.044413) (1.245069,0.043344) (1.271125,0.042312) (1.297526,0.041316) (1.324272,0.040354) (1.351364,0.039424) (1.378800,0.038525) (1.406581,0.037655) (1.434708,0.036814) (1.463180,0.036000) (1.491996,0.035212) (1.521158,0.034449) (1.550665,0.033710) (1.580517,0.032994) (1.610715,0.032299) (1.641257,0.031627) (1.672144,0.030974) (1.703377,0.030341) (1.734955,0.029727) (1.766877,0.029131) (1.799145,0.028552) (1.831758,0.027990) (1.864716,0.027445) (1.898020,0.026914) (1.931668,0.026399) (1.965661,0.025898) (2.000000,0.025411)};
\addlegendentry{$\sigma_\varepsilon^2/\sigma_u^2=0.25$}
\addplot[orange!85!black,thick] coordinates {(0.150867,0.570088) (0.151052,0.528724) (0.151607,0.490414) (0.152531,0.454976) (0.153826,0.422232) (0.155490,0.392006) (0.157524,0.364128) (0.159928,0.338434) (0.162702,0.314766) (0.165845,0.292973) (0.169358,0.272912) (0.173242,0.254446) (0.177495,0.237450) (0.182118,0.221805) (0.187110,0.207399) (0.192473,0.194129) (0.198205,0.181901) (0.204307,0.170626) (0.210779,0.160224) (0.217621,0.150620) (0.224832,0.141747) (0.232414,0.133541) (0.240365,0.125947) (0.248686,0.118913) (0.257377,0.112390) (0.266438,0.106337) (0.275869,0.100713) (0.285669,0.095484) (0.295839,0.090616) (0.306379,0.086080) (0.317289,0.081849) (0.328569,0.077899) (0.340218,0.074207) (0.352238,0.070754) (0.364627,0.067519) (0.377386,0.064487) (0.390515,0.061642) (0.404013,0.058970) (0.417882,0.056458) (0.432120,0.054094) (0.446728,0.051868) (0.461706,0.049770) (0.477054,0.047789) (0.492772,0.045920) (0.508859,0.044152) (0.525317,0.042481) (0.542144,0.040898) (0.559341,0.039398) (0.576907,0.037977) (0.594844,0.036627) (0.613150,0.035346) (0.631827,0.034128) (0.650873,0.032970) (0.670289,0.031868) (0.690074,0.030819) (0.710230,0.029819) (0.730755,0.028865) (0.751650,0.027955) (0.772915,0.027086) (0.794550,0.026255) (0.816555,0.025462) (0.838929,0.024702) (0.861674,0.023976) (0.884788,0.023280) (0.908272,0.022613) (0.932126,0.021974) (0.956349,0.021361) (0.980943,0.020772) (1.005906,0.020207) (1.031239,0.019665) (1.056942,0.019143) (1.083015,0.018642) (1.109458,0.018159) (1.136270,0.017695) (1.163452,0.017248) (1.191004,0.016817) (1.218926,0.016402) (1.247218,0.016002) (1.275880,0.015616) (1.304911,0.015243) (1.334312,0.014884) (1.364083,0.014537) (1.394224,0.014202) (1.424735,0.013878) (1.455615,0.013564) (1.486866,0.013262) (1.518486,0.012969) (1.550476,0.012685) (1.582836,0.012411) (1.615565,0.012145) (1.648665,0.011887) (1.682134,0.011638) (1.715973,0.011396) (1.750182,0.011162) (1.784761,0.010935) (1.819710,0.010714) (1.855028,0.010500) (1.890716,0.010292) (1.926774,0.010091) (1.963202,0.009895) (2.000000,0.009704)};
\addlegendentry{$\sigma_\varepsilon^2/\sigma_u^2=0.1$}
\end{semilogyaxis}
\end{tikzpicture}
\caption{Self-consistent linear-prediction weights among genuine papers. The least interior weight $t_*(\kappa)$ from~\eqref{eq:tgenuine} for three signal-to-endowment noise ratios, and the look-through threshold $L(\kappa)=\kappa/\rho$ (with $\lambda=0$). Each curve ends at the critical cost $\kappa_c$ below which no interior fixed point exists within the linear class. The plotted weights remain below $L$; these parameters illustrate loss of interior existence, not a crossing into harmful certification. Parameters: $\chi=1$, $\rho=0.1$, $a_0+\gamma^T\alpha_T r=1.5$, $b_0=1$, $S=2$.}
\label{fig:polish}
\end{figure}
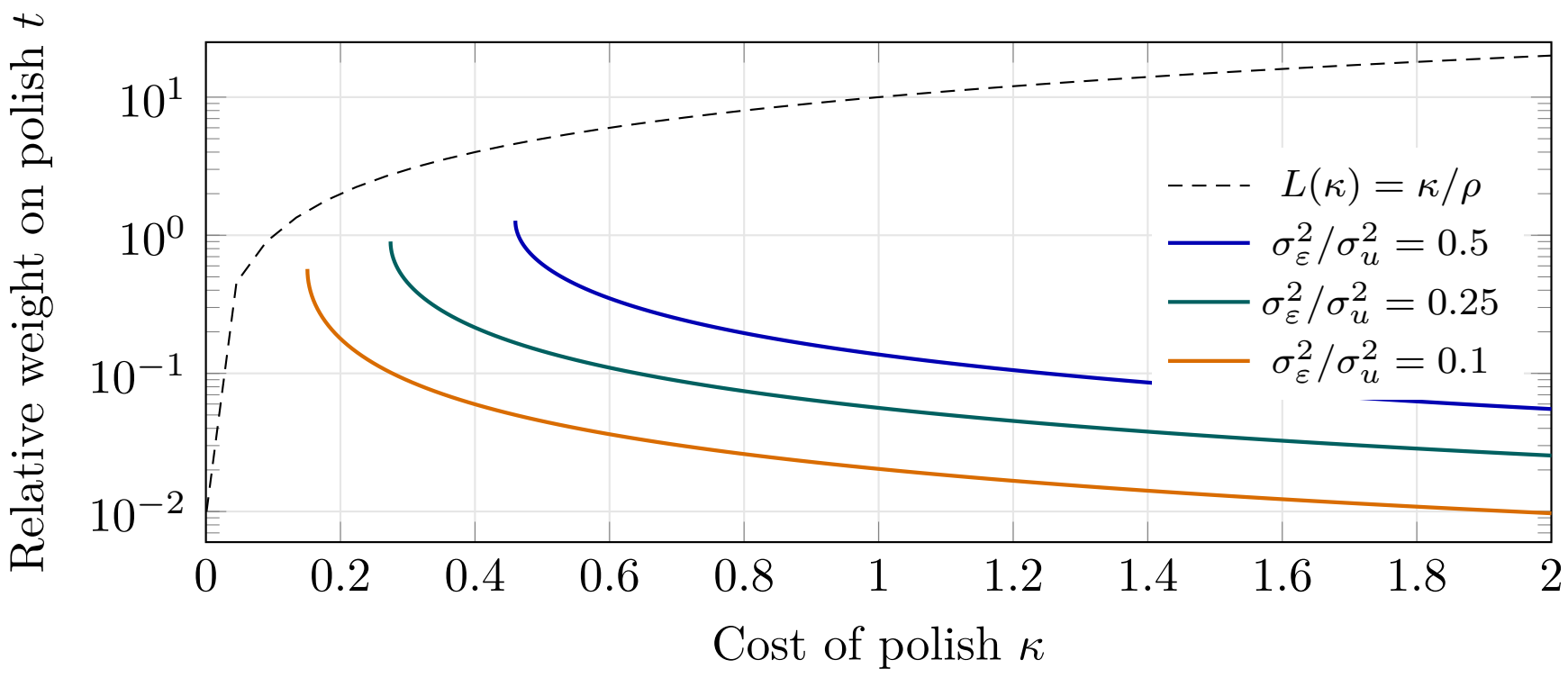

Parts~(a) and~(b) show why cheap polish need not become uninformative. In the genuine-pool benchmark, chosen presentation scales with capability while the variance $\sigma_u^2$ of the independent endowment stays fixed. The least self-consistent relative weight on polish can therefore increase as its cost falls, even as the tolerance for that weight in Theorem~\ref{thm:look}(c) shrinks. This conclusion concerns the ratio of prediction weights, not a claim that the unnormalised coefficient $w_y$ must rise. Figure~\ref{fig:polish} illustrates the increasing relative weight and the boundary of interior existence. Its curves stay below $L$ wherever an interior rule exists; the figure therefore does not show a harmful journal. Whether the self-consistent weight crosses the threshold depends on parameters. The following example shows that a crossing is possible, but it requires presentation to be far cheaper than substance and weak competition between the two for an author's time. We therefore read parts~(a) and~(b) as showing that inference need not restore the look-through condition, not that it typically fails to do so.

\begin{remark*}[A threshold crossing with substantive output]
Let $\chi=\chiu=1$, $\rho=0.01$, $a_0+\gamma^T\alpha_T r=1$, $b_0=0.1$, $S=0.1$, $\sigma_\varepsilon^2/\sigma_u^2=0.01$, and $\lambda=0$. At $\kappa=0.005$, the least fixed point of~\eqref{eq:tgenuine} is $t_*\approx0.2761<L=0.5$. The journal induces $x^{J*}/\theta\approx0.8620$, versus $x^{P*}/\theta\approx0.8163$ on a platform with the same technical reward. At $\kappa=0.004$, the least fixed point is $t_*\approx0.4584>L=0.4$, and $x^{J*}/\theta\approx0.7543<x^{P*}/\theta\approx0.7692$. Both costs are admissible and both regimes remain interior. The reversal thus concerns substantive output itself, without a presentation penalty in the output measure. To verify probability feasibility, take $\gamma=e=1$, $V=100$, $s_x=0.001$, $\pi_0'=0$, capability uniform on $[1,2]$, and independent symmetric noises $u\in\{-0.01,0.01\}$ and $\varepsilon\in\{-0.001,0.001\}$. Their variance ratio is $0.01$, and the affine endorsement probabilities at the stated equilibria lie strictly between zero and one. The calculations use the smaller positive root of~\eqref{eq:tgenuine} and~\eqref{eq:effort}. Condition~\eqref{eq:crossing} separates the two costs: its left-hand side is about $0.0033$ against $\kappa a_P\approx0.0031$ at $\kappa=0.004$, and about $0.0028$ against $0.0041$ at $\kappa=0.005$.
\end{remark*}

Theorem~\ref{thm:rational} concerns a genuine pool. When the reviewed pool also contains polished marginal papers, the sign of the linear weight can reverse.

\begin{proposition}[Polished junk and the sign of the linear weight]\label{prop:junk}
Hold $\gamma$, $\gamma^T$, $r$, $e$ and $s_x$ fixed. Suppose a share $\varphi \in (0,1)$ of the reviewed pool consists of marginal papers with no substance and presentation $y_m + u$, and the rest are genuine papers with mean capability $\bar\theta > 0$, capability variance $\Sigma$, and effort $(x,y) = \theta(a,b)$, $a,b>0$. Then
\begin{equation}\label{eq:Cmix}
\mathrm{Cov}(x,\tilde y) = (1-\varphi)\,a\bigl[b\Sigma + \varphi\bar\theta\,(b\bar\theta - y_m)\bigr],
\end{equation}
so the best linear predictor assigns polish a negative weight ($t < 0$) if and only if marginal papers are more polished than the average genuine paper, $y_m > b\bar\theta$, and sufficiently common, $\varphi > \hat\varphi := b\Sigma/\bigl(\bar\theta\,(y_m - b\bar\theta)\bigr)$.
\end{proposition}

Proposition~\ref{prop:junk} gives a covariance condition for a negative linear weight in a specified mixed pool. It is not by itself an equilibrium comparative static as the junk share changes: genuine effort and marginal presentation can also respond. For comparison with the credibility accounting of Section~\ref{sec:review}, if junk is marginal and its relative endorsement rate is $\psi_m=1$, its reviewed share is $1-\gamma$. Substituting this identity into the covariance condition gives $\gamma<1-\hat\varphi$ when $y_m>b\bar\theta$. This is a conditional comparison of the two mechanisms, not a joint derivation of $\psi_m=1$ from the unbiased signal model used here.

Theorem~\ref{thm:rational}(b) relies on presentation scaling with capability, which follows from the quadratic cost in the interior benchmark. If a common presentation ceiling binds for every genuine author, presentation carries no information about substance in a genuine pool, and the linear weight is zero (Remark~\ref{rem:saturation} in Appendix~\ref{app:ext}). With partial saturation, neither the threshold comparison nor the monotonicity in part~(b) applies without solving the corresponding constrained problem.

\subsection{The value of committing to disregard presentation}\label{sec:commit}

Theorem~\ref{thm:rational}(c)--(e) quantifies the trade-off between incentives and selection accuracy. In practice the commitment is implemented through the review process rather than by instruction, for example by a review form that elicits the contribution, its importance and the supporting evidence before any overall recommendation.

Theorem~\ref{thm:rational}(c) holds at given credibility; in a genuine pool, commitment does not change the endorsed share. Part~(d) combines part~(b) with the direct effect of $\kappa$ on $D$: as presentation becomes cheaper, the self-consistent weight weakly rises and each unit of presentation reward diverts more effort from substance, so the case for commitment strengthens on both counts. Part~(e) is the price. The selection cost is positive whenever presentation varies with capability and reading is noisy, and it vanishes as $\sigma_\varepsilon^2\to0$, when the reading signal alone identifies substance. Careful reading and commitment are thus substitutes in protecting selection accuracy: better reading lowers the self-consistent weight in~\eqref{eq:tgenuine} and the cost~\eqref{eq:selcost} of ignoring presentation. In the parameters of Figure~\ref{fig:polish} with $\sigma_\varepsilon^2/\sigma_u^2=0.25$, the least interior weight falls from about $0.45$ at $\kappa=0.3$ to $0.03$ at $\kappa=2$, and $\mathcal V_c/\theta$ falls from about $0.31$ to $0.003$.

\section{Technical screening, human attention and delay}\label{sec:review}

So far, reviewers' sensitivities and the credibility of the certificate were given. We now open the review process itself. Submissions differ in kind, AI tools filter some of them, and the rest compete for scarce human attention.

\subsection{Review technology}

Submissions to a certifying regime arrive in three streams: genuine contributions at rate $N_v$, technically flawed papers at rate $N_f$, and papers that are technically valid but marginal---plausible, polished and of negligible significance---at rate $N_m$. The AI validity screen with intensity $r$ detects a flawed paper with probability $\delta(r)$, where $\delta(0) = 0$, $\delta' \ge 0$ and $\delta < 1$, and has no false positives. By construction it cannot detect a marginal paper, which is valid. An AI-supported significance triage, with capability $\zeta \in [0,1)$, removes a fraction $\zeta$ of marginal papers before full editorial assessment; $\zeta = 0$ represents no such capability. For analytical clarity, this is an error-free benchmark: triage does not exclude genuine contributions and applies exogenously specified significance criteria rather than determining those criteria itself. Human review provides significance assessment but no additional detection of technical flaws in this benchmark. This division isolates the two services; it is an assumption rather than a general claim about human and AI capabilities. The failures of text-only assessment reported by \citet{ahmad2026} motivate studying such a case. The \emph{junk load} reaching human review and the total human load are
\begin{equation}\label{eq:load}
Z := (1 - \delta)N_f + (1 - \zeta)N_m, \qquad M := N_v + Z.
\end{equation}

Credibility is determined by what is endorsed, not by what is reviewed. For $N_v>0$, let $\bar\pi_v>0$ be the average endorsement rate of genuine contributions generated by~\eqref{eq:endorse}. A flawed or marginal paper that reaches review is erroneously endorsed at rate $\psi_f\bar\pi_v$ or $\psi_m\bar\pi_v$, respectively, where $\psi_f,\psi_m\in[0,1]$ are reduced-form relative rates. The benchmark $\psi_f=1$ represents flawed work that looks like genuine work to the significance assessor after passing technical screening. Lower $\psi_m$ represents more effective recognition of marginal work; $\psi_m=0$ means reliable recognition and $\psi_m=1$ no effective discrimination in average endorsement rates. These rates are not obtained by substituting the true substance of junk into the genuine-author rule~\eqref{eq:endorse}. The \emph{endorsed junk load} and the endorsement credibility are
\begin{equation}\label{eq:gamma}
\tilde Z := \psi_f(1 - \delta)N_f + \psi_m(1 - \zeta)N_m \le Z, \qquad
\gamma = \frac{N_v}{N_v + \tilde Z} =: \frac{N_v}{\tilde M},
\end{equation}
where $\bar\pi_v$ cancels for $N_v>0$. We use the displayed expression as a reduced-form definition at the boundary: $\gamma=0$ when $N_v=0$ and $\tilde Z>0$. It represents a journal with some erroneous endorsements but no genuine contributions, rather than extrapolating the genuine-author effort rule to a non-existent genuine pool. The case $\tilde Z=0$ is treated separately in Theorem~\ref{thm:trap}(d). Validation credibility is $\gamma^T=(N_v+N_m)/(N_v+N_m+(1-\delta)N_f)$; it does not depend on triage or on $\psi$. We hold $\psi_f$ and $\psi_m$ fixed in the comparative statics below. When $\psi_f=\psi_m=1$, $\tilde Z=Z$ and $\gamma$ equals the genuine share of the reviewed pool.

\paragraph{An implementation of endorsement leakage.} For $N_v>0$, one possible error process is that a paper of junk class $k\in\{f,m\}$ is recognised and rejected with probability $1-\psi_k$. Otherwise, the evidence perceived by the assessor has the same distribution as that of a randomly drawn genuine submission, although the underlying paper does not make the corresponding genuine contribution. Applying the genuine review rule to that perceived evidence yields average endorsement probability $\psi_k\bar\pi_v$. This construction permits a negative marginal response to polish among genuine papers and a positive leakage rate for junk. It also identifies the additional assumption: perception errors for junk need not be unbiased about its true substance. This is a feasible implementation of the reduced-form rates, not the unbiased signal model of Section~\ref{sec:bayes}. A unified model of inference and participation would have to derive $\psi_k$ jointly with review weights and pool composition.

Each paper reaching the human stage requires $h(r)\,e$ units of human time, where $h(r) > 0$ and $h' \le 0$ reflects routine checks that AI takes over. With total human attention $H$ per period, the human stage processes papers at rate $\mu(r,e) = H/(h(r)\,e)$. We model the human stage as a single-server queue with arrival rate $M$ and service rate $\mu$, so expected time in review is $T = 1/(\mu - M)$ for $\mu > M$. AI checks are assumed instantaneous, so the human stage is the bottleneck.

We treat effective review capacity as fixed over the horizon of a
submission shock. The relevant horizon is one over which submissions
can rise before the journal can expand the time available from editors
and qualified referees. The model allows capacity to differ across
review architectures through the attention budget $H$, AI support
$r$, and editorial intensity $e$. The tipping analysis holds these
inputs fixed to isolate the effect of increased submission pressure.

\paragraph{Congestion.} The queue is a tractable congestion technology rather than a literal description of every editorial workflow. It captures increasingly steep delay as submissions approach capacity; the results use only its monotonicity and its dependence on the capacity gap $\mu-M$. Leakage rates and reward slopes need not move independently in practice; as stated in Section~\ref{sec:model}, each result holds the indicated quantities fixed.

\begin{lemma}[Review frontier]\label{lem:frontier}
For given $(r, \zeta, N_v, N_f, N_m)$, an editorial intensity $e$ is attainable with delay at most $T$ if and only if
\begin{equation}\label{eq:frontier}
e \le E(r,T) := \frac{H}{h(r)\,\bigl(M + 1/T\bigr)}.
\end{equation}
\end{lemma}

With a fixed budget of human attention, more intensive assessment of each paper is possible only with longer delay or fewer papers in review. AI shifts the frontier outward by taking over routine checks and by removing papers before they reach reviewers. To see whether AI actually relieves reviewers, we must account for how cheap generation and AI screening change the number of submissions.

Flawed and marginal papers come from populations of potential generators of mass $n_f$ and $n_m$, with heterogeneous production costs distributed as $F_f(\cdot\,;\omega)$ and $F_m(\cdot\,;\omega)$, where a higher $\omega$ (cheaper generation) shifts both upward. A generator submits if the expected value of reaching human review exceeds the cost:
\begin{equation}\label{eq:generation}
N_f = n_f F_f\bigl((1-\delta)V_f;\omega\bigr), \qquad N_m = n_m F_m\bigl((1-\zeta)V_m;\omega\bigr),
\end{equation}
with $V_f, V_m > 0$. Both are decreasing in the corresponding AI capability (deterrence) and increasing in $\omega$. Let $\eta_f, \eta_m \ge 0$ denote the elasticities of $N_f$ and $N_m$ with respect to their pass probabilities $1-\delta$ and $1-\zeta$.

\begin{lemma}[Net-capacity accounting]\label{lem:capacity}
Consider a joint change in AI capabilities and submissions, $(dr, d\zeta)$ and $(dN_v, dN_f, dN_m)$, and define its net capacity gain as
\begin{equation}\label{eq:Lambda}
\begin{split}
\Lambda := {}& -\mu\frac{h'}{h}\,dr + \delta' N_f\, dr + N_m\, d\zeta\\
&- (1-\delta)\,dN_f - (1-\zeta)\,dN_m - dN_v ,
\end{split}
\end{equation}
where $dN_f$ and $dN_m$ include the deterrence responses in~\eqref{eq:generation}. At a fixed editorial intensity, $dT = -T^2\Lambda$; at a fixed delay, $d\log E = \Lambda/\mu$.
\end{lemma}

AI shortens review at the current intensity if and only if it raises the attainable intensity at the current delay, and both happen if and only if its capacity and screening effects exceed the additional human load from induced submissions. Faster or better technical checking therefore need not shorten review: if cheap generation induces submissions, in particular marginal ones that the validity screen cannot remove, faster than AI expands effective capacity, delay rises.

\subsection{Complementarity between AI and human review}

AI screening affects human review through two channels. It removes flawed papers from the pool that reviewers might endorse, which raises the credibility of endorsement, and it frees reviewer time. Both channels help science only if the review rule itself rewards substance in the sense of condition~\eqref{eq:C}.

\begin{proposition}[Technical AI and human endorsement]\label{prop:complement}
Hold the submission streams fixed.
\begin{enumerate}[label=(\alph*)]
\item \emph{Credibility channel.} $\gamma'(r) = \psi_f N_v N_f\, \delta'(r)/\tilde M^2 \ge 0$, and
\begin{equation}\label{eq:cross}
\frac{\partial^2 Q_\theta^J}{\partial r\,\partial e} = \frac{\theta\,\gamma'(r)}{D}\Bigl[(\kappa+\lambda\rho)\alpha_E + (\rho+\lambda\chi)\beta_E\Bigr].
\end{equation}
Suppose $\gamma'(r) > 0$, which holds whenever $N_v > 0$ and $\psi_f N_f\delta'(r) > 0$. Then AI screening and human endorsement are complements in producing science if and only if the look-through condition~\eqref{eq:C} holds. If the reverse strict inequality holds, they are substitutes: better screening makes a harmful endorsement more potent. At equality the cross-partial is zero. If $\gamma'(r) = 0$, the cross-partial vanishes.
\item \emph{Attention channel.} At fixed delay $T$,
\begin{equation}\label{eq:attention}
\frac{\partial \log E}{\partial r} = -\frac{h'(r)}{h(r)} + \frac{\delta'(r)N_f}{M + 1/T} \ge 0 .
\end{equation}
\item \emph{Journal output along the frontier.} With $e = E(r,T)$,
\begin{equation}\label{eq:frontierQ}
\frac{dQ_\theta^J}{dr} = \frac{\theta}{D}\Bigl\{(\kappa+\lambda\rho)\bigl(\gamma^{T\prime}\alpha_T r + \gamma^T\alpha_T\bigr) + \bigl(\gamma' E + \gamma\,\partial E/\partial r\bigr)\bigl[(\kappa+\lambda\rho)\alpha_E + (\rho+\lambda\chi)\beta_E\bigr]\Bigr\},
\end{equation}
where $\gamma^{T\prime}(r) = (N_v + N_m)N_f\delta'(r)/(N_v + N_m + (1-\delta)N_f)^2 \ge 0$. The first term, the value of better technical validation, is non-negative. The second has the sign of~\eqref{eq:C}.
\item \emph{Dilution by marginal submissions.} Suppose $N_f\delta'(r) > 0$. The screening term in~\eqref{eq:attention} is strictly decreasing in $N_m$. The credibility gain $\gamma'(r)$ is strictly decreasing in $N_m$ if $\psi_f > 0$ and $\psi_m > 0$, and independent of $N_m$ if $\psi_m = 0$. If $N_f = 0$ and $h' \equiv 0$, technical screening has no effect on human review.
\end{enumerate}
\end{proposition}

Part (a) is the complementarity: the value of endorsing significance depends on whether the endorsed pool is genuine, and AI raises that share. Part (d) explains why the complementarity may be weak in practice. The evidence in \citet{gartenberg2026} suggests that much of the recent surge consists of submissions that editors reject for lack of contribution rather than for technical error, and the agents' studies in \citet{ahmad2026} were rigorous but comparatively incremental. Such papers pass a validity screen and consume human attention, and technical AI does nothing about them. They also dilute credibility to the extent that reviewers cannot reliably recognise their lack of significance ($\psi_m > 0$).

\subsection{Validity screening versus significance triage}

If much of the surge consists of valid but marginal papers, a validity screen cannot remove them, and a different AI capability is needed. We compare the marginal value of better validity screening with that of better significance triage, both for delay and for credibility.

\begin{proposition}[Which AI capability relieves the bottleneck]\label{prop:triage}
Hold $N_v>0$, $r$ and $e$ fixed, treat $\delta$ and $\zeta$ as capability levels, and let $N_f, N_m > 0$. Delay $T = 1/(\mu - N_v - Z)$ depends on $(\delta,\zeta)$ only through the junk load $Z$, and credibility $\gamma = N_v/(N_v + \tilde Z)$ only through the endorsed junk load $\tilde Z$, with
\begin{equation}\label{eq:dZ}
\frac{\partial Z}{\partial \delta} = -N_f\,(1 + \eta_f), \quad \frac{\partial Z}{\partial \zeta} = -N_m\,(1 + \eta_m), \quad
\frac{\partial \tilde Z}{\partial \delta} = -\psi_f N_f\,(1 + \eta_f), \quad \frac{\partial \tilde Z}{\partial \zeta} = -\psi_m N_m\,(1 + \eta_m).
\end{equation}
Hence, per unit of detection probability, the marginal value of significance triage relative to validity screening is
\begin{equation}\label{eq:ratio}
\mathcal{R} := \underbrace{\frac{N_m}{N_f}}_{\text{composition}}\cdot\underbrace{\frac{1 + \eta_m}{1 + \eta_f}}_{\text{deterrence}} \quad\text{for delay,}\qquad
\mathcal{R}_\gamma := \underbrace{\frac{\psi_m}{\psi_f}}_{\text{endorsement leakage}}\,\mathcal{R} \quad\text{for credibility } (\psi_f > 0).
\end{equation}
Moreover, $\partial\gamma/\partial\delta = \psi_f N_v N_f(1+\eta_f)/\tilde M^2$ is strictly decreasing in $N_m$ if $\psi_f, \psi_m > 0$.
\end{proposition}

Proposition~\ref{prop:triage} formalises a ladder of scarcity, and its content lies in the two correction factors rather than in the composition term alone. The \emph{deterrence factor} says that a capability is worth more when the stream it filters is more elastic: each detection removes not only the paper detected but also the submissions it deters. If marginal papers are produced by low-cost mass generation, while fabricating results carries reputational and legal risks that do not scale with detection, then $\eta_m > \eta_f$ and triage is worth more than the composition alone suggests. The \emph{leakage factor} says that triage protects credibility only to the extent that reviewers would otherwise endorse marginal papers: if reviewers reliably recognise them ($\psi_m = 0$), triage is purely a capacity instrument, while if they cannot tell disclosed from hidden flaws ($\psi_f \approx 1$) and only partly recognise marginal work, $\mathcal R_\gamma < \mathcal R$. When generative AI makes papers technically clean while making plausible marginal papers cheap, $N_f$ falls and $N_m$ rises, and both ratios rise for given elasticities. This result concerns the capacity value of accurate triage; it does not establish that an AI system should independently determine which questions deserve a field's attention.

Comparing marginal products per unit of detection probability is only half of an investment rule; the other half is what each capability costs.

\begin{corollary}[Allocating an AI budget]\label{cor:invest}
Let capabilities be produced at costs $c_\delta(\delta)$ and $c_\zeta(\zeta)$, differentiable with $c_\delta', c_\zeta' > 0$, and let the journal minimise delay subject to $c_\delta(\delta) + c_\zeta(\zeta) \le K$. At any optimum with $\delta,\zeta$ interior and a binding budget,
\begin{equation}\label{eq:invest}
\frac{c_\zeta'(\zeta)}{c_\delta'(\delta)} = \mathcal R .
\end{equation}
If the journal instead maximises endorsement credibility and $\psi_f>0$, the same necessary condition holds with $\mathcal R_\gamma$ in place of $\mathcal R$.
\end{corollary}

At an interior optimum, the ratio of marginal capability costs equals the corresponding ratio of marginal benefits. A shift toward valid-but-marginal submissions changes those benefits and calls for reassessing the allocation against~\eqref{eq:invest}. Whether continued investment in validity screening is suboptimal also depends on capability costs, deterrence, leakage and the possibility of a boundary solution.

The error-free comparison makes triage attractive whenever marginal submissions consume attention. Appendix~\ref{app:false} lets triage exclude a fraction of genuine contributions. Delay then still falls, but endorsement credibility rises only if the proportional removal of endorsed junk exceeds the proportional loss of genuine papers, and the flow of genuine work retained for review falls. Faster processing alone is therefore not sufficient evidence of better certification.

\subsection{Adaptive review weights and endorsement leakage}\label{sec:adaptive}

Section~\ref{sec:bayes} determines the review weight from the composition of the reviewed pool, while the leakage rates $\psi_f,\psi_m$ above are held fixed. The two interact: the weight a review rule places on presentation determines how often polished junk is endorsed. We connect them through the implementation of leakage described in Section~\ref{sec:review}, with one change: a marginal paper that is not recognised as such keeps its own presentation.

Throughout this subsection, all junk reaching human review is marginal, $(1-\delta)N_f=0$, so validation credibility equals one in every regime. Write $Z_m:=(1-\zeta)N_m>0$. A marginal paper reaching review is recognised and rejected with probability $1-\psi^0_m$, where $\psi^0_m\in(0,1]$. The \emph{contested pool} consists of the genuine submissions and the unrecognised marginal papers; let $\varphi:=\psi^0_mZ_m/(N_v+\psi^0_mZ_m)$ be the marginal share in it. Genuine authors have effort $\theta(a,b)$ with $a,b>0$, and the capability of genuine submitters has mean $\bar\theta$ and variance $\Sigma>0$; write $\bar x:=a\bar\theta$ and $\bar y:=b\bar\theta$ for their mean substance and presentation. Reviewers observe $s$ and $\tilde y$ as in Section~\ref{sec:bayes}. An unrecognised marginal paper has no substance ($x=0$), but its reading signal has the distribution of a genuine paper's, $s=a\theta'+\varepsilon$ with $\theta'$ an independent draw from the submitters' capability distribution; its presentation is $\tilde y=y_m+u$ with $y_m\ge0$. Reviewers rank the contested pool by its best linear predictor, equivalently by $s+t\tilde y$, and endorse with probability $e(\pi_0'+s_x(s+t\tilde y))$, which we assume lies in $[0,1]$ on the relevant signal support. When $y_m=\bar y$ or $t=0$, average endorsement rates are as in the implementation of Section~\ref{sec:review} with $\psi_m=\psi^0_m$.

\begin{lemma}[Review weight and leakage in the contested pool]\label{lem:contested}
\begin{enumerate}[label=(\alph*)]
\item \emph{Leakage.} The endorsement rate of a marginal paper reaching review, relative to the mean endorsement rate of genuine papers, is
\begin{equation}\label{eq:psit}
\psi_m(t)=\psi^0_m\,\frac{\pi_0'+s_x(\bar x+t\,y_m)}{\pi_0'+s_x(\bar x+t\,\bar y)} .
\end{equation}
Hence $\psi_m(t)-\psi^0_m$ has the sign of $t\,(y_m-\bar y)$.
\item \emph{Review weight.} The best linear predictor of $x$ in the contested pool has weights
\begin{equation}\label{eq:wcontested}
\begin{gathered}
w_s=\frac{(1-\varphi)a^2\Sigma\,\bigl[\sigma_u^2+\varphi(1-\varphi)(y_m-\bar y)\,y_m\bigr]}{\Delta_c},\\
w_y=\frac{(1-\varphi)a\,\bigl[\sigma_\varepsilon^2b\Sigma+\varphi a^2b\Sigma^2-\varphi\bar\theta\,(y_m-\bar y)(a^2\Sigma+\sigma_\varepsilon^2)\bigr]}{\Delta_c},
\end{gathered}
\end{equation}
where $\Delta_c>0$ is the determinant of the covariance matrix of $(s,\tilde y)$ in the contested pool. If $w_s>0$, which holds whenever $y_m\ge\bar y$, the rule ranks by $s+t\tilde y$ with $t=w_y/w_s$. Then $t>0$ if $y_m\le\bar y$. If $y_m>\bar y$, let $\Gamma:=\bar\theta(y_m-\bar y)(a^2\Sigma+\sigma_\varepsilon^2)-a^2b\Sigma^2$; then $t<0$ if and only if $\Gamma>0$ and $\varphi>\hat\varphi_c:=\sigma_\varepsilon^2b\Sigma/\Gamma$.
\end{enumerate}
\end{lemma}

Part~(a) says that a positive weight on presentation lets junk through more often than a rule that ignores presentation if and only if the junk is more polished than the average genuine paper; a negative weight reverses this. Part~(b) is the counterpart of Proposition~\ref{prop:junk} for the contested pool. Polished junk that is still rare leaves the weight positive; once it is sufficiently common, the self-consistent weight turns negative. Combining the two parts with the effort algebra of Lemma~\ref{lem:effort} compares an adaptive review rule with commitment to $t=0$. Both determine credibility endogenously, $\gamma=N_v/(N_v+\psi_m(t)Z_m)$, and the endorsement slope $S=\gamma Ves_x$ adjusts with it.

\begin{proposition}[Adaptive review versus commitment]\label{prop:adaptive}
Fix $N_v>0$, $Z_m>0$, $\psi^0_m$ and the remaining primitives. In the \emph{adaptive} regime, the weight $t_a$ is self-consistent in the contested pool generated by the authors' best responses to $(A(\gamma_a),B(t_a))$, with $\gamma_a=N_v/(N_v+\psi_m(t_a)Z_m)$ and $w_s>0$. In the \emph{committed} regime, $t=0$ and $\gamma_c=N_v/(N_v+\psi^0_mZ_m)$. Suppose Assumption~\ref{ass:interior} holds in both regimes, and let $\bar y_a$ be mean genuine presentation in the adaptive regime.
\begin{enumerate}[label=(\alph*)]
\item If $y_m\le\bar y_a$, then $t_a>0$ and $\gamma_a\ge\gamma_c$. At equal credibility, commitment yields weakly higher net epistemic output, strictly if $\rho+\lambda\chi>0$; with endogenous credibility, the comparison of output is ambiguous.
\item If $y_m>\bar y_a$ and $t_a>0$, then $\gamma_a<\gamma_c$ and commitment yields strictly higher net epistemic output for every $\theta$.
\item If $y_m>\bar y_a$ and $t_a<0$, then $\gamma_a>\gamma_c$ and adaptive review yields strictly higher net epistemic output for every $\theta$.
\end{enumerate}
\end{proposition}

The proposition connects the incentive and credibility channels. Adaptive review protects credibility against junk that is less polished than genuine work, at the cost of rewarding presentation (case a). It harms both incentives and credibility when reviewers still place a positive weight on presentation despite the presence of polished junk (case b). If the pool leads reviewers to discount the presentation of polished junk, adaptation serves both objectives (case c). The comparisons hold at given participation; participation then responds through the journal's value $\Phi_J$, which depends on both $\gamma$ and $t$.

Which case obtains is itself an equilibrium outcome, because genuine authors' presentation responds to the review weight. Take the cost and reward parameters of Figure~\ref{fig:trap}, hold credibility at $\gamma=0.75$, and let genuine submitters have $\theta\in[1.3,2.5]$, $\sigma_\varepsilon^2=1$, $\sigma_u^2=2$ and $y_m=2$. At $\varphi=0.005$, the least self-consistent weight is $t_a\approx0.16$, genuine presentation is $\bar y_a\approx1.80<y_m$, and $\psi_m/\psi^0_m\approx1.005$ (case b). At $\varphi=0.04$, the weight has turned negative, $t_a\approx-0.14$; genuine authors reduce presentation to $\bar y_a\approx0.95$, and $\psi_m/\psi^0_m\approx0.974$ (case c). Slightly above $\varphi=0.045$ this branch reaches the corner at which genuine authors stop investing in presentation, and the least interior weight jumps to $t_a\approx0.49$ at $\varphi=0.05$, with $\bar y_a\approx2.74>y_m$ (case a). Review norms are thus partly self-fulfilling: a rule that rewards presentation makes genuine papers more polished than junk, which makes presentation informative. The jump also shows that the selected weight need not vary continuously with the composition of the pool, which matters for the participation analysis (Remark~\ref{rem:adaptive-int}).

\section{Participation, credibility and tipping}\label{sec:participation}

So far, the rate of genuine submissions was fixed. We now let authors decide whether to submit to the journal. Delay and credibility determine who submits, and who submits determines delay and credibility. We first hold credibility fixed and show that congestion alone crowds out genuine authors. We then let credibility respond to who submits and show that this feedback can trap a journal in a low-credibility equilibrium. Finally, we close the model.

Formally, a mass $n_v$ of authors with capability $\theta$ distributed on the compact support $[\theta_{\min}, \theta_{\max}]$, $\theta_{\min} > 0$, with an atomless continuous cdf $G$ satisfying $G(\theta_{\min})=0$ and $G(\theta_{\max})=1$, choose between the journal and an outside regime $o \in \{R, P\}$ with value $v_o + \theta\Phi_o$. Authors bear a cost $\tau > 0$ per unit of review delay. By Lemma~\ref{lem:effort},
\begin{equation}\label{eq:participation}
U_\theta^J = v_J - \tau T + \theta\,\Phi_J, \qquad U_\theta^o = v_o + \theta\,\Phi_o .
\end{equation}

\subsection{Fixed credibility}

\begin{assumption}\label{ass:fixed}
$\Phi_J > \Phi_o$ and $\mu > Z$.
\end{assumption}

Authors with $\theta \ge \theta^*(T) := (v_o - v_J + \tau T)/(\Phi_J - \Phi_o)$ submit to the journal, so $N_v(T) = n_v[1 - G(\theta^*(T))]$. An equilibrium is a delay $T^*$ with $T^* = 1/(\mu - Z - N_v(T^*))$. With credibility fixed, more submissions mean longer delay, and longer delay means fewer submissions, so there is exactly one equilibrium.

\begin{proposition}[Equilibrium and crowding out]\label{prop:fixed}
Under Assumption~\ref{ass:fixed}, there is a unique equilibrium $(T^*, N_v^*)$. Moreover:
\begin{enumerate}[label=(\alph*)]
\item An increase in the junk load $Z$ (for instance, through cheaper generation $\omega$) weakly raises $T^*$ and weakly lowers $N_v^*$: junk crowds out genuine authors.
\item An increase in AI capability that raises $\mu$ or lowers $Z$, holding the reward structure fixed, weakly lowers $T^*$ and weakly raises $N_v^*$.
\item If, in addition, the increase weakly raises $\Phi_J$, $N_v^*$ still weakly rises, while the effect on $T^*$ is ambiguous.
\end{enumerate}
In (a) and (b), $T^*$ moves strictly whenever the change is strict; $N_v^*$ moves strictly if, in addition, $\theta^*(T^*) \in (\theta_{\min}, \theta_{\max})$ and $G$ has a positive density there.
\end{proposition}

Junk lengthens review, longer review deters the least capable of the journal's genuine authors, and they move to the outside option. AI capabilities that expand capacity or remove junk reverse this. If they also make the journal more attractive, the induced submissions can absorb part of the capacity gain, so delay need not fall.

\begin{remark}[The outside option also depends on junk]\label{rem:outside}
Proposition~\ref{prop:fixed} holds $\Phi_o$ fixed. If the outside option is the platform, its value $\Phi_P$ depends on its validation credibility $\gamma^T_P$, which falls when more flawed papers pass its screen. Cheaper generation of flawed papers therefore also lowers $\Phi_o$, which works against the crowding out in part (a). Valid but marginal papers do not affect $\gamma^T_P$, so for junk that consists mainly of marginal papers part (a) applies unchanged.
\end{remark}

\subsection{Endogenous credibility and the credibility trap}

Proposition~\ref{prop:fixed} holds credibility fixed. But $\gamma = N_v/(N_v + \tilde Z)$ rises with $N_v$, and the journal becomes more attractive with $\gamma$ under Assumption~\ref{ass:trap}. More genuine authors raise credibility, which makes the journal more attractive to genuine authors. Because visibility is universal once every paper can be posted to a repository, we set $v_J = v_o$ and $\pi_0 = 0$; differences between the regimes then arise only from the reward slopes. To isolate the endorsement channel, we hold validation credibility $\gamma^T$ and the junk load fixed here and write $\Phi_J(\gamma)$ for the journal's value as a function of endorsement credibility; Section~\ref{sec:integrated} relaxes both restrictions. Junk is described by its load $Z$ and its \emph{endorsement leakage} $\psi := \tilde Z/Z \in [0,1]$, held fixed as $Z$ varies.

\begin{assumption}\label{ass:trap}
$v_J = v_o$ and $\pi_0 = 0$; $\Phi_J(\gamma)$ is continuous and strictly increasing in $\gamma$; and $\Phi_J(0) \le \Phi_o < \Phi_J(1)$.
\end{assumption}

The first inequality says that a journal whose endorsements carry no credibility offers capable authors nothing beyond the outside option. It holds with equality when the outside option is a platform with the journal's own screen, since a journal without credible endorsement is then exactly such a platform. The second inequality says that a fully credible journal is worth more than the outside option to some author. Strict monotonicity of $\Phi_J$ is a restriction: by the envelope theorem, $d\Phi_J/d\gamma = e\,[\alpha_E(\kappa A_J - \rho B_J) - \beta_E(\chi B_J - \rho A_J)]/D$, which is positive whenever $\beta_E \le 0$ and, for $\beta_E > 0$, whenever the journal's reward for substance outweighs its penalty on polish in this sense.

For a junk load $Z \in (0,\mu)$ and an expected rate of genuine submissions $N \in [0, \mu - Z)$, define the induced delay $T(N) = 1/(\mu - Z - N)$, the induced credibility $\gamma(N) = N/(N + \psi Z)$, and the best response
\begin{equation}\label{eq:BR}
BR(N; Z) :=
\begin{cases}
n_v\Bigl[1 - G\Bigl(\dfrac{\tau\,T(N)}{\Phi_J(\gamma(N)) - \Phi_o}\Bigr)\Bigr] & \text{if } \Phi_J(\gamma(N)) > \Phi_o,\\[2ex]
0 & \text{otherwise.}
\end{cases}
\end{equation}
An equilibrium is a fixed point $N^* = BR(N^*; Z)$. 
Because each author is negligible and takes credibility and delay as
given, such a fixed point is a Nash equilibrium of the population game:
at $N^*$ no author gains by changing her submission decision, and the
credibility and delay induced by $N^*$ are those on which the decisions
were based.
It is \emph{stable} if $BR(N) > N$ immediately below $N^*$ (where $N^* > 0$) and $BR(N) < N$ immediately above it. To discuss basins of attraction we use the adjustment dynamics $\dot N = BR(N;Z) - N$, under which submissions rise when more authors would submit than currently do. For statements about these trajectories, assume additionally that $BR(\cdot;Z)$ is locally Lipschitz on its feasible domain, ensuring a unique solution through each initial state. Locally Lipschitz $G$ and $\Phi_J$ are sufficient; the bounded capability support makes $BR$ identically zero near points where the journal advantage vanishes. This regularity holds in the numerical example. Stability from above alone does not assert a crossing, or attraction, from below.

Delay makes participation costly even for the most capable author: an author with capability $\theta_{\max}$ prefers the journal at $N$ only if $\Phi_J(\gamma(N)) - \Phi_o > \tau T(N)/\theta_{\max} \ge \tau/((\mu - Z)\theta_{\max})$. Define the \emph{participation threshold} $\tilde\gamma(Z) \in (0,1]$ by $\Phi_J(\tilde\gamma(Z)) = \Phi_o + \tau/((\mu - Z)\theta_{\max})$, with $\tilde\gamma(Z) = 1$ if the right-hand side is at least $\Phi_J(1)$, and let
\begin{equation}\label{eq:N0}
N_0(Z) := \frac{\tilde\gamma(Z)}{1 - \tilde\gamma(Z)}\,\psi Z \quad (N_0 := \infty \text{ if } \tilde\gamma(Z) = 1), \qquad
Z_{\max} := \mu - \frac{\tau}{\theta_{\max}\,(\Phi_J(1) - \Phi_o)} .
\end{equation}
The participation threshold $\tilde\gamma(Z)$ is the minimum credibility at which even the most capable author is willing to bear the delay, and $N_0(Z)$ is the number of genuine submissions needed to reach it.

\begin{theorem}[The credibility trap]\label{thm:trap}
Under Assumption~\ref{ass:trap}, with $\psi > 0$ in parts (a)--(c) and (e):
\begin{enumerate}[label=(\alph*)]
\item \emph{Collapse.} For every $Z \in (0,\mu)$, $N^* = 0$ is a stable equilibrium. Indeed, $BR(N;Z) = 0$ for all $N \le N_0(Z)$, and $N_0(Z) > 0$.
\item \emph{High credibility.} If $BR(\hat N; Z) > \hat N$ for some $\hat N$, there is an equilibrium $N^H(Z) > \hat N$; the largest such equilibrium is stable from above and satisfies $N^H(Z) > N_0(Z)$.
\item \emph{Tipping.} For $Z < Z'$, $BR(N;Z') \le BR(N;Z)$ for every $N \in [0, \mu - Z')$. There is a threshold $\bar Z \in [0, \max\{Z_{\max}, 0\}]$ such that a positive equilibrium exists for every $Z \in (0, \bar Z)$, while for $Z > \bar Z$ we have $BR(N;Z)<N$ for every feasible $N>0$, so collapse is the only equilibrium. The largest positive equilibrium $N^H(Z)$ is non-increasing on $(0,\bar Z)$, is stable from above, and satisfies $N^H(Z) > N_0(Z) \ge N_0(\bar Z/2) > 0$ for $Z \in (\bar Z/2, \bar Z)$. Its disappearance across $\bar Z$ is therefore discontinuous. The statement permits a positive equilibrium exactly at $\bar Z$ and does not require it to be stable from below.
\item \emph{No trap without endorsed junk.} If $\tilde Z = \psi Z = 0$ and $Z < \mu$, set $\gamma \equiv 1$. Then Proposition~\ref{prop:fixed} applies with $\Phi_J = \Phi_J(1)$, and its unique equilibrium is positive if and only if $Z < Z_{\max}$.
\item \emph{Basin of collapse.} For $Z \in (0,\bar Z)$ let $N^U(Z) := \min\{N > 0 : BR(N;Z) \ge N\}$. Then $N^U(Z)$ is the smallest positive equilibrium and, under the adjustment dynamics, the basin of attraction of collapse is exactly $[0, N^U(Z))$. Moreover, $N^U(Z) \ge N_0(Z)$; $N^U$ is non-decreasing on $(0,\bar Z)$; and if $Z_{\max} > 0$, then $N^U(Z) \to 0$ as $Z \downarrow 0$.
\end{enumerate}
\end{theorem}

The value of a journal's endorsement is a coordination outcome: it is high if capable authors submit, and capable authors submit only if it is high. A journal without capable authors endorses only junk, and its endorsement is then worthless. Delay reinforces the trap: even when a credibility-free journal is as good as the outside option, the cost of waiting keeps capable authors away until credibility exceeds $\tilde\gamma(Z) > 0$.

Part (e) puts the collapse equilibrium in proportion. Its existence at every positive junk load is close to definitional, since a pool without genuine papers has no credibility. What matters economically is how large a shock to genuine participation the journal can absorb, and that is the basin $[0, N^U(Z))$. When junk is scarce the basin is negligible: a journal falls into the trap only if its genuine submissions all but vanish. As junk grows, the basin widens and the largest positive equilibrium shrinks. At the final threshold $\bar Z$ the positive equilibrium disappears while bounded away from zero. In the regular example below this occurs when a stable and an unstable branch meet; the existence result does not require that particular local bifurcation geometry. A temporary drop in genuine submissions, a reputational shock or a wave of AI-generated papers, that would have been harmless at a low junk load can then push the journal into the basin of collapse, from which it does not recover when the shock passes. The model therefore predicts hysteresis (Section~\ref{sec:empirics}). AI capabilities enter through $Z$ and $\tilde Z$ (Proposition~\ref{prop:triage}): better screening and triage widen the range of junk loads in which a credible journal can exist and shrink the basin of collapse, but they cannot by themselves select the good equilibrium. That requires coordination by capable authors, editors and evaluators of research. The trap requires that reviewers endorse some junk ($\psi > 0$); a review process that never endorses junk faces congestion but not a credibility trap (part (d)).

\begin{figure}[t]
\centering
\begin{tikzpicture}
\begin{axis}[width=.482\textwidth,height=.35\textwidth,scale only axis=false,
xmin=0,xmax=1.4,ymin=0,ymax=1.1,grid=major,grid style={gray!20},
title={(a) Best responses},xlabel={Expected genuine submissions $N$},ylabel={Best response $BR(N;Z)$},
title style={font=\small},label style={font=\scriptsize},tick label style={font=\scriptsize},
legend columns=2,legend style={at={(.5,-.29)},anchor=north,font=\scriptsize,draw=none}]
\addplot[gray,densely dotted,forget plot] coordinates {(0.000000,0.000000) (1.100000,1.100000)};
\addplot[blue!70!black,thick] coordinates {(0.000000,0.000000) (0.006364,0.000000) (0.012727,0.130288) (0.019091,0.484508) (0.025455,0.660299) (0.031818,0.764911) (0.038182,0.834052) (0.044545,0.882999) (0.050909,0.919372) (0.057273,0.947397) (0.063636,0.969602) (0.070000,0.987590) (0.076364,1.000000) (0.082727,1.000000) (0.089091,1.000000) (0.095455,1.000000) (0.101818,1.000000) (0.108182,1.000000) (0.114545,1.000000) (0.120909,1.000000) (0.127273,1.000000) (0.133636,1.000000) (0.140000,1.000000) (0.146364,1.000000) (0.152727,1.000000) (0.159091,1.000000) (0.165455,1.000000) (0.171818,1.000000) (0.178182,1.000000) (0.184545,1.000000) (0.190909,1.000000) (0.197273,1.000000) (0.203636,1.000000) (0.210000,1.000000) (0.216364,1.000000) (0.222727,1.000000) (0.229091,1.000000) (0.235455,1.000000) (0.241818,1.000000) (0.248182,1.000000) (0.254545,1.000000) (0.260909,1.000000) (0.267273,1.000000) (0.273636,1.000000) (0.280000,1.000000) (0.286364,1.000000) (0.292727,1.000000) (0.299091,1.000000) (0.305455,1.000000) (0.311818,1.000000) (0.318182,1.000000) (0.324545,1.000000) (0.330909,1.000000) (0.337273,1.000000) (0.343636,1.000000) (0.350000,1.000000) (0.356364,1.000000) (0.362727,1.000000) (0.369091,1.000000) (0.375455,1.000000) (0.381818,1.000000) (0.388182,1.000000) (0.394545,1.000000) (0.400909,1.000000) (0.407273,1.000000) (0.413636,1.000000) (0.420000,1.000000) (0.426364,1.000000) (0.432727,1.000000) (0.439091,1.000000) (0.445455,1.000000) (0.451818,1.000000) (0.458182,1.000000) (0.464545,1.000000) (0.470909,1.000000) (0.477273,1.000000) (0.483636,1.000000) (0.490000,1.000000) (0.496364,1.000000) (0.502727,1.000000) (0.509091,1.000000) (0.515455,1.000000) (0.521818,1.000000) (0.528182,1.000000) (0.534545,1.000000) (0.540909,1.000000) (0.547273,1.000000) (0.553636,1.000000) (0.560000,1.000000) (0.566364,1.000000) (0.572727,1.000000) (0.579091,1.000000) (0.585455,1.000000) (0.591818,1.000000) (0.598182,1.000000) (0.604545,1.000000) (0.610909,1.000000) (0.617273,1.000000) (0.623636,1.000000) (0.630000,1.000000) (0.636364,1.000000) (0.642727,1.000000) (0.649091,1.000000) (0.655455,1.000000) (0.661818,1.000000) (0.668182,1.000000) (0.674545,1.000000) (0.680909,1.000000) (0.687273,1.000000) (0.693636,1.000000) (0.700000,1.000000) (0.706364,1.000000) (0.712727,1.000000) (0.719091,1.000000) (0.725455,1.000000) (0.731818,1.000000) (0.738182,1.000000) (0.744545,1.000000) (0.750909,1.000000) (0.757273,1.000000) (0.763636,1.000000) (0.770000,1.000000) (0.776364,1.000000) (0.782727,1.000000) (0.789091,1.000000) (0.795455,1.000000) (0.801818,1.000000) (0.808182,1.000000) (0.814545,1.000000) (0.820909,1.000000) (0.827273,1.000000) (0.833636,1.000000) (0.840000,1.000000) (0.846364,1.000000) (0.852727,1.000000) (0.859091,1.000000) (0.865455,1.000000) (0.871818,1.000000) (0.878182,1.000000) (0.884545,1.000000) (0.890909,1.000000) (0.897273,1.000000) (0.903636,1.000000) (0.910000,0.998758) (0.916364,0.995700) (0.922727,0.992558) (0.929091,0.989327) (0.935455,0.986005) (0.941818,0.982586) (0.948182,0.979068) (0.954545,0.975446) (0.960909,0.971715) (0.967273,0.967872) (0.973636,0.963910) (0.980000,0.959825) (0.986364,0.955611) (0.992727,0.951262) (0.999091,0.946772) (1.005455,0.942134) (1.011818,0.937340) (1.018182,0.932383) (1.024545,0.927255) (1.030909,0.921947) (1.037273,0.916449) (1.043636,0.910752) (1.050000,0.904844) (1.056364,0.898715) (1.062727,0.892351) (1.069091,0.885738) (1.075455,0.878864) (1.081818,0.871711) (1.088182,0.864262) (1.094545,0.856501) (1.100909,0.848405) (1.107273,0.839955) (1.113636,0.831125) (1.120000,0.821891) (1.126364,0.812224) (1.132727,0.802093) (1.139091,0.791465) (1.145455,0.780302) (1.151818,0.768563) (1.158182,0.756203) (1.164545,0.743171) (1.170909,0.729411) (1.177273,0.714862) (1.183636,0.699453) (1.190000,0.683107) (1.196364,0.665735) (1.202727,0.647239) (1.209091,0.627506) (1.215455,0.606408) (1.221818,0.583799) (1.228182,0.559511) (1.234545,0.533350) (1.240909,0.505093) (1.247273,0.474476) (1.253636,0.441193) (1.260000,0.404879) (1.266364,0.365102) (1.272727,0.321342) (1.279091,0.272971) (1.285455,0.219220) (1.291818,0.159140) (1.298182,0.091544) (1.304545,0.014929) (1.310909,0.000000) (1.317273,0.000000) (1.323636,0.000000) (1.330000,0.000000) (1.336364,0.000000) (1.342727,0.000000) (1.349091,0.000000) (1.355455,0.000000) (1.361818,0.000000) (1.368182,0.000000) (1.374545,0.000000) (1.380909,0.000000) (1.387273,0.000000) (1.393636,0.000000) (1.400000,0.000000)};
\addlegendentry{$Z=0.1$}
\addplot[only marks,mark=*,mark size=1.8pt,blue!70!black,forget plot] coordinates {(0.967644,0.967644)};
\addplot[only marks,mark=o,mark size=1.8pt,blue!70!black,forget plot] coordinates {(0.011451,0.011451)};
\addplot[teal!75!black,thick] coordinates {(0.000000,0.000000) (0.005000,0.000000) (0.010000,0.000000) (0.015000,0.000000) (0.020000,0.000000) (0.025000,0.000000) (0.030000,0.000000) (0.035000,0.000000) (0.040000,0.000000) (0.045000,0.000000) (0.050000,0.000000) (0.055000,0.000000) (0.060000,0.000000) (0.065000,0.061363) (0.070000,0.136351) (0.075000,0.201162) (0.080000,0.257705) (0.085000,0.307442) (0.090000,0.351507) (0.095000,0.390798) (0.100000,0.426030) (0.105000,0.457784) (0.110000,0.486535) (0.115000,0.512675) (0.120000,0.536530) (0.125000,0.558374) (0.130000,0.578439) (0.135000,0.596923) (0.140000,0.613995) (0.145000,0.629801) (0.150000,0.644467) (0.155000,0.658102) (0.160000,0.670804) (0.165000,0.682656) (0.170000,0.693733) (0.175000,0.704100) (0.180000,0.713817) (0.185000,0.722936) (0.190000,0.731502) (0.195000,0.739559) (0.200000,0.747143) (0.205000,0.754288) (0.210000,0.761025) (0.215000,0.767382) (0.220000,0.773384) (0.225000,0.779054) (0.230000,0.784413) (0.235000,0.789479) (0.240000,0.794270) (0.245000,0.798803) (0.250000,0.803091) (0.255000,0.807149) (0.260000,0.810989) (0.265000,0.814622) (0.270000,0.818060) (0.275000,0.821310) (0.280000,0.824384) (0.285000,0.827289) (0.290000,0.830033) (0.295000,0.832623) (0.300000,0.835065) (0.305000,0.837367) (0.310000,0.839535) (0.315000,0.841572) (0.320000,0.843485) (0.325000,0.845278) (0.330000,0.846955) (0.335000,0.848522) (0.340000,0.849980) (0.345000,0.851335) (0.350000,0.852589) (0.355000,0.853746) (0.360000,0.854808) (0.365000,0.855778) (0.370000,0.856659) (0.375000,0.857452) (0.380000,0.858161) (0.385000,0.858788) (0.390000,0.859333) (0.395000,0.859799) (0.400000,0.860188) (0.405000,0.860501) (0.410000,0.860740) (0.415000,0.860905) (0.420000,0.860999) (0.425000,0.861021) (0.430000,0.860974) (0.435000,0.860858) (0.440000,0.860673) (0.445000,0.860422) (0.450000,0.860103) (0.455000,0.859718) (0.460000,0.859268) (0.465000,0.858753) (0.470000,0.858172) (0.475000,0.857527) (0.480000,0.856818) (0.485000,0.856044) (0.490000,0.855207) (0.495000,0.854305) (0.500000,0.853339) (0.505000,0.852309) (0.510000,0.851214) (0.515000,0.850054) (0.520000,0.848830) (0.525000,0.847540) (0.530000,0.846185) (0.535000,0.844763) (0.540000,0.843274) (0.545000,0.841718) (0.550000,0.840093) (0.555000,0.838399) (0.560000,0.836636) (0.565000,0.834802) (0.570000,0.832896) (0.575000,0.830918) (0.580000,0.828866) (0.585000,0.826738) (0.590000,0.824535) (0.595000,0.822253) (0.600000,0.819892) (0.605000,0.817451) (0.610000,0.814928) (0.615000,0.812320) (0.620000,0.809626) (0.625000,0.806845) (0.630000,0.803973) (0.635000,0.801010) (0.640000,0.797953) (0.645000,0.794798) (0.650000,0.791545) (0.655000,0.788190) (0.660000,0.784730) (0.665000,0.781162) (0.670000,0.777484) (0.675000,0.773692) (0.680000,0.769782) (0.685000,0.765752) (0.690000,0.761596) (0.695000,0.757312) (0.700000,0.752894) (0.705000,0.748339) (0.710000,0.743642) (0.715000,0.738797) (0.720000,0.733800) (0.725000,0.728644) (0.730000,0.723325) (0.735000,0.717835) (0.740000,0.712168) (0.745000,0.706318) (0.750000,0.700276) (0.755000,0.694035) (0.760000,0.687587) (0.765000,0.680923) (0.770000,0.674034) (0.775000,0.666909) (0.780000,0.659538) (0.785000,0.651910) (0.790000,0.644013) (0.795000,0.635833) (0.800000,0.627358) (0.805000,0.618573) (0.810000,0.609463) (0.815000,0.600009) (0.820000,0.590196) (0.825000,0.580002) (0.830000,0.569408) (0.835000,0.558392) (0.840000,0.546929) (0.845000,0.534994) (0.850000,0.522559) (0.855000,0.509593) (0.860000,0.496064) (0.865000,0.481936) (0.870000,0.467171) (0.875000,0.451726) (0.880000,0.435556) (0.885000,0.418611) (0.890000,0.400835) (0.895000,0.382168) (0.900000,0.362543) (0.905000,0.341888) (0.910000,0.320122) (0.915000,0.297155) (0.920000,0.272886) (0.925000,0.247206) (0.930000,0.219990) (0.935000,0.191098) (0.940000,0.160374) (0.945000,0.127642) (0.950000,0.092701) (0.955000,0.055322) (0.960000,0.015245) (0.965000,0.000000) (0.970000,0.000000) (0.975000,0.000000) (0.980000,0.000000) (0.985000,0.000000) (0.990000,0.000000) (0.995000,0.000000) (1.000000,0.000000) (1.005000,0.000000) (1.010000,0.000000) (1.015000,0.000000) (1.020000,0.000000) (1.025000,0.000000) (1.030000,0.000000) (1.035000,0.000000) (1.040000,0.000000) (1.045000,0.000000) (1.050000,0.000000) (1.055000,0.000000) (1.060000,0.000000) (1.065000,0.000000) (1.070000,0.000000) (1.075000,0.000000) (1.080000,0.000000) (1.085000,0.000000) (1.090000,0.000000) (1.095000,0.000000) (1.100000,0.000000)};
\addlegendentry{$Z=0.4$}
\addplot[only marks,mark=*,mark size=1.8pt,teal!75!black,forget plot] coordinates {(0.726775,0.726775)};
\addplot[only marks,mark=o,mark size=1.8pt,teal!75!black,forget plot] coordinates {(0.065241,0.065241)};
\addplot[only marks,mark=*,black,mark size=1.8pt,forget plot] coordinates {(0.000000,0.000000)};
\end{axis}
\end{tikzpicture}\hfill
\begin{tikzpicture}
\begin{axis}[width=.482\textwidth,height=.35\textwidth,xmin=0,xmax=.8,ymin=0,ymax=1.08,
grid=major,grid style={gray!20},title={(b) Equilibria and collapse basin},
xlabel={Junk reaching review $Z$},ylabel={Genuine submissions $N^*$},
title style={font=\small},label style={font=\scriptsize},tick label style={font=\scriptsize},
legend columns=3,legend style={at={(.5,-.29)},anchor=north,font=\scriptsize,draw=none}]
\addplot[draw=none,fill=gray!21,forget plot] coordinates {(0.001000,0.000000) (0.001000,0.000105) (0.006951,0.000732) (0.012902,0.001366) (0.018853,0.002006) (0.024804,0.002653) (0.030754,0.003307) (0.036705,0.003968) (0.042656,0.004635) (0.048607,0.005310) (0.054558,0.005992) (0.060509,0.006681) (0.066460,0.007378) (0.072411,0.008082) (0.078362,0.008794) (0.084312,0.009514) (0.090263,0.010242) (0.096214,0.010978) (0.102165,0.011723) (0.108116,0.012476) (0.114067,0.013238) (0.120018,0.014009) (0.125969,0.014789) (0.131920,0.015577) (0.137871,0.016376) (0.143821,0.017184) (0.149772,0.018002) (0.155723,0.018829) (0.161674,0.019667) (0.167625,0.020516) (0.173576,0.021375) (0.179527,0.022244) (0.185478,0.023125) (0.191429,0.024018) (0.197379,0.024922) (0.203330,0.025838) (0.209281,0.026766) (0.215232,0.027706) (0.221183,0.028659) (0.227134,0.029625) (0.233085,0.030604) (0.239036,0.031597) (0.244987,0.032604) (0.250937,0.033626) (0.256888,0.034662) (0.262839,0.035713) (0.268790,0.036779) (0.274741,0.037861) (0.280692,0.038960) (0.286643,0.040075) (0.292594,0.041207) (0.298545,0.042357) (0.304496,0.043525) (0.310446,0.044712) (0.316397,0.045917) (0.322348,0.047142) (0.328299,0.048388) (0.334250,0.049654) (0.340201,0.050942) (0.346152,0.052252) (0.352103,0.053585) (0.358054,0.054941) (0.364004,0.056322) (0.369955,0.057728) (0.375906,0.059160) (0.381857,0.060618) (0.387808,0.062105) (0.393759,0.063620) (0.399710,0.065165) (0.405661,0.066740) (0.411612,0.068348) (0.417562,0.069989) (0.423513,0.071665) (0.429464,0.073376) (0.435415,0.075125) (0.441366,0.076913) (0.447317,0.078741) (0.453268,0.080612) (0.459219,0.082527) (0.465170,0.084488) (0.471121,0.086497) (0.477071,0.088558) (0.483022,0.090672) (0.488973,0.092841) (0.494924,0.095070) (0.500875,0.097361) (0.506826,0.099717) (0.512777,0.102143) (0.518728,0.104643) (0.524679,0.107220) (0.530629,0.109881) (0.536580,0.112630) (0.542531,0.115474) (0.548482,0.118420) (0.554433,0.121475) (0.560384,0.124648) (0.566335,0.127949) (0.572286,0.131389) (0.578237,0.134980) (0.584187,0.138739) (0.590138,0.142681) (0.596089,0.146827) (0.602040,0.151202) (0.607991,0.155835) (0.613942,0.160762) (0.619893,0.166029) (0.625844,0.171693) (0.631795,0.177830) (0.637745,0.184542) (0.643696,0.191974) (0.649647,0.200345) (0.655598,0.210007) (0.655698,0.210184) (0.656060,0.210827) (0.656422,0.211477) (0.656784,0.212134) (0.657145,0.212800) (0.657507,0.213473) (0.657869,0.214154) (0.658231,0.214843) (0.658593,0.215541) (0.658954,0.216248) (0.659316,0.216965) (0.659678,0.217690) (0.660040,0.218426) (0.660402,0.219172) (0.660764,0.219928) (0.661125,0.220695) (0.661487,0.221474) (0.661849,0.222264) (0.662211,0.223067) (0.662573,0.223882) (0.662934,0.224711) (0.663296,0.225554) (0.663658,0.226411) (0.664020,0.227283) (0.664382,0.228171) (0.664743,0.229076) (0.665105,0.229998) (0.665467,0.230939) (0.665829,0.231900) (0.666191,0.232880) (0.666553,0.233883) (0.666914,0.234909) (0.667276,0.235959) (0.667638,0.237035) (0.668000,0.238139) (0.668362,0.239273) (0.668723,0.240440) (0.669085,0.241641) (0.669447,0.242881) (0.669809,0.244161) (0.670171,0.245487) (0.670532,0.246862) (0.670894,0.248292) (0.671256,0.249785) (0.671618,0.251347) (0.671980,0.252988) (0.672342,0.254721) (0.672703,0.256562) (0.673065,0.258531) (0.673427,0.260657) (0.673789,0.262982) (0.674151,0.265569) (0.674512,0.268525) (0.674874,0.272058) (0.675236,0.276707) (0.675598,0.287815) (0.675598,0.288149) (0.675598,0.824402) (0.800000,0.700000) (0.800000,0.000000) (0.001000,0.000000)};
\addplot[blue!70!black,thick] coordinates {(0.001000,1.000000) (0.006951,1.000000) (0.012902,1.000000) (0.018853,1.000000) (0.024804,1.000000) (0.030754,1.000000) (0.036705,1.000000) (0.042656,1.000000) (0.048607,0.999028) (0.054558,0.995534) (0.060509,0.992004) (0.066460,0.988437) (0.072411,0.984833) (0.078362,0.981193) (0.084312,0.977516) (0.090263,0.973801) (0.096214,0.970050) (0.102165,0.966261) (0.108116,0.962436) (0.114067,0.958572) (0.120018,0.954672) (0.125969,0.950734) (0.131920,0.946758) (0.137871,0.942745) (0.143821,0.938693) (0.149772,0.934604) (0.155723,0.930478) (0.161674,0.926313) (0.167625,0.922110) (0.173576,0.917869) (0.179527,0.913590) (0.185478,0.909272) (0.191429,0.904916) (0.197379,0.900522) (0.203330,0.896089) (0.209281,0.891617) (0.215232,0.887107) (0.221183,0.882558) (0.227134,0.877970) (0.233085,0.873342) (0.239036,0.868676) (0.244987,0.863970) (0.250937,0.859225) (0.256888,0.854440) (0.262839,0.849616) (0.268790,0.844751) (0.274741,0.839847) (0.280692,0.834902) (0.286643,0.829916) (0.292594,0.824890) (0.298545,0.819823) (0.304496,0.814714) (0.310446,0.809564) (0.316397,0.804372) (0.322348,0.799138) (0.328299,0.793861) (0.334250,0.788542) (0.340201,0.783179) (0.346152,0.777773) (0.352103,0.772322) (0.358054,0.766827) (0.364004,0.761286) (0.369955,0.755700) (0.375906,0.750068) (0.381857,0.744388) (0.387808,0.738661) (0.393759,0.732885) (0.399710,0.727060) (0.405661,0.721185) (0.411612,0.715259) (0.417562,0.709282) (0.423513,0.703251) (0.429464,0.697165) (0.435415,0.691025) (0.441366,0.684827) (0.447317,0.678572) (0.453268,0.672256) (0.459219,0.665880) (0.465170,0.659440) (0.471121,0.652935) (0.477071,0.646362) (0.483022,0.639720) (0.488973,0.633006) (0.494924,0.626217) (0.500875,0.619351) (0.506826,0.612403) (0.512777,0.605371) (0.518728,0.598250) (0.524679,0.591036) (0.530629,0.583724) (0.536580,0.576309) (0.542531,0.568786) (0.548482,0.561146) (0.554433,0.553384) (0.560384,0.545490) (0.566335,0.537454) (0.572286,0.529267) (0.578237,0.520915) (0.584187,0.512383) (0.590138,0.503654) (0.596089,0.494709) (0.602040,0.485523) (0.607991,0.476066) (0.613942,0.466303) (0.619893,0.456189) (0.625844,0.445666) (0.631795,0.434659) (0.637745,0.423065) (0.643696,0.410739) (0.649647,0.397464) (0.655598,0.382887) (0.655698,0.382628) (0.656060,0.381685) (0.656422,0.380736) (0.656784,0.379779) (0.657145,0.378815) (0.657507,0.377842) (0.657869,0.376862) (0.658231,0.375873) (0.658593,0.374875) (0.658954,0.373869) (0.659316,0.372853) (0.659678,0.371828) (0.660040,0.370792) (0.660402,0.369747) (0.660764,0.368691) (0.661125,0.367624) (0.661487,0.366546) (0.661849,0.365456) (0.662211,0.364353) (0.662573,0.363238) (0.662934,0.362109) (0.663296,0.360967) (0.663658,0.359809) (0.664020,0.358637) (0.664382,0.357449) (0.664743,0.356244) (0.665105,0.355021) (0.665467,0.353780) (0.665829,0.352520) (0.666191,0.351239) (0.666553,0.349936) (0.666914,0.348610) (0.667276,0.347259) (0.667638,0.345882) (0.668000,0.344478) (0.668362,0.343043) (0.668723,0.341576) (0.669085,0.340074) (0.669447,0.338534) (0.669809,0.336953) (0.670171,0.335327) (0.670532,0.333651) (0.670894,0.331919) (0.671256,0.330126) (0.671618,0.328263) (0.671980,0.326321) (0.672342,0.324287) (0.672703,0.322145) (0.673065,0.319875) (0.673427,0.317448) (0.673789,0.314822) (0.674151,0.311934) (0.674512,0.308677) (0.674874,0.304843) (0.675236,0.299892) (0.675598,0.288483) (0.675598,0.288149)};
\addlegendentry{High credibility}
\addplot[blue!70!black,dashed,thick] coordinates {(0.001000,0.000105) (0.006951,0.000732) (0.012902,0.001366) (0.018853,0.002006) (0.024804,0.002653) (0.030754,0.003307) (0.036705,0.003968) (0.042656,0.004635) (0.048607,0.005310) (0.054558,0.005992) (0.060509,0.006681) (0.066460,0.007378) (0.072411,0.008082) (0.078362,0.008794) (0.084312,0.009514) (0.090263,0.010242) (0.096214,0.010978) (0.102165,0.011723) (0.108116,0.012476) (0.114067,0.013238) (0.120018,0.014009) (0.125969,0.014789) (0.131920,0.015577) (0.137871,0.016376) (0.143821,0.017184) (0.149772,0.018002) (0.155723,0.018829) (0.161674,0.019667) (0.167625,0.020516) (0.173576,0.021375) (0.179527,0.022244) (0.185478,0.023125) (0.191429,0.024018) (0.197379,0.024922) (0.203330,0.025838) (0.209281,0.026766) (0.215232,0.027706) (0.221183,0.028659) (0.227134,0.029625) (0.233085,0.030604) (0.239036,0.031597) (0.244987,0.032604) (0.250937,0.033626) (0.256888,0.034662) (0.262839,0.035713) (0.268790,0.036779) (0.274741,0.037861) (0.280692,0.038960) (0.286643,0.040075) (0.292594,0.041207) (0.298545,0.042357) (0.304496,0.043525) (0.310446,0.044712) (0.316397,0.045917) (0.322348,0.047142) (0.328299,0.048388) (0.334250,0.049654) (0.340201,0.050942) (0.346152,0.052252) (0.352103,0.053585) (0.358054,0.054941) (0.364004,0.056322) (0.369955,0.057728) (0.375906,0.059160) (0.381857,0.060618) (0.387808,0.062105) (0.393759,0.063620) (0.399710,0.065165) (0.405661,0.066740) (0.411612,0.068348) (0.417562,0.069989) (0.423513,0.071665) (0.429464,0.073376) (0.435415,0.075125) (0.441366,0.076913) (0.447317,0.078741) (0.453268,0.080612) (0.459219,0.082527) (0.465170,0.084488) (0.471121,0.086497) (0.477071,0.088558) (0.483022,0.090672) (0.488973,0.092841) (0.494924,0.095070) (0.500875,0.097361) (0.506826,0.099717) (0.512777,0.102143) (0.518728,0.104643) (0.524679,0.107220) (0.530629,0.109881) (0.536580,0.112630) (0.542531,0.115474) (0.548482,0.118420) (0.554433,0.121475) (0.560384,0.124648) (0.566335,0.127949) (0.572286,0.131389) (0.578237,0.134980) (0.584187,0.138739) (0.590138,0.142681) (0.596089,0.146827) (0.602040,0.151202) (0.607991,0.155835) (0.613942,0.160762) (0.619893,0.166029) (0.625844,0.171693) (0.631795,0.177830) (0.637745,0.184542) (0.643696,0.191974) (0.649647,0.200345) (0.655598,0.210007) (0.655698,0.210184) (0.656060,0.210827) (0.656422,0.211477) (0.656784,0.212134) (0.657145,0.212800) (0.657507,0.213473) (0.657869,0.214154) (0.658231,0.214843) (0.658593,0.215541) (0.658954,0.216248) (0.659316,0.216965) (0.659678,0.217690) (0.660040,0.218426) (0.660402,0.219172) (0.660764,0.219928) (0.661125,0.220695) (0.661487,0.221474) (0.661849,0.222264) (0.662211,0.223067) (0.662573,0.223882) (0.662934,0.224711) (0.663296,0.225554) (0.663658,0.226411) (0.664020,0.227283) (0.664382,0.228171) (0.664743,0.229076) (0.665105,0.229998) (0.665467,0.230939) (0.665829,0.231900) (0.666191,0.232880) (0.666553,0.233883) (0.666914,0.234909) (0.667276,0.235959) (0.667638,0.237035) (0.668000,0.238139) (0.668362,0.239273) (0.668723,0.240440) (0.669085,0.241641) (0.669447,0.242881) (0.669809,0.244161) (0.670171,0.245487) (0.670532,0.246862) (0.670894,0.248292) (0.671256,0.249785) (0.671618,0.251347) (0.671980,0.252988) (0.672342,0.254721) (0.672703,0.256562) (0.673065,0.258531) (0.673427,0.260657) (0.673789,0.262982) (0.674151,0.265569) (0.674512,0.268525) (0.674874,0.272058) (0.675236,0.276707) (0.675598,0.287815) (0.675598,0.288149)};
\addlegendentry{Unstable}
\addplot[black,thick] coordinates {(0.000000,0.000000) (0.800000,0.000000)};
\addlegendentry{Collapse}
\addplot[gray,densely dotted,forget plot] coordinates {(0.675598,0.000000) (0.675598,0.288149)};
\node[font=\scriptsize,anchor=east] at (axis cs:0.662598,.39) {$\bar Z\approx0.676$};
\end{axis}
\end{tikzpicture}
\caption{The credibility trap in a numerical example. Panel (a): best responses $BR(N;Z)$ for two junk loads, with the $45^\circ$ line; equilibria are its intersections with each curve, filled if stable and hollow if unstable. Panel (b): equilibrium genuine submissions as the junk load rises; the high-credibility branch (solid) meets the unstable branch (dashed) at $\bar Z \approx 0.68$, beyond which only collapse remains. The shaded region is the basin of attraction of collapse, $[0,N^U(Z))$ for $Z<\bar Z$ and the whole domain beyond. Parameters: $\chi = \kappa = 1$, $\rho = 0.1$, $a_0 = b_0 = 1$, $\gamma^T = 1$, $\alpha_T r = 0.5$, $\alpha_E e = 2$, $\beta_E e = 0.25$, junk consisting of marginal papers with $\psi_m = 1$, outside option $P$ with the same screen ($\Phi_o = \Phi_P \approx 1.49$), $\tau = 1$, $\mu = 1.5$, $n_v = 1$, $\theta \sim U[0.5, 2.5]$.}
\label{fig:trap}
\end{figure}
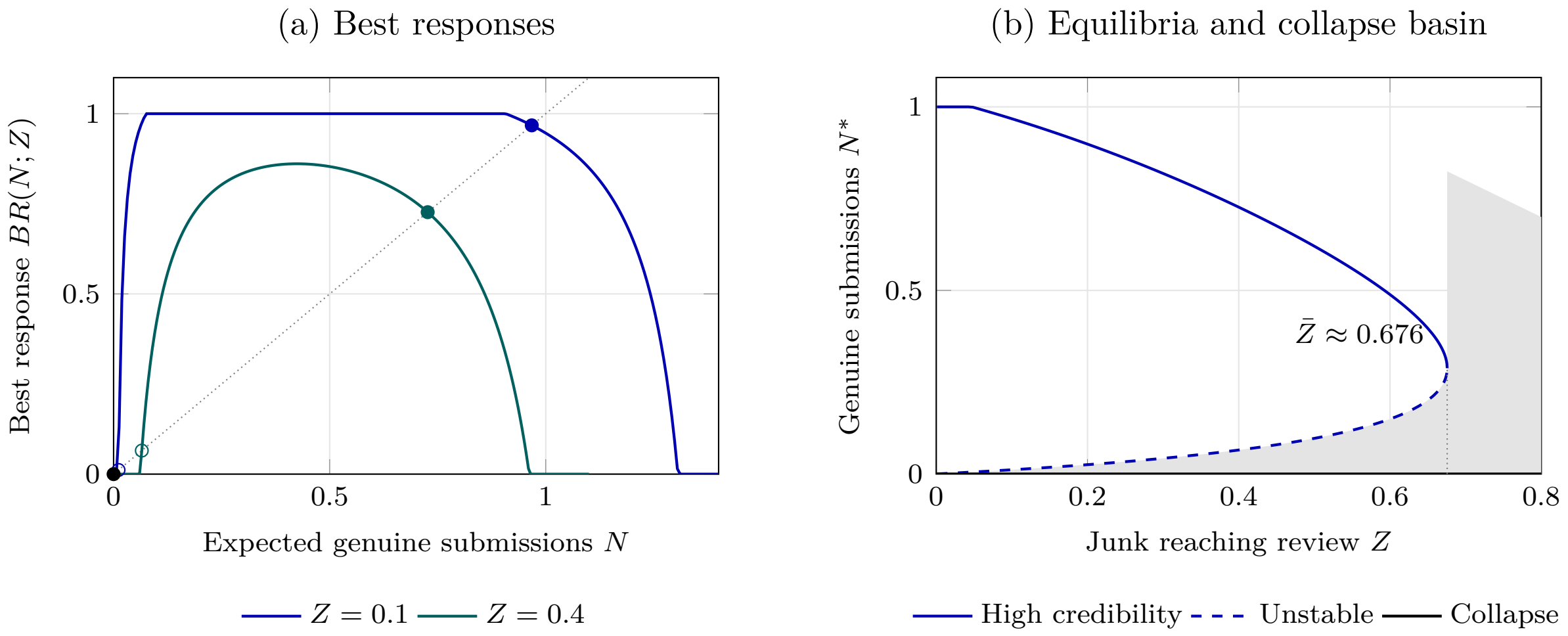

\paragraph{Numerical illustration.} Junk in the example consists of marginal papers, which are valid ($\gamma^T=1$) and have the same average endorsement rate as genuine contributions ($\psi=\psi_m=1$). This leakage can be implemented by the perception-error construction in Section~\ref{sec:review}; it is not inferred from the unbiased signal model of Section~\ref{sec:bayes}, and Section~\ref{sec:adaptive} shows how it changes when the review weight adapts. Figure~\ref{fig:trap} uses $A_J(\gamma)=1.5+2\gamma$ and $B_J(\gamma)=1-0.25\gamma$. Thus reviewers penalise additional polish within the genuine production technology even though they sometimes mistake marginal work for genuine contributions. The function $\Phi_J$ is strictly increasing, and Assumption~\ref{ass:interior} holds for all $\gamma\in[0,1]$. The outside option has the same screen, so $\Phi_J(0)=\Phi_o\approx1.49<\Phi_J(1)\approx6.21$: a zero-credibility journal differs from that option only by its waiting cost, while the trap requires the participation--credibility feedback. At $Z=0.1$, equilibria are $N^*\in\{0,\ 0.01,\ 0.97\}$; at $Z=0.4$, $N^*\in\{0,\ 0.07,\ 0.73\}$. The basin of collapse grows from $[0,0.011)$ to $[0,0.065)$, compared with lower bounds $N_0\approx0.011$ and $0.058$. The high-credibility equilibrium falls from $0.97$ to $0.73$, then disappears at $\bar Z\approx0.68$, where genuine submissions are still about $0.29$. A small further increase in junk sends them to zero.

\subsection{An integrated equilibrium}\label{sec:integrated}

The analysis so far holds different objects fixed in different places: Section~\ref{sec:review} fixes submissions, and Theorem~\ref{thm:trap} fixes the junk load and validation credibility. We now close the model. Genuine authors choose effort (Lemma~\ref{lem:effort}) and whether to submit; flawed and marginal papers are generated as in~\eqref{eq:generation}, but generators may now value the journal's credibility; endorsement credibility, validation credibility and delay follow from who is in the pool. The review architecture $(r,e,\delta,\zeta)$, the reviewers' weights and the leakage rates $\psi_f,\psi_m$ are held fixed, as is the outside option.

\begin{assumption}\label{ass:integrated}
(i) $v_J = v_o$ and $\pi_0 = 0$. (ii) The generators' values $V_f(\gamma)$ and $V_m(\gamma)$ are continuous and non-decreasing in $\gamma$, so that the junk loads $Z(\gamma)$ and $\tilde Z(\gamma)$ obtained from~\eqref{eq:load}, \eqref{eq:gamma} and~\eqref{eq:generation} are continuous and non-decreasing; $\tilde Z(0) > 0$ and $Z(0) < \mu$. (iii) The journal's value $\Phi_J(\gamma,\gamma^T)$ from Lemma~\ref{lem:effort} is continuous on $[0,1]\times[0,1]$, and $\Phi_J(0,\gamma^T) \le \Phi_o$ for all $\gamma^T$.
\end{assumption}

Given an expected rate of genuine submissions $N \ge 0$, credibility must be consistent with the junk it attracts: $\gamma = N/(N + \tilde Z(\gamma))$. Because the right-hand side is non-increasing in $\gamma$, this has a unique solution $\gamma(N)$, which is continuous and non-decreasing in $N$ with $\gamma(0) = 0$ (Appendix). Validation credibility is then $\gamma^T(N) = (N + N_m)/(N + N_m + (1-\delta)N_f)$ evaluated at $\gamma(N)$, the delay is $T(N) = 1/(\mu - N - Z(\gamma(N)))$ on the interval $\mathcal D = [0,\bar N)$ where $N + Z(\gamma(N)) < \mu$, and the best response $BR(N)$ is~\eqref{eq:BR} with $\Phi_J(\gamma(N),\gamma^T(N))$ in place of $\Phi_J(\gamma(N))$. An \emph{integrated equilibrium} is a fixed point $N^* = BR(N^*)$.

\begin{proposition}[Integrated equilibrium]\label{prop:integrated}
Under Assumption~\ref{ass:integrated}:
\begin{enumerate}[label=(\alph*)]
\item $BR$ is continuous on $\mathcal D$, and $N^* = 0$ is a stable integrated equilibrium: there is $N_0^{\mathrm{int}} > 0$ with $BR(N) = 0$ for all $N \in [0, N_0^{\mathrm{int}}]$.
\item If $BR(\hat N) > \hat N$ for some $\hat N \in \mathcal D$, a positive integrated equilibrium exists; the largest one exceeds $N_0^{\mathrm{int}}$ and is stable from above.
\item Let generation cost shift with $\omega\ge0$, with Assumption~\ref{ass:integrated} holding throughout the parameter range. Suppose a higher $\omega$ weakly raises the induced load $Z(\gamma(N;\omega);\omega)$ and weakly lowers the induced value $\Phi_J(\gamma(N;\omega),\gamma^T(N;\omega))$ at every $N$ in the common feasible domain. Then $BR(N;\omega)$ is non-increasing in $\omega$, and the set of $\omega$ admitting a positive equilibrium is a lower interval. Define $\bar\omega$ as its supremum, with $\bar\omega=0$ for the empty set and allowing $\bar\omega=+\infty$. A positive equilibrium exists for $0\le\omega<\bar\omega$; for $\omega>\bar\omega$, $BR(N;\omega)<N$ at every feasible $N>0$, so only collapse remains. The largest positive equilibrium is non-increasing wherever it exists. For any $0\le\omega_1<\bar\omega$, it is bounded below by $N_0^{\mathrm{int}}(\omega_1)>0$ on $[\omega_1,\bar\omega)$. Thus, if $0<\bar\omega<\infty$, the maximal equilibrium falls discontinuously to zero across this boundary. A finite boundary is ensured if there are finite $\omega_L<\omega_H$ with $BR(\hat N;\omega_L)>\hat N$ at some feasible $\hat N>0$ and $BR(N;\omega_H)<N$ at every feasible $N>0$; then $\bar\omega\in[\omega_L,\omega_H]$. Monotonicity alone does not ensure a finite boundary.
\end{enumerate}
\end{proposition}

Proposition~\ref{prop:integrated} preserves the credibility trap of Theorem~\ref{thm:trap} when effort, both credibilities, delay, participation and the junk load adjust jointly, conditional on fixed reviewer weights and leakage rates. Collapse remains stable when some junk is endorsed even at zero credibility. A positive equilibrium exists under the participation condition in part (b), and, if cheaper generation eventually eliminates every positive equilibrium, the maximal equilibrium falls discontinuously at a finite boundary. The proposition does not assert that every cost-reduction path reaches such a boundary or that adjustment selects the maximal equilibrium.

The monotonicity conditions in part (c) must be checked on the induced load and value, rather than only on generation costs. More marginal papers can raise validation credibility because marginal papers are valid, potentially offsetting a loss of endorsement credibility. One sufficient case is $\alpha_T r=0$, common positive leakage $\psi_f=\psi_m=\psi$, a journal value non-decreasing in $\gamma$, and generation shifts that raise each junk stream at every fixed $\gamma$. The inner fixed point then implies that $\gamma(N;\omega)$ falls and, for $N>0$, the induced load $Z=N(1-\gamma)/(\psi\gamma)$ rises; at $N=0$ the load rises directly. Even in this case a finite tipping boundary needs a sufficiently large shock.

For example, retain the parameters of Figure~\ref{fig:trap}, use marginal junk only, and let $Z(\omega)=0.1+0.1(1-e^{-\omega})$. This is an admissible special case with junk values constant in credibility and a generation cost shift that increases submissions. The load rises but stays below $0.2$. Throughout that range $BR(0.5;\omega)=1>0.5$, so a positive equilibrium persists for every finite $\omega$. The example distinguishes the possibility of tipping from its inevitability.

\begin{remark}[Marginal papers from genuine authors]\label{rem:genuine-marginal}
The model draws marginal papers from a separate population of generators. If genuine authors also produce them, two cases differ.
(i) \emph{Quantity margin.} Genuine authors may submit marginal side papers in addition to their main work. If an author's decision to do so does not depend on her own genuine submission choice, and the expected value of a side paper depends on the journal only through its pass probability and credibility, these authors form a generator population with value $V_m(\gamma)$ in~\eqref{eq:generation}. Assumption~\ref{ass:integrated}(ii) then covers them, and Proposition~\ref{prop:integrated} applies unchanged.
(ii) \emph{Substitution margin.} Authors of low capability may submit a marginal paper instead of a genuine one. The same cutoffs then determine $N_v$ and part of $N_m$, so the junk load depends on the rate of genuine submissions directly, not only through credibility. Higher credibility raises the returns to both kinds of submission, and the induced load need not be monotone in $N$. Theorem~\ref{thm:trap}, which holds the junk load fixed, does not apply, and the monotonicity conditions of Proposition~\ref{prop:integrated}(c) must be verified for the joint cutoff structure. We do not analyse this case.
Theorems~\ref{thm:look} and~\ref{thm:rational} are unaffected in both cases, because they concern the effort of authors who submit genuine work. Substitution also strengthens the logic of Proposition~\ref{prop:aggregate}: other things equal, an author who replaces a genuine paper by a marginal one lowers aggregate output by more than one who moves to the outside option.
\end{remark}

\begin{remark}[Adaptive review in the integrated model]\label{rem:adaptive-int}
Proposition~\ref{prop:integrated} holds review weights and leakage rates fixed. Suppose instead that the review weight adapts as in Section~\ref{sec:adaptive}, so that leakage $\psi_m(t(N))$ and the journal's value $\Phi_J(\gamma,\gamma^T,t)$ depend on a selected weight $t(N)$. Parts (a) and (b) continue to hold, with the same proofs, if the selected inner solution $(\gamma(N),t(N))$ is continuous in $N$, $t(N)$ remains in a compact set, and the endorsed junk load is bounded away from zero near $N=0$. Then $\gamma(N)\to0$ as $N\downarrow0$, and $\Phi_J(0,\gamma^T,t)$ does not depend on $t$, because the endorsement slope $S$ vanishes at $\gamma=0$. Part~(c) can fail. When polished junk is common in the contested pool, more junk can lower the review weight and hence leakage (Lemma~\ref{lem:contested}), so the induced endorsed load need not rise with generation volume; and the example in Section~\ref{sec:adaptive} shows that the selected weight can jump. The tipping result then requires verifying the monotonicity and continuity conditions of part~(c) directly.
\end{remark}

\subsection{Reputation for substance}\label{sec:reputation}

Endorsement credibility counts only how many endorsed papers are genuine. Readers may also judge a certificate by how much substance the genuine papers behind it contain. A journal whose reviewers reward presentation endorses genuine papers with less substance, and its certificate can then be worth less even if no junk is endorsed. This subsection adds that channel and shows how it couples the incentive condition of Theorem~\ref{thm:look} with the participation analysis of Theorem~\ref{thm:trap}.

Let $J$ and $P$ apply the same screen with the same validation credibility, so that $A_P=a_0+\gamma^T\alpha_T r$ and $B_P=b_0$, and let $a_P:=(\kappa A_P-\rho B_P)/D>0$ be substantive effort per unit of capability on the platform. Set $\pi_0=0$ and let the journal endorse with probability~\eqref{eq:endorse} with $s_y=t s_x$. Genuine authors exert effort $(x,y)=\theta(a,b)$. Because the endorsement probability $e s_x\theta(a+tb)$ is proportional to capability, the mean substance of endorsed genuine papers is
\[
\bar x_E=\frac{\mathbb E[x\,\pi_J]}{\mathbb E[\pi_J]}=a\,\frac{\mathbb E[\theta^2]}{\mathbb E[\theta]},
\]
with expectations over genuine submitters. Readers value an endorsement at $\gamma V\,\bar x_E/\bar x_P=\gamma V a/a_P$, where $\bar x_P=a_P\,\mathbb E[\theta^2]/\mathbb E[\theta]$ is the mean substance the same authors would produce on the platform. The certificate's reputation is thus the substance it adds for a given pool of authors; absolute publication quality also reflects selection, which Section~\ref{sec:aggregate} treats separately. The benchmark of the previous sections values the certificate at $\bar x_E=\bar x_P$. Each author is negligible and takes $\bar x_E$ as given. With
\[
\sigma:=\frac{\gamma eVs_x}{a_P},\qquad u(t):=\frac{\sigma(\kappa-\rho t)}{D},
\]
the journal's reward slopes become $A_J=A_P+\sigma a$ and $B_J=B_P+\sigma t a$. A \emph{reputational equilibrium} is a level $a$ that authors' best responses to these slopes reproduce. The benchmark slopes are $A_J^0=A_P+\sigma a_P$ and $B_J^0=B_P+\sigma t a_P$, with substantive effort $a_J^0=a_P(1+u(t))$, and $\Phi^0_J(\gamma)$ and $\Phi^R_J(\gamma)$ denote the journal's value with the benchmark and the reputational slopes.

\begin{proposition}[Reputation for substance]\label{prop:reputation}
Let $\gamma>0$ and $u(t)<1$ (see Remark~\ref{rem:rep-existence}), and suppose that Assumption~\ref{ass:interior} holds at the benchmark and reputational slopes.
\begin{enumerate}[label=(\alph*)]
\item \emph{Equilibrium substance.} The reputational equilibrium is unique, $a_J=a_P/(1-u(t))$. The certificate's reputation $\bar x_E/\bar x_P=1/(1-u(t))$ exceeds one if and only if $t<\kappa/\rho$.
\item \emph{The incentive condition is unchanged.} $Q^J_\theta-Q^P_\theta=\theta\sigma a_J\bigl[(\kappa+\lambda\rho)-(\rho+\lambda\chi)t\bigr]/D$, so the journal improves on the platform if and only if $t<L(\kappa,\chi)$, as in Theorem~\ref{thm:look}(a).
\item \emph{Amplification.} $a_J-a_J^0=a_P\,u(t)^2/(1-u(t))\ge0$ and $da_J/dt=(da_J^0/dt)/(1-u(t))^2$. For weights $t_c<t'$ with $u(t_c),u(t')\in(0,1)$, the substantive effort gained by committing to $t_c$ rather than $t'$ exceeds its benchmark value by the factor $1/[(1-u(t_c))(1-u(t'))]>1$.
\item \emph{Participation and tipping.} Let $t\ge0$, and let the hypotheses hold for every $\gamma\in(0,1]$. Then $\Phi^R_J(0)=\Phi^0_J(0)=\Phi_P$, $\Phi^R_J$ is continuous and strictly increasing, and, for $\gamma>0$, $\Phi^R_J(\gamma)<\Phi^0_J(\gamma)$ if $t>\kappa/\rho$ and $\Phi^R_J(\gamma)>\Phi^0_J(\gamma)$ if $t<\kappa/\rho$. Consider the participation model of Theorem~\ref{thm:trap} with the platform as outside option. If $t>\kappa/\rho$, then $BR^R\le BR^0$ pointwise, so the tipping threshold is weakly lower, $\bar Z^R\le\bar Z^0$, and wherever both are defined the basin of collapse is weakly larger, $N^{U,R}(Z)\ge N^{U,0}(Z)$, and the largest positive equilibrium weakly smaller, $N^{H,R}(Z)\le N^{H,0}(Z)$. All inequalities reverse if $t<\kappa/\rho$.
\item \emph{Self-consistent weights.} In a genuine pool, let $r:=\sigma_\varepsilon^2/\sigma_u^2$. Every interior self-consistent linear review rule is $t=rb/a$, where $(a,b)$ solves
\[
(D-\kappa\sigma)\,a+\rho\sigma r\,b=\kappa A_P-\rho B_P,\qquad \rho\sigma\,a+(D-\chi\sigma r)\,b=\chi B_P-\rho A_P .
\]
If $\Delta_R:=(D-\kappa\sigma)(D-\chi\sigma r)-\rho^2\sigma^2r\ne0$, there is at most one interior self-consistent weight.
\end{enumerate}
\end{proposition}

The reputation channel does not move the threshold: the certificate's value scales both reward slopes by the same equilibrium substance, so the sign of the comparison with the platform is still governed by condition~\eqref{eq:C}. What changes is the magnitude and the link to participation. Below $\kappa/\rho$, the threshold for substantive effort alone ($L$ at $\lambda=0$), a certificate that induces more substance is worth more, which induces still more substance; the feedback amplifies the gain from a contribution-first rule. Above it, a review norm that rewards presentation lowers the substance behind the certificate and therefore its value to authors. Within the incentive channel this dampens the harm, because the devalued certificate also rewards presentation less. In the participation channel it lowers the journal's value at every level of credibility, relative to a certificate valued by its genuine share alone, so a smaller shock suffices to push the journal into the basin of collapse, and collapse becomes the only equilibrium at a lower junk load. The incentive condition of Theorem~\ref{thm:look} and the credibility trap of Theorem~\ref{thm:trap} are thus no longer separate: a review norm that weakens incentives for substance also makes the certificate more fragile than its credibility alone would indicate. This is a comparison across reader valuations at a given norm, not across norms (Remark~\ref{rem:norms}). Part~(e) shows a further consequence. Because the certificate's value moves with substance, the self-consistency condition becomes linear rather than quadratic, and the multiplicity of review norms in Theorem~\ref{thm:rational} disappears in this benchmark.

\begin{remark}[Existence of the reputational equilibrium]\label{rem:rep-existence}
The hypothesis $u(t)<1$ is a substantive restriction. For $t\ge0$, $u(t)\le u(0)=\sigma\kappa/D$, so $\sigma\kappa<D$ suffices; in part~(d), where the hypotheses must hold for every $\gamma\in(0,1]$, this requires $eVs_x\kappa<Da_P$. If $u(t)\ge1$, no reputational equilibrium exists in the linear model: the fixed-point condition $a(1-u(t))=a_P$ has no positive solution, because the feedback from substance to the certificate's value and back to substance has a gain of at least one. The restriction fails for strong certificates. In the parameters of Figure~\ref{fig:trap}, $eVs_x=2$ and $a_P\approx1.41$, so $u(0)\approx1.43$ at $\gamma=1$; at $t=0$ the hypothesis holds only for $\gamma<0.70$, and Proposition~\ref{prop:reputation}(d) does not cover that example. A bounded valuation removes the restriction. If readers value an endorsement at $\gamma V\,h(\bar x_E/\bar x_P)$ with $h$ continuous, non-negative and bounded, the equilibrium condition for $z:=a/a_P$ becomes $z=1+u(t)\,h(z)$, which has a solution for every $t\le\kappa/\rho$ by the intermediate value theorem. The linear case $h(z)=z$ is the unbounded one.
\end{remark}

\begin{remark}[Proposition~\ref{prop:reputation} does not rank review norms]\label{rem:norms}
Part~(d) compares, at a given weight $t$, a certificate valued by the substance behind it with one valued by its genuine share alone. It does not compare a journal whose reviewers reward presentation with one whose reviewers do not. Such a comparison runs the other way in the implementation used here. With $s_x$ fixed, a higher $t$ raises the reward slope on presentation, and by the envelope theorem $\partial\Phi/\partial B=y/\theta>0$. Under the count-based valuation, $A_J^0$ does not depend on $t$ and $B_J^0=B_P+\sigma ta_P$ rises with it, so $\Phi_J^0(\gamma)$ is strictly increasing in $t$ for every $\gamma>0$ as long as Assumption~\ref{ass:interior} holds; by the argument of part~(d), $BR^0$ is then pointwise non-decreasing in $t$, and the tipping threshold $\bar Z^0$ weakly rises. With the reputational valuation the effect is not signed analytically, but in the cost and platform parameters of Figure~\ref{fig:trap} with $\sigma=0.4$ (so $u(0)\approx0.40$), $\Phi_J^R$ rises from about $3.29$ at $t=0$ to $23.1$ at $t=\kappa/\rho=10$ and $33.2$ at $t=15$, against $\Phi_P\approx1.49$. Rewarding presentation is privately attractive to authors, which is also why the rule with the stronger incentive for substance attracts fewer genuine authors in Section~\ref{sec:policy-equilibrium}. Raising $t$ at fixed $s_x$ also raises every author's endorsement probability. An economically meaningful ranking of norms by their fragility therefore requires a normalisation, for instance a fixed acceptance rate, which we leave to future work.
\end{remark}

With linear reader valuation, the channel amplifies and couples but does not create additional equilibria in effort. A reader valuation that is convex in substance, for instance a certificate that is worth little unless the endorsed work clears a threshold of significance, could generate a second, low-substance equilibrium in which authors polish because the certificate is weak and the certificate is weak because authors polish. We leave that case to future work.

\subsection{Comparing review rules with endogenous participation}\label{sec:policy-equilibrium}

An editor choosing whether to commit to a contribution-first rule changes both the incentives authors face and the pool from which reviewers learn. We compare commitment to $t_c=0$ with an adaptive rule $t_a=w_y/w_s$ in a special case that joins Section~\ref{sec:adaptive} to the participation model. All low-contribution submissions are technically valid. We hold their inflow $Z_m>0$ and the journal's review capacity $\mu$ fixed across policies, and use an outside platform with the same validation screen. Thus $\gamma^T=1$ in both venues. The fixed inflow isolates the response of genuine authors to a change in the review rule. We set $v_J=v_o$ and $\pi_0'=0$, write $S_0=Ve s_x$, and measure substantive work with $\lambda=0$.

At a positive equilibrium under policy $i\in\{a,c\}$, authors with $\theta\ge \theta_i^*$ submit to the journal. Their capability distribution determines $\bar x_i$ and $\bar y_i$ through the effort coefficients $(a_i,b_i)$ in Lemma~\ref{lem:effort}, and enters the prediction weights in Lemma~\ref{lem:contested}. Let $\psi_i$ denote marginal papers' relative endorsement rate from~\eqref{eq:psit}. We solve
\begin{equation}\label{eq:policy-eq}
\begin{gathered}
N_i=n_v[1-G(\theta_i^*)],\quad
\theta_i^*(\Phi_J(\gamma_i,t_i)-\Phi_P)
  = \frac{\tau}{\mu-Z_m-N_i},\qquad
\gamma_i=\frac{N_i}{N_i+\psi_i Z_m},
\\[-1pt]
t_c=0,\quad \psi_c=\psi_m^0,\qquad
t_a=\frac{w_y}{w_s},\quad w_s>0 .
\end{gathered}
\end{equation}
The indifference equation applies at an interior cutoff. The weight $t_a$ is computed on the pool generated by that same cutoff and the effort choices it induces. A policy equilibrium therefore fixes both the review rule and authors' response to it.

\paragraph{Comparing the outcomes.}
The credibility ranking is exact: $\gamma_a>\gamma_c$ if and only if $N_a\psi_c>N_c\psi_a$. A rule can attract fewer genuine authors and still earn a more credible endorsement if it reduces the endorsement of marginal papers sufficiently. Let $a_o$ denote substantive effort per unit of capability on the outside platform, and define $M_i=n_v\int_{\theta_i^*}^{\theta_{\max}}\theta\,dG(\theta)$. Aggregate substantive work across the two venues is $X_i=a_o n_v\mathbb E[\theta]+(a_i-a_o)M_i$. Hence
\begin{equation}\label{eq:policy-diff}
\begin{split}
a_a-a_c &= \frac{S_0}{D}\left[\kappa(\gamma_a-\gamma_c)-\rho\gamma_a t_a\right],\\
X_a-X_c &= (a_a-a_c)M_a+(a_c-a_o)(M_a-M_c).
\end{split}
\end{equation}
The first term in the output difference is the change in substantive effort for the authors who choose the journal under adaptation. The second is the change due to participation, valued at the committed rule's effort gain over the platform. This gain satisfies $a_c-a_o=\kappa S_0\gamma_c/D>0$ in the interior benchmark. In particular, if $y_m>\bar y_a$, $t_a>0$, and $N_a\le N_c$, then $\psi_a>\psi_c$, $\gamma_a<\gamma_c$, and commitment yields higher aggregate substantive work. If $y_m>\bar y_a$, $t_a<0$, and $N_a\ge N_c$, adaptation yields higher credibility and higher aggregate substantive work. When participation moves in the opposite direction, the two margins must be compared explicitly.

\paragraph{A numerical illustration.} Appendix~\ref{app:policy-computation} computes both policy equilibria along locally stable positive-participation branches of a calibrated example. The ranking reverses as presentation becomes cheaper: where the adaptive weight on presentation is positive, commitment yields more aggregate substantive work and a more credible endorsement, and where it turns negative, adaptation does; in both cases the preferred rule attracts slightly fewer genuine authors.

\subsection{Average quality versus aggregate output}\label{sec:aggregate}

Finally, we distinguish the quality of what a journal publishes from the output of science as a whole. Journals and their evaluators typically observe the former; the two can move in opposite directions. Aggregate net epistemic output counts both regimes and subtracts a social cost $\eta_{\mathrm{harm}} \ge 0$ for each flawed paper that passes the technical screen:
\begin{equation}\label{eq:aggregate}
Q = n_v\Bigl[q_o \int_{\theta_{\min}}^{\theta^*} \theta\, dG(\theta) + q_J \int_{\theta^*}^{\theta_{\max}} \theta\, dG(\theta)\Bigr] - \eta_{\mathrm{harm}}\,(1-\delta)N_f .
\end{equation}

\begin{proposition}[Average quality versus aggregate output]\label{prop:aggregate}
Hold credibility fixed and suppose $q_J > q_o$, $q_J>0$, $\theta^* \in (\theta_{\min}, \theta_{\max})$ and $G$ has a positive density $g$ at $\theta^*$.
\begin{enumerate}[label=(\alph*)]
\item Holding $N_f$ fixed, $\partial Q/\partial \theta^* = -n_v(q_J - q_o)\,\theta^* g(\theta^*) < 0$, while the average capability of the journal's genuine authors, $\mathbb{E}[\theta \mid \theta \ge \theta^*]$, and hence their average output $q_J\,\mathbb{E}[\theta \mid \theta \ge \theta^*]$, is strictly increasing in $\theta^*$.
\item Consider an increase in the junk load $dZ > 0$ that comes with $dN_f \ge 0$. By Proposition~\ref{prop:fixed}(a) it raises $T^*$ and hence $\theta^*$, and
\[
dQ = -n_v(q_J - q_o)\,\theta^* g(\theta^*)\,d\theta^* - \eta_{\mathrm{harm}}\,(1-\delta)\,dN_f < 0 ,
\]
while the average capability of the journal's genuine authors rises. If the increase consists only of marginal papers, the second term vanishes and the decline in $Q$ operates through crowding out alone.
\end{enumerate}
\end{proposition}

Average quality is thus a misleading metric when generation is cheap: congestion selects the journal's genuine authors toward stronger ones, so the journal can look better while the system produces less. Two qualifications apply. First, Proposition~\ref{prop:aggregate} isolates the selection effect. Junk that is sometimes endorsed lowers credibility and, when~\eqref{eq:C} holds, $q_J$; the average output of genuine papers rises only if selection dominates, while the decline in $Q$ is reinforced. Second, the journal's endorsed pool also contains the undetected flawed papers, so the average quality of everything it endorses need not rise. With endogenous credibility, the same logic applies until the threshold $\bar Z$, at which point the decline in aggregate output becomes discontinuous.

\section{Empirical implications}\label{sec:empirics}

The model yields conditional implications for evaluation, effort allocation and participation. Editorial data of the kind analysed by \citet{gartenberg2026} can inform these mechanisms: submission records, textual measures of presentation, decision codes, turnaround times and author histories. Testing them requires separating changes in the review rule from changes in the submitted pool and in the technology of research. P1--P6 state implications under the conditions of the corresponding results; P7 is an auxiliary hypothesis.

\paragraph{P1. The review weight on presentation depends on the pool (Theorem~\ref{thm:rational}, Proposition~\ref{prop:junk}, Lemma~\ref{lem:contested}).} Among genuine papers whose presentation scales with capability, the self-consistent weight on presentation is positive. In a pool with polished marginal papers, it turns negative only once they are sufficiently common. A test would classify marginal and substantive contributions with assessments independent of the focal recommendation, measure their presentation, and examine whether the evaluation rule changes with the resulting covariance. A regression of acceptance on presentation and noisy substance proxies does not by itself identify $s_y/s_x$, and high desk-rejection rates of heavily AI-written submissions \citep{gartenberg2026} do not identify a negative response to polish, because such submissions may differ in substance.

\paragraph{P2. Cheaper polish changes effort and the relative value of certification (Lemma~\ref{prop:cheap}).} Holding reward slopes, credibility and resource competition fixed, a fall in $\kappa$ raises presentation and, when $\rho>0$, lowers substantive effort, even at a journal that looks through polish. Cheaper polish increases the advantage of a regime that induces less presentation effort. Exogenous access to writing tools could provide variation, but such tools may also reduce execution costs or the competition between tasks, so a test needs measures of both substantive activity and presentation and evidence on which cost margins changed.

\paragraph{P3. The relative value of triage rises with marginal submissions (Propositions~\ref{prop:complement}(d) and~\ref{prop:triage}).} If cheaper generation shifts submissions from flawed toward valid but marginal papers, the composition term $N_m/N_f$ rises, and with it the delay benefit of significance triage relative to validity screening; the credibility benefit also depends on $\psi_m/\psi_f$. A rising share of desk rejections for limited contribution is consistent with this shift only if screening criteria and the recording of rejection reasons are comparable over time.

\paragraph{P4. Turnaround, crowding out and selection (Propositions~\ref{prop:fixed} and~\ref{prop:aggregate}).} At fixed credibility, rewards and capacity, a higher junk load lengthens delay and weakly reduces genuine participation; selection raises the mean capability of the remaining genuine authors while aggregate net output falls. Author seniority, prior acceptances and citations are possible proxies for capability, not direct observations of $\theta$. A test would follow submission choices and output across venues while measuring review capacity and the outside option.

\paragraph{P5. Nonlinear responses and history dependence (Theorem~\ref{thm:trap}, Proposition~\ref{prop:integrated}).} An increase in submission pressure can eliminate a positive equilibrium discontinuously. A later reduction in junk can restore the existence of a positive equilibrium without restoring participation from the collapse state. Policy changes that affect submission loads, such as the arXiv submission cap of October 2026 \citep{the2026} or new submission fees, can motivate tests of asymmetric responses, which must distinguish credibility feedback from persistent changes in capacity, author composition and outside options. If readers also value the substance behind a certificate, this valuation amplifies the response to such shocks at venues whose review rewards presentation beyond $\kappa/\rho$ and dampens it below (Proposition~\ref{prop:reputation}(d)). The prediction holds the review norm fixed; it does not imply that venues rewarding presentation respond more strongly than venues that do not (Remark~\ref{rem:norms}).

\paragraph{P6. Leakage and the review weight (Proposition~\ref{prop:adaptive}).} If review rules adapt, low-contribution papers should be endorsed more often than their recognition-failure baseline only where they are more polished than genuine submissions and presentation receives positive weight. Across fields or venues with comparable recognition rates and conditional on the presentation of genuine submissions, the relative acceptance rate of polished low-contribution papers should therefore rise with the estimated weight on presentation, whereas that of unpolished low-contribution papers should fall. Fixed leakage, the benchmark of Section~\ref{sec:participation}, predicts no such interaction. Estimating it requires classifying both accepted and rejected papers; audits of accepted papers alone do not supply the relevant denominators.

\paragraph{P7. Auxiliary hypothesis: agreement on validity and significance.} If a shared technical standard constrains assessments more tightly than judgments of significance, separate scores may exhibit greater agreement on validity. The Discovery Stack pilot reports greater consistency in Quality than in Impact scores \citep{mcgargill2026}. This is an additional hypothesis about the assessment technology, not a variance ranking derived from the model, and greater agreement does not establish greater accuracy.

Linking these data to model parameters requires further identification. Submission responses to exogenous changes in screening could identify the deterrence elasticities in~\eqref{eq:ratio} if other participation incentives are held fixed. Post-publication impact is an imperfect measure of whether a paper was marginal when reviewed. Identifying $s_y/s_x$ requires variation that isolates the evaluation response to polish from substance and from changes in who submits.

\subsection{Motivating evidence from ICLR}\label{sec:iclr}

Reviews and decisions from the main track of the International Conference on Learning Representations (ICLR) for 2024--2026 provide a descriptive benchmark for the distinction between substance and presentation. The review form records separate scores for soundness, presentation and contribution alongside an overall rating. Using 150,634 reviews of 38,256 submissions compiled by Paper Copilot \citep{yang2026papercopilot}, we regress the standardised overall rating on the three sub-scores and reviewer confidence with manuscript fixed effects, so that estimates come from differences across reviewers of the same paper, and we estimate acceptance logits on mean sub-scores. Appendix~\ref{app:iclr} describes the data and reports the estimates in Table~\ref{tab:iclr-review}.

Conditional on soundness, contribution and confidence, a one-point higher presentation score is associated with about 0.2 standard deviations higher overall ratings in 2024 and 2025; the presentation coefficient is 37--41\% as large as the contribution coefficient, and about one-third as large in the acceptance logits. The ratio rises slightly from 2024 to 2025 in the rating regressions ($p=0.064$) and falls in the acceptance logits, so the data leave Theorem~\ref{thm:rational}(b) untested. Over the three years, submissions rise from 7,404 to 19,814.

These associations do not identify the structural coefficient $s_y$. Sub-scores and recommendations come from the same reviewer, so a reviewer-specific overall impression can drive both; the contribution score captures only part of what the model calls substance; and the presentation score can reward clarity that helps readers assess the evidence rather than the residual polish represented by $y$, an ambiguity that the language-editing experiment of \citet{feld2024} shares. Assessing Theorem~\ref{thm:look}(a) would require a causal estimate of the response to polish and information on the cost parameters that determine $L$. A sharper test of P1 and P6 would combine these scores with paper-level measures of AI-assisted writing and with contribution assessments independent of the focal recommendation, exploiting variation across primary areas and years in the prevalence of polished low-contribution submissions.

\section{Implications for review design}\label{sec:conclusion}

Cheap production changes both the incentives a certificate creates and the composition of the pool behind it. In our benchmark, a journal, or any certifier that bundles validation with accountable endorsement, improves net epistemic output relative to the same technical validation platform when its relative reward to presentation is below a cost-dependent threshold, and this condition extends to threshold and quota rules. Best linear prediction need not respect the threshold, and endorsement leakage and delay can undermine the participation that sustains credibility. The results identify conditions for these mechanisms, not an unconditional prediction that AI reduces scientific progress or that journals collapse.

\paragraph{Separate the basis for endorsement from presentation.}
Separate assessments of validity and significance make editorial objectives easier to inspect \citep{mcgargill2026}. The model directs attention to how those assessments and presentation enter the final recommendation. Theorem~\ref{thm:rational}(c)--(e) implies that the case for a contribution-first review form, in which reviewers state the contribution, its importance and the supporting evidence before recommending a decision, strengthens as presentation becomes cheaper; its cost is a loss of selection accuracy that is small when reading is careful. Proposition~\ref{prop:adaptive} qualifies this recommendation. Where low-value submissions are already more polished than genuine work and common, an adaptive rule that penalises presentation can protect both incentives and credibility; when reviewers still reward polish and low-contribution submissions are more polished than genuine work, commitment dominates at a given rate of genuine participation. When participation adjusts, the gain in credibility and substantive effort can come with fewer genuine submissions, as in Figure~\ref{fig:policy-equilibrium}. Editors therefore need to track the composition of the submission pool and what is endorsed alongside submission volumes. When clarity limits evaluation, the review should identify the claims or evidence that cannot be assessed.

\paragraph{Allocate screening resources according to the objective and the errors.}
Technical failures and limited contribution should be tracked as distinct reasons for screening decisions. A larger marginal stream raises the relative capacity value of significance triage, adjusted for deterrence, and its credibility value depends on whether reviewers would otherwise endorse those submissions (Proposition~\ref{prop:triage} and Corollary~\ref{cor:invest}). With false exclusions, faster processing can coexist with reduced credibility and fewer genuine contributions retained for review (Appendix~\ref{app:false}), so a delay objective should be paired with a minimum-retention constraint and audits of both types of screening error. AI can support overlap checks and the examination of novelty claims, while editors remain responsible for significance criteria and for reviewing contested exclusions.

\paragraph{Monitor endorsed junk and participation, not publication averages.}
The relevant load is not only what reaches reviewers but also what they mistakenly endorse. Monitoring both separates congestion from dilution of the certificate. Endorsed junk expands the basin of collapse. When readers value the substance behind the certificate, a review norm that rewards presentation beyond $\kappa/\rho$ expands it further than the endorsed share alone would indicate (Proposition~\ref{prop:reputation}), so credibility measured by the genuine share overstates the robustness of such a journal. A loss of positive participation is discontinuous when it occurs; a reduction in junk can restore the existence of a good equilibrium without selecting it. Submission fees and quotas should therefore be assessed alongside their effects on participation and alternative venues \citep{lopezlira2026,lawford2026,fourie2026}. Average capability among genuine authors can rise while aggregate output falls (Proposition~\ref{prop:aggregate}), so average quality of published work is a misleading indicator.

\paragraph{Test changes to the review process.}
A pilot could compare contribution-first and conventional review forms, measuring how clearly reviewers identify the contribution and how recommendations change. A complementary experiment would assign evaluators to versions of a paper that differ in surface polish while independently verifying that claims, evidence and comprehensibility are preserved, extending \citet{feld2024}. Triage pilots should audit false exclusions, including contributions using unfamiliar methods or addressing emerging topics.

\paragraph{Extensions.}
The inference results use best linear prediction and, for the derivation of self-consistent weights, an affine endorsement link; they do not characterise an unrestricted Bayesian ranking or an optimal editorial policy. Section~\ref{sec:policy-equilibrium} illustrates the joint adjustment of review weights, leakage, effort and participation on selected positive-equilibrium branches at fixed marginal inflow. A broader comparison would account for the response of the marginal stream to policy and derive an equilibrium selection rule when multiple self-consistent review norms exist. Heterogeneous delay costs and competition among journals could change selection and redistribution across venues. A reader valuation that is non-linear in the substance behind the certificate could add a low-substance equilibrium to the collapse and high-credibility states (Section~\ref{sec:reputation}); a bounded valuation would also remove the existence restriction of the linear reputation channel (Remark~\ref{rem:rep-existence}). Ranking review norms by their fragility requires a normalisation such as a fixed acceptance rate (Remark~\ref{rem:norms}). Genuine authors who substitute marginal for genuine submissions link the junk load to participation directly; the integrated model does not cover this case (Remark~\ref{rem:genuine-marginal}). An explicit information structure would be needed to determine whether global-games reasoning \citep{morris1998} yields a unique switching point. The ICLR evidence remains descriptive and confined to one conference.

\appendix
\section{Proofs}\label{app:proofs}

\begin{proof}[Proof of Lemma~\ref{lem:effort}]
$U_\theta^\ell$ is strictly concave because its Hessian is $-\theta^{-1}\bigl(\begin{smallmatrix}\chi & \rho\\ \rho & \kappa\end{smallmatrix}\bigr)$ with $\chi > 0$ and $D > 0$. The first-order conditions $\chi x + \rho y = \theta A_\ell$ and $\rho x + \kappa y = \theta B_\ell$ have the unique solution~\eqref{eq:effort}; Assumption~\ref{ass:interior} makes both components positive. Substituting into~\eqref{eq:Q} gives $q_\ell$. Writing $z = (x,y)^\top$, $a = (A_\ell, B_\ell)^\top$ and $K$ for the cost matrix, the maximum of $a^\top z - z^\top K z/(2\theta)$ is $\tfrac{\theta}{2}\, a^\top K^{-1} a$ with $K^{-1} = D^{-1}\bigl(\begin{smallmatrix}\kappa & -\rho\\ -\rho & \chi\end{smallmatrix}\bigr)$, which yields $\Phi_\ell$; positivity follows from positive definiteness of $K^{-1}$.
\end{proof}

\begin{proof}[Proof of Lemma~\ref{prop:regimes}]
By~\eqref{eq:qPhi}, $Q_\theta^m - Q_\theta^\ell = \tfrac{\theta}{D}\bigl[(\kappa + \lambda\rho)(A_m - A_\ell) - (\rho + \lambda\chi)(B_m - B_\ell)\bigr]$ and $\theta/D > 0$. Since $\kappa + \lambda\rho > 0$ and $\rho + \lambda\chi \ge 0$, the bracket is positive whenever $A_m>A_\ell$ and $B_m\le B_\ell$. If $A_m=A_\ell$ and $B_m<B_\ell$, it is positive exactly when $\rho+\lambda\chi>0$. It is non-positive when $A_m\le A_\ell$ and $B_m\ge B_\ell$. If $\rho+\lambda\chi=0$, its sign is the sign of $A_m-A_\ell$.
\end{proof}

\begin{proof}[Proof of Theorem~\ref{thm:look}]
(a) Since $e_P = 0$, $r_J = r_P$ and $\gamma^T_J = \gamma^T_P$, \eqref{eq:slopes} gives $A_J - A_P = \gamma\alpha_E e$ and $B_J - B_P = -\gamma\beta_E e$. Substitute into the expression in the previous proof. Since $\gamma e > 0$, the sign is that of $(\kappa+\lambda\rho)\alpha_E + (\rho+\lambda\chi)\beta_E = V\bigl[(\kappa+\lambda\rho)s_x - (\rho+\lambda\chi)s_y\bigr]$. If $\rho + \lambda\chi > 0$, divide by $V s_x (\rho + \lambda\chi) > 0$ to obtain the threshold form; if $\rho + \lambda\chi = 0$, the condition holds for all $s_y$ and $L = \infty$.

(b) At an interior maximiser of the differentiable function $U_\theta^J$, the gradient vanishes:
$\chi x+\rho y=\theta(A_P+g_x)$ and $\rho x+\kappa y=\theta(B_P+g_y)$. At $P$, Lemma~\ref{lem:effort} gives the unique solution of $\chi x+\rho y=\theta A_P$ and $\rho x+\kappa y=\theta B_P$. Subtracting, the effort differences solve $K(\Delta x,\Delta y)^\top=\theta(g_x,g_y)^\top$, where $K$ is the cost matrix with inverse $D^{-1}\bigl(\begin{smallmatrix}\kappa&-\rho\\-\rho&\chi\end{smallmatrix}\bigr)$. This gives the first two expressions in~\eqref{eq:link}, and $\Delta Q=\Delta x-\lambda\Delta y$ gives the third. If $g_x>0$ and $\rho+\lambda\chi>0$, dividing the bracket by $g_x(\rho+\lambda\chi)>0$ yields the threshold form.

For the single-index case, write $\eta:=\varepsilon+tu$ and let $P:\mathbb R\to[0,1]$ be the non-decreasing endorsement rule applied to $s+t\tilde y=x+ty+\eta$. Then $\Pi(x,y)=\mathbb E[P(x+ty+\eta)]=:\Pi_1(x+ty)$. Since $P$ is monotone and bounded, it differs from a right-continuous version only at countably many points, which have probability zero under the density $f_\eta$ of $\eta$. The right-continuous version can be written as $P(w)=P(-\infty)+\int\mathbb 1\{w\ge c\}\,dP(c)$ with a finite measure $dP$, so $\Pi_1(z)=P(-\infty)+\int[1-F_\eta(c-z)]\,dP(c)$. Because $f_\eta$ is bounded and continuous, dominated convergence gives $\Pi_1'(z)=\int f_\eta(c-z)\,dP(c)$. Hence $g_x=\gamma V\Pi_1'$ and $g_y=t\,\gamma V\Pi_1'$ at the optimum, and $g_y/g_x=t$ whenever $\Pi_1'>0$.

(c) $\partial L/\partial\kappa = 1/(\rho+\lambda\chi) > 0$ and $\partial L/\partial\chi = -\lambda(\kappa+\lambda\rho)/(\rho+\lambda\chi)^2 \le 0$, with $\partial\chi/\partial\chi_X = 1$. If $s_y \le 0$, then $s_y/s_x \le 0 < L$. If $s_y>0$, $s_y/s_x < L$ is equivalent to $\kappa + \lambda\rho > (\rho+\lambda\chi)s_y/s_x$, that is, $\kappa > \hat\kappa$. Finally, $\partial\hat\kappa/\partial\chi_X = \lambda s_y/s_x \ge 0$ when $s_y > 0$. The extension to single-index rules follows from part~(b).
\end{proof}

\begin{proof}[Proof of Lemma~\ref{prop:cheap}]
(a) Differentiate~\eqref{eq:effort}, using $\partial D/\partial\kappa = \chi$ and $\partial D/\partial\chi = \kappa$. For example,
$\partial x^*/\partial\kappa = \theta\bigl[A_\ell D - (\kappa A_\ell - \rho B_\ell)\chi\bigr]/D^2 = \theta\rho(\chi B_\ell - \rho A_\ell)/D^2 = \rho y^*/D$, and similarly for the others; $\partial/\partial\chi_X = \partial/\partial\chi$. The derivatives of $Q$ follow by linearity, and the signs use $x^*, y^* > 0$. The bound follows from $x^* \le \theta\kappa A_\ell/D$ and $D \ge \chiu\kappa - \rho^2 > 0$.
(b) Apply~\eqref{eq:dkappa}--\eqref{eq:dchi} to each regime and subtract. The expression for $y^{J*} - y^{P*}$ in the text follows from~\eqref{eq:effort}.
\end{proof}

\begin{proof}[Proof of Lemma~\ref{lem:projection}]
The reading noise $\varepsilon$ is independent of $(x,\tilde y)$ and has mean zero. Thus $\mathrm{Var}(s)=\mathcal V_x+\sigma_\varepsilon^2$, $\mathrm{Cov}(s,\tilde y)=C$, $\mathrm{Cov}(x,s)=\mathcal V_x$ and $\mathrm{Cov}(x,\tilde y)=C$. The weights of the best linear predictor solve the normal equations
\[
\begin{pmatrix} \mathcal V_x + \sigma_\varepsilon^2 & C\\ C & \mathcal V_y\end{pmatrix}\begin{pmatrix} w_s\\ w_y\end{pmatrix} = \begin{pmatrix} \mathcal V_x\\ C\end{pmatrix}.
\]
The determinant is $\Delta_{\mathrm{pred}} = (\mathcal V_x\mathcal V_y - C^2) + \sigma_\varepsilon^2 \mathcal V_y > 0$, because $\mathcal V_x\mathcal V_y > C^2$ and $\mathcal V_y > 0$. Cramer's rule gives $w_s = (\mathcal V_x\mathcal V_y - C^2)/\Delta_{\mathrm{pred}} > 0$ and $w_y = \bigl((\mathcal V_x + \sigma_\varepsilon^2)C - C\mathcal V_x\bigr)/\Delta_{\mathrm{pred}} = \sigma_\varepsilon^2 C/\Delta_{\mathrm{pred}}$. Since $w_s > 0$, ranking by $\hat x$ is ranking by $s + t\tilde y$ with $t = w_y/w_s$. For an author with effort $(x,y)$, $\mathbb E[s + t\tilde y] = x + t y$ because $\varepsilon$ and $u$ have mean zero, so the expected endorsement probability is $e(\pi_0' + s_x x + t s_x y)$, which is~\eqref{eq:endorse} with $\pi_0=\pi_0'$ and $s_y=t s_x$. The sign statement follows from $\mathcal V_x\mathcal V_y - C^2 > 0$.
\end{proof}

\begin{proof}[Proof of Theorem~\ref{thm:rational}]
(a) Under $s_y = t s_x$, \eqref{eq:slopes} gives $A = a_0 + \gamma^T\alpha_T r + S$ and $B = b_0 - \gamma\beta_E e = b_0 + \gamma V e\, t s_x = B(t)$. By Lemma~\ref{lem:effort}, a genuine author of capability $\theta$ chooses $(x,y) = \theta(a,b)$ with $a = (\kappa A - \rho B(t))/D$ and $b = (\chi B(t) - \rho A)/D$, both positive at an interior rule. In the genuine pool, $\mathcal V_x = a^2\Sigma$, $C = \mathrm{Cov}(a\theta, b\theta + u) = ab\Sigma$ and $\mathcal V_y = b^2\Sigma + \sigma_u^2$, so $\mathcal V_x\mathcal V_y - C^2 = a^2\Sigma\sigma_u^2 > 0$. Lemma~\ref{lem:projection} gives $t = \sigma_\varepsilon^2 ab\Sigma/(a^2\Sigma\sigma_u^2) = (\sigma_\varepsilon^2/\sigma_u^2)\,b/a$, which is~\eqref{eq:tgenuine}, and $t > 0$ because $a,b > 0$.

(b) Fix an admissible $\kappa$ and let $\bar t(\kappa) := (\kappa A - \rho b_0)/(\rho S)$ (with $\bar t = \infty$ if $\rho = 0$) be the weight at which $\kappa A = \rho B(t)$. If $\bar t(\kappa)\le 0$, no non-negative weight is interior and the feasible interval is empty. Otherwise, for $t \in [0,\bar t(\kappa))$ the denominator of $f$ is positive, and the numerator $\chi B(t) - \rho A \ge \chi b_0 - \rho A > 0$; hence the interior rules are exactly the fixed points of $f(\cdot;\kappa)$ on that interval. On a non-empty feasible interval $f(\cdot;\kappa)$ is continuous and increasing, with $f(0;\kappa) > 0$. Let $\mathcal F(\kappa) := \{t \in [0,\bar t(\kappa)) : f(t;\kappa) \le t\}$. If $\mathcal F(\kappa)$ is non-empty, the function $f(t;\kappa) - t$ is positive at $0$ and non-positive somewhere, so by the intermediate value theorem it has a root, and its least root is $t_*(\kappa) = \min\mathcal F(\kappa)$ (the infimum is attained by continuity and is a root because $f(t;\kappa) - t > 0$ to its left). Conversely, if $\mathcal F(\kappa)$ is empty there is no interior rule. Now let $\kappa' > \kappa$. Then $\bar t(\kappa') \ge \bar t(\kappa)$, with strict inequality if $\rho>0$, and for $t \in [0,\bar t(\kappa))$, the denominator of $f$ is strictly larger at $\kappa'$ while the numerator is unchanged and positive, so $f(t;\kappa') < f(t;\kappa)$. Hence $\mathcal F(\kappa) \subseteq \mathcal F(\kappa')$. This shows that $\mathcal K = \{\kappa : \mathcal F(\kappa) \ne \emptyset\}$ is an up-set, hence an interval unbounded above once it is non-empty, and that $t_*(\kappa') = \min\mathcal F(\kappa') \le \min\mathcal F(\kappa) = t_*(\kappa)$. Non-emptiness: as $\kappa \to \infty$, $f(t;\kappa) \to 0$ uniformly on bounded $t$-sets while the upper endpoint $\bar t(\kappa)$ is either infinite or tends to infinity, so $f(1;\kappa) \le 1$ for $\kappa$ large.

(c) In a genuine pool there is no junk, so the rule does not affect $\gamma$, and $A$ is the same under both rules, while $B(t)=b_0+St$. By Lemma~\ref{lem:effort}, $Q_\theta^J=\theta[(\kappa+\lambda\rho)A-(\rho+\lambda\chi)B]/D$ whenever Assumption~\ref{ass:interior} holds, which it does at $t_*$ by interiority of the rule and at $t_c$ by hypothesis. Hence $Q_\theta^J(t_c)-Q_\theta^J(t_*)=-\theta(\rho+\lambda\chi)[B(t_c)-B(t_*)]/D=\theta(\rho+\lambda\chi)S(t_*-t_c)/D$, which is non-negative because $t_c\le t_*$ and $S>0$, and positive under the stated strict conditions.

(d) By part~(a), $t_*(\kappa)>0$, and by part~(b), $t_*$ is non-increasing on $\mathcal K$. The function $D(\kappa)=\chi\kappa-\rho^2$ is positive and strictly increasing on the admissible range. Hence $t_*(\kappa)/D(\kappa)$ is strictly decreasing, and $\mathcal V_c=\theta(\rho+\lambda\chi)S\,t_*/D$ is non-increasing, strictly decreasing if $\rho+\lambda\chi>0$.

(e) Under $t_c=0$, authors choose $\theta(a_c,b_c)$ with $a_c,b_c>0$. As in the proof of part~(a), $\mathrm{Var}(s)=a_c^2\Sigma+\sigma_\varepsilon^2$, $\mathrm{Cov}(x,s)=a_c^2\Sigma$, $\mathrm{Cov}(s,\tilde y)=\mathrm{Cov}(x,\tilde y)=a_cb_c\Sigma$ and $\mathrm{Var}(\tilde y)=b_c^2\Sigma+\sigma_u^2$. Ranking by $s$ is ranking by the best linear predictor based on $s$ alone, whose slope $a_c^2\Sigma/(a_c^2\Sigma+\sigma_\varepsilon^2)$ is positive. By the partial-regression formula, adding $\tilde y$ reduces the mean squared error by the squared partial covariance of $x$ and $\tilde y$ given $s$, divided by the partial variance of $\tilde y$ given $s$. The partial covariance is $a_cb_c\Sigma-a_c^2\Sigma\cdot a_cb_c\Sigma/\mathrm{Var}(s)=a_cb_c\Sigma\sigma_\varepsilon^2/\mathrm{Var}(s)$, and the partial variance is $b_c^2\Sigma+\sigma_u^2-a_c^2b_c^2\Sigma^2/\mathrm{Var}(s)=\sigma_u^2+b_c^2\Sigma\sigma_\varepsilon^2/\mathrm{Var}(s)$. Their ratio is~\eqref{eq:selcost}, which is positive because $a_c,b_c,\Sigma,\sigma_\varepsilon^2,\sigma_u^2>0$.

(f) Since $A=A_P+S$ and $B(L)=b_0+S\kappa/\rho$,
\[
\kappa A-\rho B(L)=\kappa A_P-\rho b_0=Da_P>0,\qquad
\chi B(L)-\rho A=\chi b_0-\rho A_P+\frac{S(\chi\kappa-\rho^2)}{\rho}=D\Bigl(b_P+\frac{S}{\rho}\Bigr).
\]
The first identity gives $L<\bar t(\kappa)$, so $[0,L]$ lies in the feasible interval of part~(b), and $f(L;\kappa)=(\sigma_\varepsilon^2/\sigma_u^2)(b_P+S/\rho)/a_P$. Under the conditions of part~(b), $f(0;\kappa)>0$ and $f(\cdot;\kappa)$ is continuous on $[0,L]$. If $f(L;\kappa)\le L$, the intermediate value theorem gives a root of $f(t;\kappa)-t$ in $(0,L]$, so $t_*(\kappa)\le L$. Hence $t_*(\kappa)>L$ implies $f(L;\kappa)>L=\kappa/\rho$, which is~\eqref{eq:crossing} after multiplication by $\rho a_P>0$.
\end{proof}

\begin{proof}[Proof of Proposition~\ref{prop:junk}]
Let $\mathbb 1_g$ indicate a genuine paper. Then $x = \mathbb 1_g\, a\theta$ and $\tilde y = \mathbb 1_g\, b\theta + (1 - \mathbb 1_g)y_m + u$. Since $x = 0$ for marginal papers, $\mathbb E[x\tilde y] = (1-\varphi)ab\,\mathbb E[\theta^2] = (1-\varphi)ab(\Sigma + \bar\theta^2)$, while $\mathbb E[x] = (1-\varphi)a\bar\theta$ and $\mathbb E[\tilde y] = (1-\varphi)b\bar\theta + \varphi y_m$. Hence
$\mathrm{Cov}(x,\tilde y) = (1-\varphi)a\bigl[b\Sigma + b\bar\theta^2 - (1-\varphi)b\bar\theta^2 - \varphi\bar\theta y_m\bigr] = (1-\varphi)a\bigl[b\Sigma + \varphi\bar\theta(b\bar\theta - y_m)\bigr]$. The variance condition of Lemma~\ref{lem:projection} holds because $u$ is independent of $x$ with $\sigma_u^2 > 0$, so by the lemma $t$ has the sign of this covariance. Since $(1-\varphi)a > 0$, $t < 0$ if and only if $\varphi\bar\theta(y_m - b\bar\theta) > b\Sigma$, which requires $y_m > b\bar\theta$ and is then equivalent to $\varphi > \hat\varphi$.
\end{proof}

\begin{proof}[Proof of Lemma~\ref{lem:frontier}]
$T \le \bar T$ if and only if $\mu \ge M + 1/\bar T$, and since $\mu = H/(he)$ this is $e \le H/\bigl(h(M + 1/\bar T)\bigr)$.
\end{proof}

\begin{proof}[Proof of Lemma~\ref{lem:capacity}]
From $T = 1/(\mu - M)$, $dT = T^2(dM - d\mu)$. At fixed $e$, $d\mu = -\mu(h'/h)\,dr$, and by~\eqref{eq:load}, $dM = -\delta' N_f\,dr - N_m\,d\zeta + (1-\delta)\,dN_f + (1-\zeta)\,dN_m + dN_v$. Hence $dM - d\mu = -\Lambda$. From~\eqref{eq:frontier}, $d\log E = -(h'/h)\,dr - dM/(M + 1/T)$, and $M + 1/T = \mu$, so $d\log E = \Lambda/\mu$.
\end{proof}

\begin{proof}[Proof of Proposition~\ref{prop:complement}]
(a) Differentiate~\eqref{eq:gamma}, with $\tilde Z$ depending on $r$ through $\delta$, to obtain $\gamma' = -N_v \tilde Z'/\tilde M^2 = \psi_f N_v N_f\delta'/\tilde M^2$. From~\eqref{eq:slopes}, $\partial^2 A_J/\partial r\,\partial e = \alpha_E\gamma'$ and $\partial^2 B_J/\partial r\,\partial e = -\beta_E\gamma'$; substitute into the expression for $Q_\theta^J$ in~\eqref{eq:qPhi}. If $\gamma' > 0$, the sign of~\eqref{eq:cross} is that of~\eqref{eq:C}. (b) $\log E = \log H - \log h - \log(M + 1/T)$, and $\partial M/\partial r = -\delta' N_f$; differentiate. (c) Differentiating $\gamma^T = (N_v+N_m)/(N_v+N_m+(1-\delta)N_f)$ gives $\gamma^{T\prime}$. Along the frontier, $dA_J/dr = \gamma^{T\prime}\alpha_T r + \gamma^T\alpha_T + \alpha_E(\gamma' E + \gamma\,\partial E/\partial r)$ and $dB_J/dr = -\beta_E(\gamma' E + \gamma\,\partial E/\partial r)$; substitute and collect terms. Since $\gamma' \ge 0$ and $\partial E/\partial r \ge 0$ by (b), the second term has the sign of~\eqref{eq:C}. (d) $M$ is strictly increasing in $N_m$, so $\delta' N_f/(M + 1/T)$ is strictly decreasing in $N_m$ whenever $N_f\delta' > 0$. $\tilde M$ is strictly increasing in $N_m$ if $\psi_m > 0$ and independent of $N_m$ if $\psi_m = 0$, which gives the claims for $\gamma' = \psi_f N_v N_f\delta'/\tilde M^2$. If $N_f = 0$ and $h' \equiv 0$, both channels vanish.
\end{proof}

\begin{proof}[Proof of Proposition~\ref{prop:triage}]
Write $p_f = 1-\delta$ and $p_m = 1-\zeta$, so $Z = p_f N_f(p_f) + p_m N_m(p_m)$. Then $d(p_f N_f)/dp_f = N_f + p_f N_f' = N_f(1 + \eta_f)$, so $\partial Z/\partial\delta = -N_f(1+\eta_f)$; likewise for $\zeta$. Since $\tilde Z = \psi_f p_f N_f(p_f) + \psi_m p_m N_m(p_m)$ with constant $\psi_f, \psi_m$, the derivatives of $\tilde Z$ follow in the same way. $T = 1/(\mu - N_v - Z)$ depends on $(\delta,\zeta)$ only through $Z$ and is strictly increasing in it, so the ratio of the marginal effects of $\zeta$ and $\delta$ on $T$ is the ratio of $\partial Z/\partial\zeta$ to $\partial Z/\partial\delta$, which is $\mathcal{R}$. Likewise, $\gamma = N_v/(N_v + \tilde Z)$ depends on them only through $\tilde Z$, which gives $\mathcal{R}_\gamma$. Finally, $\partial\gamma/\partial\delta = -N_v(\partial\tilde Z/\partial\delta)/\tilde M^2 = \psi_f N_v N_f(1+\eta_f)/\tilde M^2$, and $\tilde M$ is strictly increasing in $N_m$ when $\psi_m > 0$.
\end{proof}

\begin{proof}[Proof of Corollary~\ref{cor:invest}]
Minimising $T$ is equivalent to minimising $Z$, since $T$ is strictly increasing in $Z$. The constraint gradient $(c_\delta', c_\zeta')$ is non-zero, so at an interior optimum with a binding budget the Karush--Kuhn--Tucker conditions give a multiplier $\nu > 0$ with $-\partial Z/\partial\delta = \nu c_\delta'(\delta)$ and $-\partial Z/\partial\zeta = \nu c_\zeta'(\zeta)$. Dividing and using~\eqref{eq:dZ} gives $c_\zeta'/c_\delta' = N_m(1+\eta_m)/(N_f(1+\eta_f)) = \mathcal R$. Maximising $\gamma$ is equivalent to minimising $\tilde Z$, and the same argument with the derivatives of $\tilde Z$ in~\eqref{eq:dZ} gives $\mathcal R_\gamma$.
\end{proof}

\begin{proof}[Proof of Lemma~\ref{lem:contested}]
(a) A marginal paper reaching review is recognised and rejected with probability $1-\psi^0_m$. Otherwise its expected endorsement probability is $e(\pi_0'+s_x(\mathbb E[s]+t\,y_m))$ with $\mathbb E[s]=a\bar\theta=\bar x$, because $\varepsilon$ and $u$ have mean zero. The mean endorsement probability of genuine papers is $e(\pi_0'+s_x(\bar x+t\bar y))=\bar\pi_v>0$. The ratio gives~\eqref{eq:psit}. The numerator minus the denominator of the fraction equals $s_xt(y_m-\bar y)$, and the denominator is positive, so $\psi_m(t)-\psi^0_m$ has the sign of $t(y_m-\bar y)$.

(b) Let $\mathbb 1_g$ indicate a genuine paper, with $\Pr(\mathbb 1_g=1)=1-\varphi$. Then $x=\mathbb 1_g\,a\theta$, $s=a\theta^\circ+\varepsilon$, where $\theta^\circ$ equals $\theta$ for genuine papers and the independent draw $\theta'$ for marginal papers, so that $\theta^\circ$ has the capability distribution and is independent of $\mathbb 1_g$; and $\tilde y=\mathbb 1_g\,b\theta+(1-\mathbb 1_g)y_m+u$. Direct computation gives
\[
\mathrm{Var}(s)=a^2\Sigma+\sigma_\varepsilon^2,\quad \mathrm{Cov}(x,s)=(1-\varphi)a^2\Sigma,\quad \mathrm{Cov}(s,\tilde y)=(1-\varphi)ab\Sigma,
\]
\[
\mathrm{Cov}(x,\tilde y)=(1-\varphi)a\bigl[b\Sigma+\varphi\bar\theta(\bar y-y_m)\bigr],\quad \mathrm{Var}(\tilde y)=(1-\varphi)b^2\Sigma+\varphi(1-\varphi)(\bar y-y_m)^2+\sigma_u^2 .
\]
For example, $\mathrm{Cov}(x,s)=\mathbb E[\mathbb 1_ga^2\theta^2]-(1-\varphi)a\bar\theta\cdot a\bar\theta=(1-\varphi)a^2\Sigma$, because $x=0$ for marginal papers. The covariance $\mathrm{Cov}(x,\tilde y)$ is~\eqref{eq:Cmix} with $\bar y=b\bar\theta$, and $\mathrm{Var}(\tilde y)$ follows from the law of total variance. The covariance matrix of $(s,\tilde y)$ is the sum of the covariance matrix of their noise-free parts and $\mathrm{diag}(\sigma_\varepsilon^2,\sigma_u^2)$, so its determinant satisfies $\Delta_c\ge\sigma_\varepsilon^2\sigma_u^2>0$. Cramer's rule applied to the normal equations gives
$w_s=[\mathrm{Var}(\tilde y)\mathrm{Cov}(x,s)-\mathrm{Cov}(s,\tilde y)\mathrm{Cov}(x,\tilde y)]/\Delta_c$ and
$w_y=[\mathrm{Var}(s)\mathrm{Cov}(x,\tilde y)-\mathrm{Cov}(s,\tilde y)\mathrm{Cov}(x,s)]/\Delta_c$. Substituting,
\[
w_s=\frac{(1-\varphi)a^2\Sigma}{\Delta_c}\Bigl\{\mathrm{Var}(\tilde y)-(1-\varphi)b\bigl[b\Sigma+\varphi\bar\theta(\bar y-y_m)\bigr]\Bigr\}
=\frac{(1-\varphi)a^2\Sigma}{\Delta_c}\bigl[\sigma_u^2+\varphi(1-\varphi)(y_m-\bar y)y_m\bigr],
\]
using $(\bar y-y_m)^2-\bar y(\bar y-y_m)=(y_m-\bar y)y_m$, and
\[
w_y=\frac{(1-\varphi)a}{\Delta_c}\Bigl\{(a^2\Sigma+\sigma_\varepsilon^2)\bigl[b\Sigma+\varphi\bar\theta(\bar y-y_m)\bigr]-(1-\varphi)a^2b\Sigma^2\Bigr\},
\]
which is~\eqref{eq:wcontested}. If $y_m\ge\bar y$, then $y_m\ge0$ implies that the bracket in $w_s$ is at least $\sigma_u^2>0$. When $w_s>0$, ranking by the predictor is ranking by $s+t\tilde y$ with $t=w_y/w_s$, which has the sign of the bracket in $w_y$, because $(1-\varphi)a>0$. If $y_m\le\bar y$, that bracket is at least $\sigma_\varepsilon^2b\Sigma+\varphi a^2b\Sigma^2>0$. If $y_m>\bar y$, the bracket equals $\sigma_\varepsilon^2b\Sigma-\varphi\Gamma$; since $\sigma_\varepsilon^2b\Sigma>0$, it is negative if and only if $\Gamma>0$ and $\varphi>\hat\varphi_c$.
\end{proof}

\begin{proof}[Proof of Proposition~\ref{prop:adaptive}]
With all junk marginal, validation credibility equals one in both regimes, so $A=a_0+\alpha_Tr+S$ and $B=b_0+St$ with $S=\gamma Ves_x$, which is positive because $N_v>0$. By Lemma~\ref{lem:effort}, $Q_\theta^J=\theta[(\kappa+\lambda\rho)A-(\rho+\lambda\chi)B]/D$ in each regime, so
\begin{equation}\label{eq:adaptdiff}
Q_\theta^{c}-Q_\theta^{a}=\frac{\theta}{D}\Bigl[(\kappa+\lambda\rho)(S_c-S_a)+(\rho+\lambda\chi)S_at_a\Bigr].
\end{equation}
Credibility $\gamma=N_v/(N_v+\psi_mZ_m)$ is strictly decreasing in leakage, and $S$ is proportional to $\gamma$. In the committed regime $\psi_m=\psi_m(0)=\psi^0_m$ by~\eqref{eq:psit}.
(a) If $y_m\le\bar y_a$, Lemma~\ref{lem:contested}(b), applied to the adaptive pool, gives $t_a>0$, and Lemma~\ref{lem:contested}(a) gives $\psi_m(t_a)\le\psi^0_m$, hence $\gamma_a\ge\gamma_c$. At equal credibility the first term in~\eqref{eq:adaptdiff} vanishes and the second is non-negative, positive if $\rho+\lambda\chi>0$. With endogenous credibility the first term is non-positive and the second non-negative.
(b) If $y_m>\bar y_a$ and $t_a>0$, then $\psi_m(t_a)>\psi^0_m$, so $\gamma_a<\gamma_c$ and $S_c>S_a$. Both terms in~\eqref{eq:adaptdiff} are non-negative and the first is positive.
(c) If $y_m>\bar y_a$ and $t_a<0$, then $\psi_m(t_a)<\psi^0_m$, so $\gamma_a>\gamma_c$ and $S_a>S_c$. Both terms are non-positive and the first is negative.
\end{proof}

\begin{proof}[Proof of Proposition~\ref{prop:fixed}]
Let $c := Z < \mu$ and define, for $N \in [0, \mu - c)$, the delay $T(N) = 1/(\mu - c - N)$, which is continuous and strictly increasing with $T(N) \to \infty$ as $N \uparrow \mu - c$. The demand $N_v(T)$ is continuous and non-increasing in $T$ because $\theta^*$ is increasing in $T$ (as $\tau > 0$ and $\Phi_J > \Phi_o$) and $G$ is continuous. An equilibrium is a root of $\phi(N) := N - N_v(T(N))$ on $[0,\mu - c)$. The function $\phi$ is continuous and strictly increasing, with $\phi(0) \le 0$ and $\phi(N) \to \mu - c > 0$ as $N \uparrow \mu - c$. Hence there is a unique root $N_v^*$, and $T^* = T(N_v^*)$.
For the comparative statics of $T^*$, note that $T^*$ is the unique solution of $T = \Psi(T) := 1/(\mu - c - N_v(T))$ on $\mathcal{T} := \{T > 0 : \mu - c - N_v(T) > 0\}$. Because $N_v$ is non-increasing, $\mathcal{T}$ is an interval unbounded above, $\Psi$ is non-increasing on it, and $T - \Psi(T)$ is strictly increasing, negative near the lower end of $\mathcal{T}$ and positive for large $T$.
(a) An increase in $c$ raises $T(N)$ pointwise and lowers $N_v(T(N))$, so $\phi$ shifts up and $N_v^*$ falls weakly. $\Psi$ rises strictly pointwise (and $\mathcal{T}$ shrinks), so $T^*$ rises, strictly if the increase in $c$ is strict.
(b) An increase in $\mu$ or a decrease in $c$ lowers $T(N)$ pointwise, so $\phi$ shifts down and $N_v^*$ rises weakly; $\Psi$ falls strictly pointwise, so $T^*$ falls, strictly if the change is strict.
(c) A weak increase in $\Phi_J$ weakly lowers $\theta^*(T)$ wherever $\theta^*(T) \ge \theta_{\min} > 0$, so $N_v(T)$ rises weakly pointwise and $\phi$ shifts down again; $N_v^*$ rises weakly. Its effect on $\Psi$ is upward, opposite to (b), so the net effect on $T^*$ is ambiguous.
For strictness of $N_v^*$: if $\theta^*(T^*)$ is interior and $g(\theta^*(T^*)) > 0$, a strict shift moves $N_v(T(N))$ strictly in a neighbourhood of $N_v^*$, so the root of $\phi$ moves strictly.
\end{proof}

\begin{proof}[Proof of Theorem~\ref{thm:trap}]
(a) Since $\Phi_J(\tilde\gamma(Z)) > \Phi_o \ge \Phi_J(0)$, $\tilde\gamma(Z) > 0$, and with $\psi Z > 0$ this gives $N_0(Z) > 0$. For $N \le N_0(Z)$, $\gamma(N) = N/(N+\psi Z) \le \tilde\gamma(Z)$, so $\Phi_J(\gamma(N)) - \Phi_o \le \tau/((\mu - Z)\theta_{\max}) \le \tau T(N)/\theta_{\max}$, using $T(N) \ge 1/(\mu - Z)$. (If $\tilde\gamma(Z) = 1$, the same inequality holds for every $\gamma \in [0,1]$ by definition.) If $\Phi_J(\gamma(N)) \le \Phi_o$, then $BR(N;Z) = 0$ by~\eqref{eq:BR}. Otherwise the argument of $G$ in~\eqref{eq:BR} is at least $\theta_{\max}$, so again $BR(N;Z) = 0$: even the most capable author prefers the outside option. Thus $0 = BR(0;Z)$, and $BR(N) = 0 < N$ on $(0, N_0(Z)] \cap (0, \mu - Z)$, so the equilibrium is stable.

(b) Since $BR(\hat N) > \hat N \ge 0$, we have $\Phi_J(\gamma(\hat N)) > \Phi_o$, and because $\gamma(N)$ and $\Phi_J$ are increasing, $\Phi_J(\gamma(N)) > \Phi_o$ for all $N \ge \hat N$. On $[\hat N, \mu - Z)$, $BR$ is therefore continuous, as a composition of continuous functions. As $N \uparrow \mu - Z$, $T(N) \to \infty$ while $\Phi_J(\gamma(N)) - \Phi_o$ stays bounded, so the argument of $G$ tends to infinity and $BR(N) \to 0 < \mu - Z$. By the intermediate value theorem, $BR(N) - N$ has a root in $(\hat N, \mu - Z)$. The set of roots is closed in this interval and bounded away from $\mu - Z$, so a largest root $N^H$ exists; $BR(N) < N$ for all $N > N^H$, which is stability from above. By (a), any positive root exceeds $N_0(Z)$.

(c) Let $Z < Z'$ and $N \in [0, \mu - Z')$. Credibility is weakly lower and delay strictly higher at $Z'$, so $BR(N;Z')\le BR(N;Z)$ by~\eqref{eq:BR}; feasibility at $Z'$ also implies feasibility at $Z$. Let $\mathcal E:=\{Z\in(0,\mu):BR(N;Z)=N\text{ for some }N>0\}$ and set $\bar Z=\sup\mathcal E$, with $\bar Z=0$ if $\mathcal E$ is empty. If $Z'\in\mathcal E$ has a positive root $N'$, then $BR(N';Z)\ge N'$ for every $Z<Z'$. Either equality already gives a root at $Z$, or the negative sign of $BR(N;Z)-N$ near $\mu-Z$ and continuity give a root above $N'$. Hence $\mathcal E$ is a lower interval. For every $Z<\bar Z$ a positive root exists, while for $Z>\bar Z$ no positive root exists. In the latter case $BR(N;Z)<N$ at every feasible $N>0$: a weak reverse inequality would itself yield a root, by the same continuity argument. This includes tangential roots in the existence region rather than excluding them by genericity. If $Z\ge Z_{\max}$, part (a) gives $BR\equiv0$, so $\bar Z\le\max\{Z_{\max},0\}<\mu$. Whenever positive roots exist, they are bounded away from both zero by part (a) and the capacity boundary, so there is a largest one. Above it $BR(N;Z)<N$, which gives stability from above. If $Z'>Z$, any candidate $N>N^H(Z)$ in the domain at $Z'$ satisfies $BR(N;Z')\le BR(N;Z)<N$, so $N^H(Z')\le N^H(Z)$. Finally $N_0(Z)$ is non-decreasing in $Z$, because both $\tilde\gamma(Z)$ and $\psi Z$ are non-decreasing. Every positive root exceeds $N_0(Z)$, yielding the stated positive lower bound near a non-zero $\bar Z$ and the discontinuity across the threshold.

(d) If $\psi Z = 0$, endorsement credibility is $1$ for every $N > 0$, and we extend this to $N = 0$ by continuity. Then $\Phi_J = \Phi_J(1) > \Phi_o$, and with $Z < \mu$ Assumption~\ref{ass:fixed} holds, so Proposition~\ref{prop:fixed} applies. Its unique equilibrium is positive if and only if $N_v(T(0)) > 0$, that is, $\theta^*(1/(\mu - Z)) = \tau/((\mu - Z)(\Phi_J(1) - \Phi_o)) < \theta_{\max}$, which is equivalent to $Z < Z_{\max}$.

(e) First, $BR(\cdot;Z)$ is continuous on $[0,\mu - Z)$: on the set where $\Phi_J(\gamma(N)) > \Phi_o$ it is a composition of continuous functions, and as $\Phi_J(\gamma(N)) - \Phi_o \downarrow 0$ the argument of $G$ diverges, so $BR \to 0$, matching the value on the complement. For $Z < \bar Z$ the set $\mathcal U := \{N \in (0,\mu - Z) : BR(N;Z) \ge N\}$ is non-empty by part (c), closed in $(0,\mu - Z)$ by continuity, and disjoint from $(0, N_0(Z)]$ by (a). Hence its infimum $N^U(Z) \ge N_0(Z) > 0$ is attained, $BR(N^U) \ge N^U$, and since $BR(N) < N$ for $N < N^U$, continuity gives $BR(N^U) \le N^U$. So $N^U$ is an equilibrium, and it is the smallest positive one. Under $\dot N = BR(N) - N$, any solution starting in $(0, N^U)$ is strictly decreasing while it remains there and is bounded below by $0$; it converges to a rest point in $[0, N^U)$, and the only one is $0$. Local Lipschitz continuity gives uniqueness of the adjustment trajectory. A solution starting at or above $N^U$ therefore cannot cross the rest point $N^U$, so it does not converge to $0$. The basin of collapse is therefore $[0, N^U(Z))$. Monotonicity: for $Z < Z' < \bar Z$, $BR(\cdot;Z') \le BR(\cdot;Z)$ on the common domain by (c), so $\mathcal U(Z') \subseteq \mathcal U(Z)$ and $N^U(Z') \ge N^U(Z)$. Limit: suppose $Z_{\max} > 0$, so $\Phi_J(1) - \Phi_o > \tau/(\mu\theta_{\max})$ and $BR_0(N) := n_v\bigl[1 - G\bigl(\tau/((\mu - N)(\Phi_J(1) - \Phi_o))\bigr)\bigr]$ satisfies $BR_0(0) > 0$. By continuity of $BR_0$, fix $\nu \in (0,\mu)$ with $BR_0(\nu) > \nu$. As $Z \downarrow 0$, $\gamma(\nu) = \nu/(\nu + \psi Z) \to 1$ and $T(\nu) \to 1/(\mu - \nu)$, so $BR(\nu;Z) \to BR_0(\nu)$ by continuity of $\Phi_J$ and $G$. Hence $BR(\nu;Z) > \nu$ for all small $Z$, so $N^U(Z) \le \nu$. Since $\nu$ can be taken arbitrarily small, $N^U(Z) \to 0$.
\end{proof}

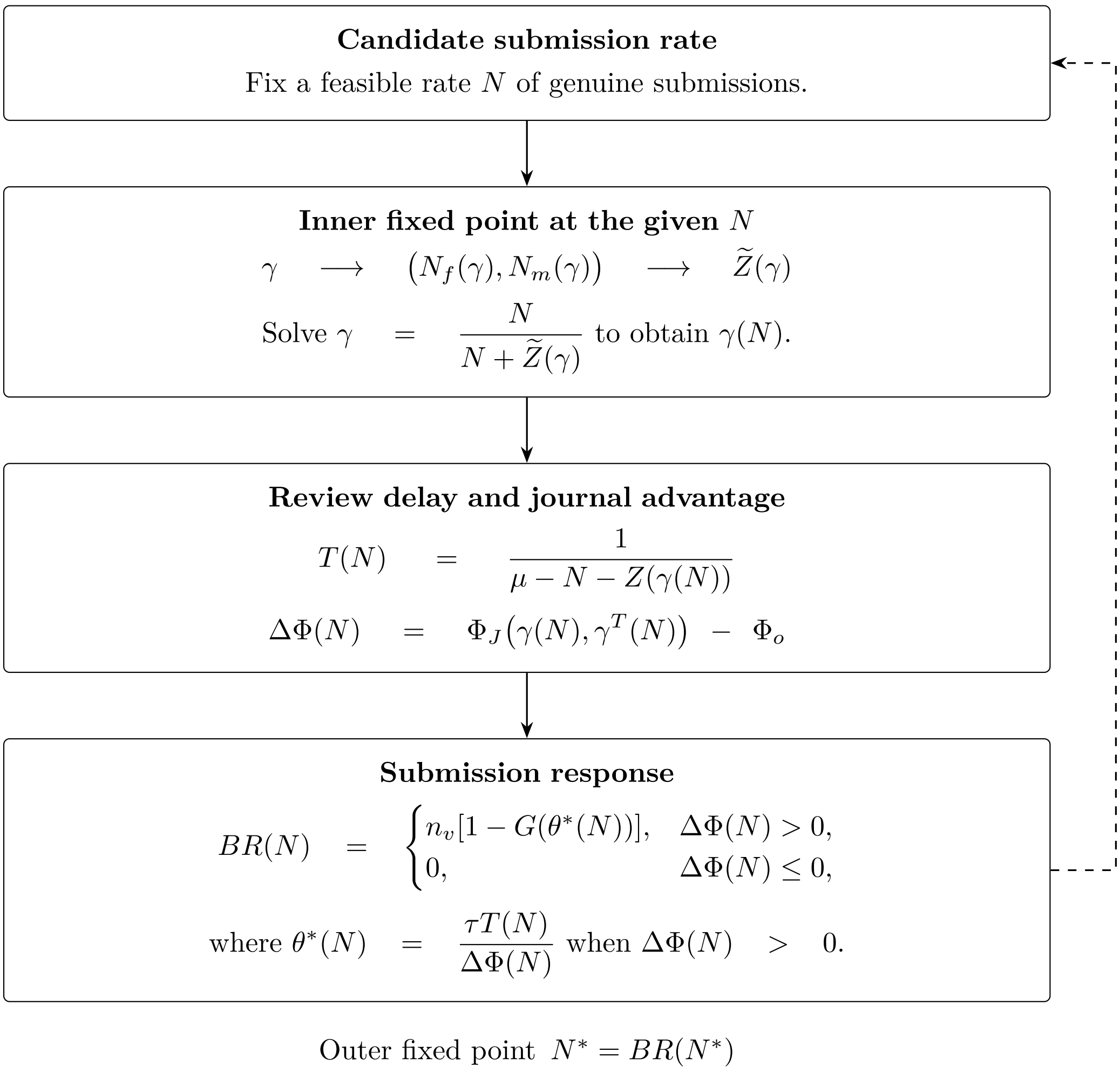
\begin{figure}[htbp]
\centering
\begin{tikzpicture}[
  node distance=8mm,
  block/.style={
    draw, rounded corners=2pt, align=center,
    text width=.77\linewidth, inner xsep=8pt, inner ysep=8pt,
    font=\small
  },
  flow/.style={-{Stealth[length=2mm]}, semithick},
  feedback/.style={flow, dashed}
]

\node[block] (input) {
  \textbf{Candidate submission rate}\\[3pt]
  Fix a feasible rate \(N\) of genuine submissions.
};

\node[block, below=of input] (credibility) {
  \textbf{Inner fixed point at the given \(N\)}\\[4pt]
  \(\displaystyle
    \gamma \longrightarrow
    \bigl(N_f(\gamma),N_m(\gamma)\bigr)
    \longrightarrow \widetilde Z(\gamma)
  \)\\[5pt]
  Solve \(\displaystyle \gamma=\frac{N}{N+\widetilde Z(\gamma)}\)
  to obtain \(\gamma(N)\).
};

\node[block, below=of credibility] (conditions) {
  \textbf{Review delay and journal advantage}\\[5pt]
  \(\displaystyle
    T(N)=\frac{1}{\mu-N-Z(\gamma(N))}
  \)\\[6pt]
  \(\displaystyle
    \Delta\Phi(N)=\Phi_J\bigl(\gamma(N),\gamma^T(N)\bigr)-\Phi_o
  \)
};

\node[block, below=of conditions] (response) {
  \textbf{Submission response}\\[5pt]
  \(\displaystyle
  BR(N)=
  \begin{cases}
    n_v[1-G(\theta^*(N))], & \Delta\Phi(N)>0,\\
    0,                    & \Delta\Phi(N)\le 0,
  \end{cases}
  \)\\[6pt]
  where \(\displaystyle \theta^*(N)=\frac{\tau T(N)}{\Delta\Phi(N)}\)
  when \(\Delta\Phi(N)>0\).
};

\draw[flow] (input) -- (credibility);
\draw[flow] (credibility) -- (conditions);
\draw[flow] (conditions) -- (response);

\coordinate (right) at ([xshift=8mm]response.east);
\draw[feedback] (response.east) -- (right) |- (input.east);

\node[below=3mm of response, font=\small] {
  Outer fixed point \(\,N^*=BR(N^*)\)
};

\end{tikzpicture}
\caption{Computing the integrated equilibrium. For each candidate
submission rate \(N\), solve for endorsement credibility and compute the
conditions faced by authors. The dashed arrow denotes the outer
equilibrium condition.}
\label{fig:integrated-equilibrium}
\end{figure}

\begin{proof}[Proof of Proposition~\ref{prop:integrated}]
\emph{Credibility given $N$.} For $N > 0$, the map $\mathcal H_N(\gamma) := N/(N + \tilde Z(\gamma))$ is continuous and non-increasing on $[0,1]$ with values in $(0,1)$, because $\tilde Z$ is non-decreasing and $\tilde Z \ge \tilde Z(0) > 0$. Hence $\gamma - \mathcal H_N(\gamma)$ is continuous and strictly increasing, negative at $0$ and positive at $1$, and has a unique root $\gamma(N) \in (0,1)$. If $N' > N$, then $\mathcal H_{N'} \ge \mathcal H_N$ pointwise, so $\gamma - \mathcal H_{N'}(\gamma) \le \gamma - \mathcal H_N(\gamma)$ and $\gamma(N') \ge \gamma(N)$. Continuity of $\gamma(\cdot)$ on $(0,\infty)$ follows because it is the root of a jointly continuous function that is strictly increasing in $\gamma$. At $N = 0$, $\gamma(0) = 0$, and $0 \le \gamma(N) \le N/(N + \tilde Z(0)) \to 0$ as $N \downarrow 0$, so $\gamma$ is continuous on $[0,\infty)$. The induced junk streams $N_f(\gamma(N))$, $N_m(\gamma(N))$ and load $Z(\gamma(N))$ are continuous and non-decreasing in $N$; validation credibility $\gamma^T(N)$ is continuous because its denominator is at least $N + \tilde Z(0) > 0$. The total load $N + Z(\gamma(N))$ is continuous and strictly increasing, equals $Z(0) < \mu$ at $N = 0$ and exceeds $\mu$ for $N > \mu$, so $\mathcal D = [0,\bar N)$ for a unique $\bar N \in (0,\mu]$, and $T(N) \to \infty$ as $N \uparrow \bar N$.

(a) Continuity of $BR$ on $\mathcal D$ follows as in the proof of Theorem~\ref{thm:trap}(e), with $\Phi_J(\gamma(N),\gamma^T(N))$ continuous in $N$. Since $\Phi_J$ is uniformly continuous on $[0,1]^2$ and $\Phi_J(0,\cdot) \le \Phi_o$, there is $\bar\gamma > 0$ such that $\Phi_J(\gamma,\gamma^T) - \Phi_o < \tau/(\mu\theta_{\max})$ for all $\gamma \le \bar\gamma$ and all $\gamma^T$. By continuity of $\gamma(\cdot)$ at $0$, there is $N_0^{\mathrm{int}} > 0$ with $\gamma(N) \le \bar\gamma$ on $[0,N_0^{\mathrm{int}}]$. For such $N$, $\Phi_J(\gamma(N),\gamma^T(N)) - \Phi_o < \tau/(\mu\theta_{\max}) \le \tau T(N)/\theta_{\max}$, because $T(N) \ge 1/\mu$. As in Theorem~\ref{thm:trap}(a), this gives $BR(N) = 0$ on $[0,N_0^{\mathrm{int}}]$, so $N^* = 0$ is an equilibrium and $BR(N) < N$ on $(0,N_0^{\mathrm{int}}]$.

(b) $BR(\hat N) - \hat N > 0$, and as $N \uparrow \bar N$, $T(N) \to \infty$ while $\Phi_J$ is bounded, so $BR(N) \to 0$ and $BR(N) - N \to -\bar N < 0$. By continuity there is a root in $(\hat N,\bar N)$. The roots form a closed subset of $[\hat N,\bar N)$ bounded away from $\bar N$, so a largest root exists, and $BR(N) < N$ above it. Every positive root exceeds $N_0^{\mathrm{int}}$ by (a).

(c) By continuity at the capacity boundary, the load ordering on common domains implies $\mathcal D(\omega')\subseteq\mathcal D(\omega)$ for $\omega'>\omega$: otherwise the later load would reach $\mu$ at an earlier boundary strictly inside its feasible domain. On the common domain, higher load and lower journal value imply $BR(N;\omega')\le BR(N;\omega)$. Define $\mathcal E_{\mathrm{int}}=\{\omega\ge0:BR(N;\omega)=N\text{ for some feasible }N>0\}$. A positive root $N'$ at $\omega'$ is feasible at every smaller $\omega$, with $BR(N';\omega)\ge N'$. Equality gives a root; strict inequality and the negative sign near the capacity boundary give one above $N'$ by continuity. Hence $\mathcal E_{\mathrm{int}}$ is a lower interval. Its extended-real supremum gives existence below $\bar\omega$ and non-existence above it. Above the threshold, even $BR(N;\omega)\ge N$ at a positive feasible $N$ would yield a root by the same argument. Roots are bounded away from zero by (a) and from the capacity boundary, so a largest root exists. The best-response ordering makes it non-increasing, as in Theorem~\ref{thm:trap}(c).

Fix $\omega_1<\bar\omega$ and choose $N_0^{\mathrm{int}}(\omega_1)>0$ inside its feasible domain as in (a). A positive root $N\le N_0^{\mathrm{int}}(\omega_1)$ at any $\omega\ge\omega_1$ would satisfy $N=BR(N;\omega)\le BR(N;\omega_1)=0$, a contradiction. This uniform positive lower bound proves the discontinuity at any finite boundary. The endpoint conditions put $\omega_L$ in $\mathcal E_{\mathrm{int}}$ and $\omega_H$ outside it, bounding the threshold. The existence set includes tangential roots, including any at the threshold; no genericity exception is needed.
\end{proof}

\begin{proof}[Proof of Proposition~\ref{prop:reputation}]
(a) By Lemma~\ref{lem:effort}, $D a=\kappa A_J-\rho B_J=\kappa A_P-\rho B_P+\sigma a(\kappa-\rho t)=Da_P+Du(t)a$. Since $u(t)<1$, the unique solution is $a_J=a_P/(1-u(t))$, and $\bar x_E/\bar x_P=a_J/a_P$. Because $\sigma>0$ and $D>0$, $u(t)>0$ if and only if $t<\kappa/\rho$, and $1/(1-u)>1$ if and only if $u>0$.

(b) $A_J-A_P=\sigma a_J$ and $B_J-B_P=\sigma t a_J$. Substituting into the expression in the proof of Lemma~\ref{prop:regimes} gives the displayed difference. Since $\theta\sigma a_J/D>0$, its sign is that of $(\kappa+\lambda\rho)-(\rho+\lambda\chi)t$, which is the sign of~\eqref{eq:C} at $s_y=ts_x$.

(c) $a_J-a_J^0=a_P[1/(1-u)-1-u]=a_Pu^2/(1-u)$. Since $du/dt=-\sigma\rho/D$, $da_J/dt=-a_P\sigma\rho/[D(1-u)^2]$ and $da_J^0/dt=-a_P\sigma\rho/D$. Finally, $a_J(t_c)-a_J(t')=a_P[u(t_c)-u(t')]/[(1-u(t_c))(1-u(t'))]$, whereas the benchmark gives $a_P[u(t_c)-u(t')]$; the ratio exceeds one when both $u$ values lie in $(0,1)$.

(d) By the envelope argument in the proof of Lemma~\ref{lem:effort}, $\partial\Phi/\partial A=a$ and $\partial\Phi/\partial B=b$ per unit of capability, both positive on the interior cone $\{\kappa A>\rho B,\ \chi B>\rho A\}$. The cone is convex, so the gradient is positive along any segment between interior slope pairs. At $\gamma=0$, $\sigma=0$ and both value functions equal $\Phi_P$. The reputational slope increments are $\sigma a_J=\sigma a_P/(1-\sigma(\kappa-\rho t)/D)$ and $t\sigma a_J$. As functions of $\sigma\propto\gamma$, $\sigma a_J$ is continuous with derivative $a_P/(1-u)^2>0$, and $t\ge0$, so both slopes are continuous and non-decreasing in $\gamma$, with $A_J$ strictly increasing; hence $\Phi^R_J$ is continuous and strictly increasing. For $\gamma>0$, $A_J-A_J^0=\sigma(a_J-a_P)$ and $B_J-B_J^0=\sigma t(a_J-a_P)$. If $t>\kappa/\rho$, then $a_J<a_P$ by (a), so both differences are non-positive and the first is negative, which gives $\Phi^R_J<\Phi^0_J$; the case $t<\kappa/\rho$ is symmetric. Both value functions equal $\Phi_P=\Phi_o$ at $\gamma=0$ and are strictly increasing, so Assumption~\ref{ass:trap} holds for each. In~\eqref{eq:BR}, a lower value of the journal weakly raises the argument of $G$ or sets the response to zero, so $\Phi^R_J\le\Phi^0_J$ implies $BR^R\le BR^0$ at every feasible $N$. The comparisons of $\bar Z$, $N^U$ and $N^H$ then follow exactly as the comparisons across junk loads in the proof of Theorem~\ref{thm:trap}(c) and~(e): a positive root of $BR^R$ yields $BR^0(N';Z)\ge N'$ and hence a positive root of $BR^0$; the set $\{N>0:BR^R(N)\ge N\}$ is contained in the corresponding set for $BR^0$; and above the largest root of $BR^0$, $BR^R\le BR^0<N$.

(e) The derivation in the proof of Theorem~\ref{thm:rational}(a) uses only $(x,y)=\theta(a,b)$ in a genuine pool, so every interior self-consistent rule satisfies $t=rb/a$. Then $\sigma ta=\sigma rb$, and the first-order conditions $Da=\kappa(A_P+\sigma a)-\rho(B_P+\sigma rb)$ and $Db=\chi(B_P+\sigma rb)-\rho(A_P+\sigma a)$ are the displayed linear system. If its determinant $\Delta_R$ is non-zero, $(a,b)$ and hence $t$ are unique.
\end{proof}

\begin{proof}[Proof of Proposition~\ref{prop:aggregate}]
(a) With credibility, $N_f$ and $\delta$ fixed, Leibniz's rule in~\eqref{eq:aggregate} gives $\partial Q/\partial\theta^*=-n_v(q_J-q_o)\theta^*g(\theta^*)<0$. For $m(t):=\mathbb E[\theta\mid\theta\ge t]$, differentiation gives $m'(t)=g(t)(m(t)-t)/(1-G(t))$. This is positive at the interior cutoff because $g(\theta^*)>0$ and $m(\theta^*)>\theta^*$; multiplication by $q_J>0$ preserves the sign.

(b) Proposition~\ref{prop:fixed}(a) raises $T^*$ and hence $\theta^*$ when $dZ>0$. Differentiating~\eqref{eq:aggregate} with credibility and $\delta$ fixed gives the displayed $dQ<0$: the selection term is negative by (a), and the harm term is non-positive because $dN_f\ge0$. Average capability rises by (a). For marginal-only growth $dN_f=0$.
\end{proof}

\section{Extensions}\label{app:ext}

\subsection{Joint reductions in presentation cost and resource competition}\label{app:jointcost}

\begin{proposition*}[Joint reductions in presentation cost and resource competition]
Hold the reward slopes, $\chi$ and $\lambda$ fixed, and consider a differentiable technology path $\kappa(z),\rho(z)$ along which $D>0$ and the effort choices remain interior. Write $a=-\kappa'(z)\ge0$ and $b=-\rho'(z)\ge0$. Then
\begin{align}
\frac{d x^*}{dz}
&=\frac{(\kappa y^*-\rho x^*)b-\rho y^*a}{D},\label{eq:jointcostx}\\
\frac{d Q}{dz}
&=\frac{\bigl[(\kappa+\lambda\rho)y^*-(\rho+\lambda\chi)x^*\bigr]b
-(\rho+\lambda\chi)y^*a}{D}.\label{eq:jointcostQ}
\end{align}
Thus cheaper presentation lowers substantive effort along this path if and only if $\rho y^*a>(\kappa y^*-\rho x^*)b$. The fixed-$\rho$ crowding-out result is recovered at $b=0$; when $\rho=0$ remains fixed, substantive effort is unchanged, and when also $\lambda=0$, so is $Q$.
\end{proposition*}

\begin{proof}
Differentiate the first-order conditions $\chi x^*+\rho y^*=\theta A$ and $\rho x^*+\kappa y^*=\theta B$ along the path. Inverting the cost matrix gives $dx^*=[(\rho x^*-\kappa y^*)d\rho+\rho y^*d\kappa]/D$ and $dy^*=[(\rho y^*-\chi x^*)d\rho-\chi y^*d\kappa]/D$. Substitute $d\kappa/dz=-a$, $d\rho/dz=-b$ and $dQ=dx^*-\lambda dy^*$.
\end{proof}

This result separates the gain from reducing the resources absorbed by presentation from the incentive to produce more of it. Neither a decline in substantive effort nor a decline in the net-output proxy is an unconditional consequence of cheaper writing. The empirical relevance of the fixed-$\rho$ benchmark depends on which cost margins the technology changes.

\subsection{Saturated presentation}\label{app:saturation}

\begin{remark}[Saturated polish]\label{rem:saturation}
Theorem~\ref{thm:rational}(b) relies on presentation scaling with capability, $y=b\theta$, which follows from the quadratic cost~\eqref{eq:cost} in the interior benchmark. Suppose instead that chosen presentation is bounded by $y\le\bar y$. If the cap binds for every genuine author in a genuine-only reviewed pool, then $\tilde y=\bar y+u$ and is uncorrelated with substance, so Lemma~\ref{lem:projection} gives $t=0$. This weight satisfies the threshold of Theorem~\ref{thm:look}(a); the output comparison under a cap additionally depends on which constraints bind in the two regimes. In particular, if the same cap binds in both $J$ and $P$, their technical rewards are the same, and substantive effort is positive in both, the first-order condition gives $x_\theta^{\ell*}=(\theta A_\ell-\rho\bar y)/\chi$. Thus, with $t=0$, $Q_\theta^J-Q_\theta^P=\theta\gamma e\alpha_E/\chi>0$. This conditional capped comparison follows because presentation is identical in both regimes. With partial saturation, or different binding constraints across regimes, neither that expression nor the monotonicity result in Theorem~\ref{thm:rational}(b) applies without solving the corresponding constrained problem.
\end{remark}

\subsection{Significance triage with false exclusions}\label{app:false}

The error-free comparison makes triage attractive whenever marginal submissions consume attention. This local extension shows why faster processing alone is an insufficient objective when triage can exclude genuine contributions. Hold genuine arrivals $N_v>0$, their effort distribution, review capacity and the leakage rates fixed. Let a fraction $a(\zeta)\in[0,1)$ of genuine papers be incorrectly excluded, where $a$ is differentiable and $a'\ge0$. Assume these exclusions are independent of capability and potential review signals and outcomes within the genuine stream, so the mean substantive effort of surviving genuine papers remains $\bar x>0$ and their average endorsement propensity is unchanged. Write
\[
N_g=(1-a)N_v,\qquad
T_a=\frac{1}{\mu-N_g-Z},\qquad
\gamma_a=\frac{N_g}{N_g+\tilde Z},\qquad
X_g=N_g\bar x.
\]
Here $Z$ and $\tilde Z$ are as in Proposition~\ref{prop:triage}, including deterrence. $X_g$ measures the flow of genuine substantive work retained for review, not field-wide output: an excluded paper may still circulate elsewhere. Technical validation precedes triage as in the benchmark.

\begin{proposition*}[Capacity relief with false exclusions]
Suppose $\mu>N_g+Z$, $N_m>0$ and the derivatives in Proposition~\ref{prop:triage} exist. Then
\begin{align}
\frac{\partial T_a}{\partial\zeta}
&=-T_a^2\bigl[a'N_v+N_m(1+\eta_m)\bigr]<0,\label{eq:false-delay}\\
\frac{\partial\gamma_a}{\partial\zeta}
&=\frac{N_g\psi_mN_m(1+\eta_m)-a'N_v\tilde Z}{(N_g+\tilde Z)^2},\label{eq:false-credibility}\\
\frac{\partial X_g}{\partial\zeta}&=-a'N_v\bar x\le0.\label{eq:false-retention}
\end{align}
When $\tilde Z>0$, triage raises endorsement credibility if and only if
\begin{equation}\label{eq:false-condition}
\frac{\psi_mN_m(1+\eta_m)}{\tilde Z}
>\frac{a'}{1-a}.
\end{equation}
Thus delay can fall even when endorsement credibility and the flow of genuine work retained for review both fall. The error-free benchmark is recovered by setting $a\equiv0$.
\end{proposition*}
\begin{proof}
Differentiate $N_g$ to obtain $\partial N_g/\partial\zeta=-a'N_v$. Proposition~\ref{prop:triage} gives $\partial Z/\partial\zeta=-N_m(1+\eta_m)$ and $\partial\tilde Z/\partial\zeta=-\psi_mN_m(1+\eta_m)$. Substitution into the derivative of $1/(\mu-N_g-Z)$ gives~\eqref{eq:false-delay}. The quotient rule gives~\eqref{eq:false-credibility}; differentiating $X_g$ gives~\eqref{eq:false-retention}. Divide the numerator in~\eqref{eq:false-credibility} by $N_g\tilde Z>0$ to obtain~\eqref{eq:false-condition}.
\end{proof}

The left-hand side of~\eqref{eq:false-condition} is the proportional removal of endorsed junk; the right-hand side is the proportional loss of genuine papers. This separates a capacity gain from an improvement in the certificate. For example, if reviewers already reject all marginal work ($\psi_m=0$), some flawed papers are endorsed ($\tilde Z>0$), and $a'>0$, stronger significance triage reduces credibility even though it shortens the queue. More generally, an editor can pair the delay objective in Corollary~\ref{cor:invest} with a minimum-retention constraint $X_g\ge\underline X_g$ and audit both types of screening error. Such a constraint does not determine a complete welfare optimum, but it prevents measured delay improvements from being treated as sufficient evidence of better certification. Endogenous participation, effort and capability-dependent false exclusions would require extending this local comparison.

\section{Numerical comparison of review rules}\label{app:policy-computation}

This appendix reports the numerical comparison of adaptive review and contribution-first commitment summarised in Section~\ref{sec:policy-equilibrium}.

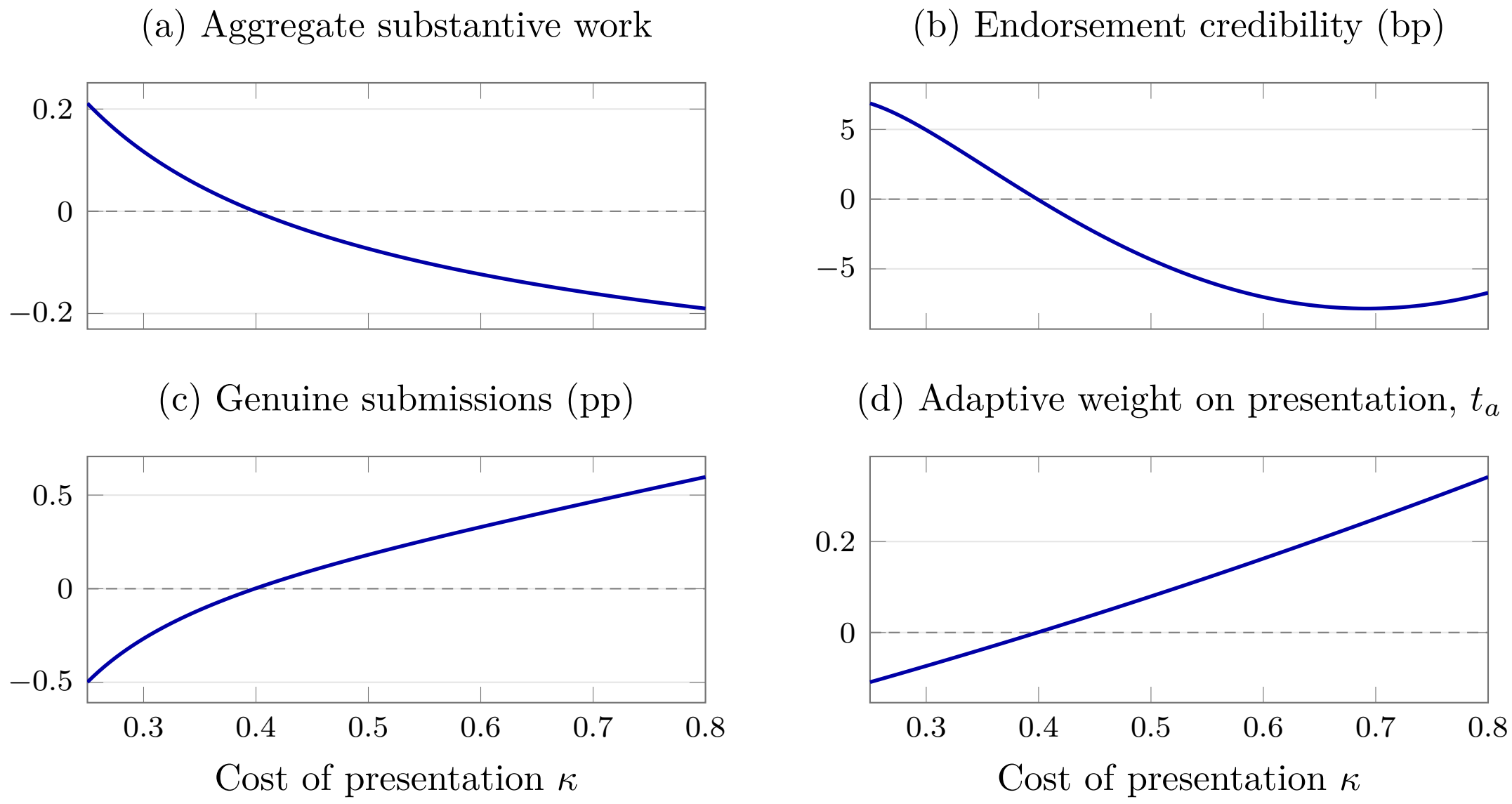
\begin{figure}[tbp]
\centering
\begin{tikzpicture}
\begin{groupplot}[
group style={group size=2 by 2,horizontal sep=1.55cm,vertical sep=1.2cm},
width=.465\textwidth,height=.245\textwidth,
xmin=.25,xmax=.8,xtick={.3,.4,.5,.6,.7,.8},
title style={font=\small},tick label style={font=\scriptsize},
label style={font=\small},axis line style={black!55},tick style={black!55},
ymajorgrids,grid style={gray!20},
scaled ticks=false,
]
\nextgroupplot[title={(a) Aggregate substantive work},xticklabels={}]
\addplot[gray,densely dashed,forget plot] coordinates {(.25,0) (.8,0)};
\addplot[blue!65!black,line width=1pt,no marks] coordinates {(0.250,0.21114831) (0.255,0.19999508) (0.260,0.18928527) (0.265,0.17899227) (0.270,0.16909151) (0.275,0.15956025) (0.280,0.15037747) (0.285,0.14152367) (0.290,0.13298076) (0.295,0.12473194) (0.300,0.11676158) (0.305,0.10905513) (0.310,0.10159903) (0.315,0.09438061) (0.320,0.08738807) (0.325,0.08061033) (0.330,0.07403705) (0.335,0.06765853) (0.340,0.06146569) (0.345,0.05544998) (0.350,0.04960338) (0.355,0.04391834) (0.360,0.03838777) (0.365,0.03300497) (0.370,0.02776363) (0.375,0.02265781) (0.380,0.01768187) (0.385,0.01283051) (0.390,0.00809870) (0.395,0.00348168) (0.400,-0.00102506) (0.405,-0.00542578) (0.410,-0.00972454) (0.415,-0.01392517) (0.420,-0.01803135) (0.425,-0.02204653) (0.430,-0.02597401) (0.435,-0.02981694) (0.440,-0.03357831) (0.445,-0.03726097) (0.450,-0.04086762) (0.455,-0.04440087) (0.460,-0.04786318) (0.465,-0.05125691) (0.470,-0.05458432) (0.475,-0.05784755) (0.480,-0.06104868) (0.485,-0.06418966) (0.490,-0.06727239) (0.495,-0.07029867) (0.500,-0.07327023) (0.505,-0.07618874) (0.510,-0.07905576) (0.515,-0.08187285) (0.520,-0.08464145) (0.525,-0.08736297) (0.530,-0.09003876) (0.535,-0.09267012) (0.540,-0.09525830) (0.545,-0.09780448) (0.550,-0.10030983) (0.555,-0.10277546) (0.560,-0.10520243) (0.565,-0.10759177) (0.570,-0.10994447) (0.575,-0.11226149) (0.580,-0.11454374) (0.585,-0.11679211) (0.590,-0.11900746) (0.595,-0.12119062) (0.600,-0.12334238) (0.605,-0.12546351) (0.610,-0.12755475) (0.615,-0.12961683) (0.620,-0.13165044) (0.625,-0.13365624) (0.630,-0.13563489) (0.635,-0.13758701) (0.640,-0.13951321) (0.645,-0.14141409) (0.650,-0.14329020) (0.655,-0.14514209) (0.660,-0.14697031) (0.665,-0.14877537) (0.670,-0.15055776) (0.675,-0.15231798) (0.680,-0.15405650) (0.685,-0.15577376) (0.690,-0.15747022) (0.695,-0.15914630) (0.700,-0.16080243) (0.705,-0.16243900) (0.710,-0.16405640) (0.715,-0.16565503) (0.720,-0.16723524) (0.725,-0.16879740) (0.730,-0.17034187) (0.735,-0.17186897) (0.740,-0.17337905) (0.745,-0.17487242) (0.750,-0.17634939) (0.755,-0.17781027) (0.760,-0.17925536) (0.765,-0.18068494) (0.770,-0.18209930) (0.775,-0.18349870) (0.780,-0.18488341) (0.785,-0.18625370) (0.790,-0.18760982) (0.795,-0.18895200) (0.800,-0.19028050)};
\nextgroupplot[title={(b) Endorsement credibility (bp)},xticklabels={}]
\addplot[gray,densely dashed,forget plot] coordinates {(.25,0) (.8,0)};
\addplot[blue!65!black,line width=1pt,no marks] coordinates {(0.250,6.87017599) (0.255,6.73601649) (0.260,6.58467996) (0.265,6.41822422) (0.270,6.23846161) (0.275,6.04699457) (0.280,5.84524523) (0.285,5.63448018) (0.290,5.41583122) (0.295,5.19031294) (0.300,4.95883763) (0.305,4.72222796) (0.310,4.48122778) (0.315,4.23651150) (0.320,3.98869209) (0.325,3.73832800) (0.330,3.48592919) (0.335,3.23196240) (0.340,2.97685566) (0.345,2.72100234) (0.350,2.46476465) (0.355,2.20847673) (0.360,1.95244740) (0.365,1.69696252) (0.370,1.44228720) (0.375,1.18866765) (0.380,0.93633291) (0.385,0.68549634) (0.390,0.43635698) (0.395,0.18910077) (0.400,-0.05609839) (0.405,-0.29907762) (0.410,-0.53968419) (0.415,-0.77777475) (0.420,-1.01321459) (0.425,-1.24587699) (0.430,-1.47564261) (0.435,-1.70239900) (0.440,-1.92604005) (0.445,-2.14646562) (0.450,-2.36358105) (0.455,-2.57729686) (0.460,-2.78752837) (0.465,-2.99419542) (0.470,-3.19722204) (0.475,-3.39653624) (0.480,-3.59206973) (0.485,-3.78375774) (0.490,-3.97153876) (0.495,-4.15535442) (0.500,-4.33514925) (0.505,-4.51087058) (0.510,-4.68246834) (0.515,-4.84989495) (0.520,-5.01310521) (0.525,-5.17205614) (0.530,-5.32670691) (0.535,-5.47701870) (0.540,-5.62295465) (0.545,-5.76447973) (0.550,-5.90156069) (0.555,-6.03416596) (0.560,-6.16226559) (0.565,-6.28583118) (0.570,-6.40483583) (0.575,-6.51925405) (0.580,-6.62906174) (0.585,-6.73423612) (0.590,-6.83475569) (0.595,-6.93060017) (0.600,-7.02175048) (0.605,-7.10818868) (0.610,-7.18989793) (0.615,-7.26686249) (0.620,-7.33906764) (0.625,-7.40649967) (0.630,-7.46914585) (0.635,-7.52699439) (0.640,-7.58003446) (0.645,-7.62825608) (0.650,-7.67165016) (0.655,-7.71020848) (0.660,-7.74392362) (0.665,-7.77278898) (0.670,-7.79679874) (0.675,-7.81594786) (0.680,-7.83023204) (0.685,-7.83964772) (0.690,-7.84419204) (0.695,-7.84386287) (0.700,-7.83865872) (0.705,-7.82857881) (0.710,-7.81362301) (0.715,-7.79379181) (0.720,-7.76908635) (0.725,-7.73950839) (0.730,-7.70506027) (0.735,-7.66574494) (0.740,-7.62156595) (0.745,-7.57252738) (0.750,-7.51863390) (0.755,-7.45989071) (0.760,-7.39630358) (0.765,-7.32787878) (0.770,-7.25462312) (0.775,-7.17654389) (0.780,-7.09364893) (0.785,-7.00594653) (0.790,-6.91344549) (0.795,-6.81615507) (0.800,-6.71408502)};
\nextgroupplot[title={(c) Genuine submissions (pp)},xlabel={Cost of presentation $\kappa$}]
\addplot[gray,densely dashed,forget plot] coordinates {(.25,0) (.8,0)};
\addplot[blue!65!black,line width=1pt,no marks] coordinates {(0.250,-0.49909528) (0.255,-0.46994222) (0.260,-0.44244216) (0.265,-0.41643871) (0.270,-0.39179418) (0.275,-0.36838686) (0.280,-0.34610883) (0.285,-0.32486403) (0.290,-0.30456671) (0.295,-0.28514011) (0.300,-0.26651530) (0.305,-0.24863020) (0.310,-0.23142881) (0.315,-0.21486044) (0.320,-0.19887910) (0.325,-0.18344303) (0.330,-0.16851415) (0.335,-0.15405772) (0.340,-0.14004197) (0.345,-0.12643778) (0.350,-0.11321844) (0.355,-0.10035939) (0.360,-0.08783798) (0.365,-0.07563336) (0.370,-0.06372624) (0.375,-0.05209879) (0.380,-0.04073447) (0.385,-0.02961796) (0.390,-0.01873502) (0.395,-0.00807240) (0.400,0.00238223) (0.405,0.01264038) (0.410,0.02271279) (0.415,0.03260947) (0.420,0.04233982) (0.425,0.05191262) (0.430,0.06133610) (0.435,0.07061799) (0.440,0.07976553) (0.445,0.08878556) (0.450,0.09768447) (0.455,0.10646831) (0.460,0.11514277) (0.465,0.12371320) (0.470,0.13218467) (0.475,0.14056197) (0.480,0.14884959) (0.485,0.15705183) (0.490,0.16517271) (0.495,0.17321607) (0.500,0.18118553) (0.505,0.18908452) (0.510,0.19691632) (0.515,0.20468401) (0.520,0.21239053) (0.525,0.22003867) (0.530,0.22763109) (0.535,0.23517032) (0.540,0.24265874) (0.545,0.25009866) (0.550,0.25749225) (0.555,0.26484157) (0.560,0.27214862) (0.565,0.27941527) (0.570,0.28664332) (0.575,0.29383449) (0.580,0.30099041) (0.585,0.30811266) (0.590,0.31520271) (0.595,0.32226201) (0.600,0.32929191) (0.605,0.33629372) (0.610,0.34326868) (0.615,0.35021799) (0.620,0.35714278) (0.625,0.36404415) (0.630,0.37092315) (0.635,0.37778076) (0.640,0.38461796) (0.645,0.39143565) (0.650,0.39823471) (0.655,0.40501598) (0.660,0.41178027) (0.665,0.41852835) (0.670,0.42526096) (0.675,0.43197880) (0.680,0.43868256) (0.685,0.44537289) (0.690,0.45205039) (0.695,0.45871569) (0.700,0.46536934) (0.705,0.47201189) (0.710,0.47864387) (0.715,0.48526578) (0.720,0.49187810) (0.725,0.49848130) (0.730,0.50507582) (0.735,0.51166208) (0.740,0.51824048) (0.745,0.52481142) (0.750,0.53137527) (0.755,0.53793238) (0.760,0.54448309) (0.765,0.55102772) (0.770,0.55756660) (0.775,0.56410001) (0.780,0.57062824) (0.785,0.57715156) (0.790,0.58367024) (0.795,0.59018451) (0.800,0.59669462)};
\nextgroupplot[title={(d) Adaptive weight on presentation, $t_a$},xlabel={Cost of presentation $\kappa$}]
\addplot[gray,densely dashed,forget plot] coordinates {(.25,0) (.8,0)};
\addplot[blue!65!black,line width=1pt,no marks] coordinates {(0.250,-0.10904975) (0.255,-0.10553101) (0.260,-0.10200309) (0.265,-0.09846584) (0.270,-0.09491913) (0.275,-0.09136281) (0.280,-0.08779678) (0.285,-0.08422094) (0.290,-0.08063519) (0.295,-0.07703945) (0.300,-0.07343364) (0.305,-0.06981768) (0.310,-0.06619152) (0.315,-0.06255509) (0.320,-0.05890834) (0.325,-0.05525120) (0.330,-0.05158365) (0.335,-0.04790562) (0.340,-0.04421708) (0.345,-0.04051800) (0.350,-0.03680832) (0.355,-0.03308803) (0.360,-0.02935708) (0.365,-0.02561545) (0.370,-0.02186311) (0.375,-0.01810003) (0.380,-0.01432619) (0.385,-0.01054157) (0.390,-0.00674613) (0.395,-0.00293988) (0.400,0.00087723) (0.405,0.00470520) (0.410,0.00854406) (0.415,0.01239382) (0.420,0.01625449) (0.425,0.02012608) (0.430,0.02400863) (0.435,0.02790212) (0.440,0.03180659) (0.445,0.03572203) (0.450,0.03964846) (0.455,0.04358589) (0.460,0.04753433) (0.465,0.05149379) (0.470,0.05546427) (0.475,0.05944578) (0.480,0.06343833) (0.485,0.06744192) (0.490,0.07145657) (0.495,0.07548227) (0.500,0.07951902) (0.505,0.08356684) (0.510,0.08762573) (0.515,0.09169568) (0.520,0.09577671) (0.525,0.09986881) (0.530,0.10397199) (0.535,0.10808624) (0.540,0.11221157) (0.545,0.11634797) (0.550,0.12049546) (0.555,0.12465401) (0.560,0.12882365) (0.565,0.13300435) (0.570,0.13719613) (0.575,0.14139897) (0.580,0.14561289) (0.585,0.14983786) (0.590,0.15407389) (0.595,0.15832098) (0.600,0.16257912) (0.605,0.16684831) (0.610,0.17112854) (0.615,0.17541980) (0.620,0.17972209) (0.625,0.18403541) (0.630,0.18835974) (0.635,0.19269509) (0.640,0.19704143) (0.645,0.20139877) (0.650,0.20576710) (0.655,0.21014641) (0.660,0.21453669) (0.665,0.21893793) (0.670,0.22335013) (0.675,0.22777326) (0.680,0.23220733) (0.685,0.23665232) (0.690,0.24110823) (0.695,0.24557503) (0.700,0.25005272) (0.705,0.25454129) (0.710,0.25904073) (0.715,0.26355102) (0.720,0.26807216) (0.725,0.27260412) (0.730,0.27714690) (0.735,0.28170048) (0.740,0.28626485) (0.745,0.29083999) (0.750,0.29542590) (0.755,0.30002255) (0.760,0.30462993) (0.765,0.30924803) (0.770,0.31387683) (0.775,0.31851632) (0.780,0.32316647) (0.785,0.32782728) (0.790,0.33249872) (0.795,0.33718078) (0.800,0.34187345)};
\end{groupplot}
\end{tikzpicture}
\caption{Adaptive review and contribution-first commitment with endogenous participation. Panels (a)--(c) report adaptive minus committed outcomes. Positive values favour adaptation for the outcome shown. Panel (d) reports the adaptive weight, with the dashed line marking the committed weight of zero. Credibility differences are in basis points (bp), and submission differences are in percentage points (pp) of the genuine author population. Parameters and the equilibrium calculation are in Section~\ref{app:policy-details}.}
\label{fig:policy-equilibrium}
\end{figure}

Figure~\ref{fig:policy-equilibrium} varies the cost of presentation $\kappa$ along locally stable positive-participation branches, computed as described in the next subsection. At $\kappa=0.60$, marginal papers are more polished than genuine submissions and the adaptive rule still puts a positive weight on presentation ($t_a\simeq 0.163$). Commitment raises aggregate substantive work by $0.123$ and endorsement credibility by $7.02$ basis points, while reducing genuine participation by $0.33$ percentage points. At $\kappa=0.30$, the adaptive weight is negative ($t_a\simeq -0.073$). Adaptation raises aggregate substantive work by $0.117$ and credibility by $4.96$ basis points, while drawing $0.27$ percentage points fewer genuine authors. In each case the rule with the stronger incentive for substantive work also endorses relatively fewer marginal papers. It attracts fewer genuine authors because the other rule offers a higher private return to submitting at the margin.

In this mixed pool the sign of the adaptive weight changes near $\kappa=0.399$. Lemma~\ref{lem:contested} explains the reversal. Changing $\kappa$ changes genuine authors' presentation and substantive effort as well as the composition of the contested pool. The predictor eventually discounts presentation on the selected branch. The comparison holds along this branch at fixed marginal inflow and fixed $\rho$. The branch and crossing are properties of the example, and Theorem~\ref{thm:rational}(b) pertains to a genuine-only pool with credibility held fixed. Appendix~\ref{app:jointcost} gives the separate condition for substantive effort to fall when both presentation costs and resource competition decline.

\subsection{Parameters and computation}\label{app:policy-details}

Use $G=U[0.5,2.5]$, $n_v=1$, $Z_m=0.22$, $\psi_m^0=0.60$, $\mu=1.22$, $\tau=0.13$, $\chi=1$, $\rho=0.17$, $a_0+\alpha_T r=1.5$, $b_0=1$, $e=1$, $V=40$, $s_x=0.05$, and $y_m=2.3$. The independent reading and presentation noises are symmetric two-point variables $\varepsilon\in\{-0.6,0.6\}$ and $u\in\{-0.7,0.7\}$. Each noise outcome is equiprobable. The endorsement probability is the affine score in Section~\ref{sec:adaptive} clipped to $[0,1]$ outside its feasible range.

For $0<N<1$, the cutoff is $\theta^*=2.5-2N$, and genuine submitters' capability mean and variance are $2.5-N$ and $N^2/3$. With $A=1.5+2\gamma$ and $B=1+2\gamma t$, compute $(a,b)$ and $\Phi_J$ using Lemma~\ref{lem:effort}. Use $A_P=1.5$ and $B_P=1$ to compute $\Phi_P$ and $a_o$. Set $(\bar x,\bar y)=(a,b)(2.5-N)$ and $\varphi=\psi_m^0 Z_m/(N+\psi_m^0 Z_m)$. The residuals for the adaptive rule, evaluated using Lemma~\ref{lem:contested}, are
\begin{align*}
F_1 &= (1-N)(2.5-2N)(\Phi_J-\Phi_P)-\tau,\\
F_2 &= \gamma\left[N(\bar x+t\bar y)+\psi_m^0 Z_m(\bar x+t y_m)\right]-N(\bar x+t\bar y),\\
F_3 &= t-w_y/w_s .
\end{align*}
For commitment, set $t=0$ and $\gamma=N/(N+\psi_m^0 Z_m)$, and solve $F_1=0$ for the larger interior root. For adaptation, start the simultaneous root solve at $(N,\gamma,t)=(0.969,0.88,-0.13)$ at $\kappa=0.25$. Follow the positive-participation branch in steps of $0.005$ using the preceding root as the starting point, up to $\kappa=0.80$. At each plotted point the absolute residuals are below $10^{-9}$. We checked positive efforts, $w_s>0$ for adaptation, feasibility of the endorsement probabilities for both streams, and local stability for the participation adjustment $\dot N=BR(N)-N$. For the latter check, differentiate $F_2=F_3=0$ locally with respect to $N$ to obtain the response of $(\gamma,t)$, and use it to evaluate whether $BR'(N)<1$. For commitment, differentiate the explicit credibility formula. Both branches pass this check at every plotted point. Aggregate work is $X_i=1.5a_o+(a_i-a_o)N_i(2.5-N_i)$.

For genuine authors, the computed affine optimum is also optimal under clipping. To check this, replace clipping to $[0,1]$ by clipping only from below at zero, which weakly raises expected utility at any effort pair. With the four joint noise states, the expected positive part of the affine score is the maximum over subsets of the four states of the corresponding sums of scores divided by four. For each subset, maximise the associated affine utility less the quadratic cost over $x,y\ge0$. The resulting indirect utility is affine in $\theta$. Checking the two endpoints of the capability support suffices to compare it to the full-subset optimum. At all plotted points the full subset dominates and its optimal efforts have raw probabilities in $[0,1]$ for every state, yielding equality with the clipped objective. This check covers authors who might deviate to the journal as well as those who submit in equilibrium.

\section{ICLR data and estimates}\label{app:iclr}

We use reviews and decisions from the main track of the International Conference on Learning Representations (ICLR) for 2024--2026 to describe how presentation assessments covary with recommendations and decisions. The review form records separate scores for soundness, presentation and contribution, alongside an overall rating. These data provide a descriptive benchmark for the distinction between substance and presentation. They do not identify the causal effect of polish or test the effects of cheaper generation.

\paragraph{Data and specification.}
We use the public OpenReview records compiled by Paper Copilot \citep{yang2026papercopilot}. After removing eight repeated submission rows and three repeated review rows, the data contain 38,890 submissions and 150,634 reviews; all reviews have the scores required for our specification and cover 38,256 distinct submissions. The annual submission counts include withdrawals and desk rejections. ICLR reset the 2026 scores to their pre-rebuttal state following the OpenReview security incident \citep{iclr2026retrospective}, so we use 2024--2025 for comparisons over time and report 2026 as a separate cross-section. For reviewer $j$ of manuscript $i$, we estimate separately by year
\begin{equation}\label{eq:iclr-review}
\mathrm{Overall}_{ij}
= \alpha_i + \beta_S\,\mathrm{Soundness}_{ij}
+ \beta_P\,\mathrm{Presentation}_{ij}
+ \beta_C\,\mathrm{Contribution}_{ij}
+ \beta_H\,\mathrm{Confidence}_{ij}
+ \varepsilon_{ij},
\end{equation}
with the overall rating standardised within year. The manuscript fixed effect $\alpha_i$ absorbs common characteristics of each submission, so estimates come from differences across reviewers of the same paper. Standard errors are clustered by manuscript, and the ratio $\hat w=\hat\beta_P/\hat\beta_C$ uses the delta method; it describes associations per point on the reported scales, not a scale-invariant preference parameter. We also estimate annual acceptance logits on the manuscript's mean sub-scores with primary-area fixed effects, using papers with an accept or reject decision.

\begin{table}[htbp]
\centering
\caption{Review scores and acceptance decisions at ICLR}
\label{tab:iclr-review}
\small
\begin{tabular}{@{}lrrr@{}}
\toprule
 & 2024 & 2025 & 2026$^{\dagger}$ \\
\midrule
Unique submissions & 7,404 & 11,672 & 19,814 \\
\addlinespace
\multicolumn{4}{@{}l}{\textit{A. Overall rating with manuscript fixed effects}} \\
Soundness & 0.322 (0.009) & 0.319 (0.007) & 0.390 (0.006) \\
Presentation & 0.195 (0.008) & 0.204 (0.006) & 0.244 (0.005) \\
Contribution & 0.524 (0.009) & 0.493 (0.007) & 0.605 (0.006) \\
Presentation / contribution & 0.372 (0.017) & 0.413 (0.014) & 0.403 (0.010) \\
Reviews & 28,028 & 46,748 & 75,858 \\
Manuscripts & 7,262 & 11,520 & 19,474 \\
\addlinespace
\multicolumn{4}{@{}l}{\textit{B. Acceptance logit with primary-area fixed effects}} \\
Mean soundness & 1.591 (0.126) & 1.949 (0.104) & 1.431 (0.078) \\
Mean presentation & 1.026 (0.100) & 0.894 (0.081) & 1.025 (0.066) \\
Mean contribution & 2.743 (0.122) & 2.674 (0.099) & 2.658 (0.077) \\
Presentation / contribution & 0.374 (0.041) & 0.334 (0.034) & 0.386 (0.028) \\
Decided manuscripts & 5,698 & 8,614 & 13,696 \\
Acceptance rate in sample & 39.7\% & 43.0\% & 39.1\% \\
\bottomrule
\end{tabular}
\par\smallskip
\begin{minipage}{0.94\textwidth}
\footnotesize
\textit{Notes.} Standard errors are in parentheses. Panel A controls for reviewer confidence and clusters errors by manuscript. Panel B reports log-odds coefficients and model-based standard errors. Ratio standard errors use the delta method. $^{\dagger}$ The 2026 scores were reset to their pre-rebuttal values.
\end{minipage}
\end{table}

\paragraph{Presentation and evaluation.}
Table~\ref{tab:iclr-review} shows a positive association between presentation and the overall recommendation. Among reviewers of the same manuscript, a one-point higher presentation score is associated with 0.195 and 0.204 standard deviations higher overall ratings in 2024 and 2025, conditional on soundness, contribution and confidence; the presentation coefficient is 37--41\% as large as the contribution coefficient. In the acceptance logits, presentation is positively associated with the final decision conditional on the other mean sub-scores, with a coefficient about one-third as large as that on contribution. From 2024 to 2025, $\hat w$ rises by 0.041 (standard error 0.022, two-sided $p=0.064$, treating the annual estimates as independent), mainly because the contribution coefficient falls, while the corresponding ratio in the acceptance logits falls from 0.374 to 0.334. The year comparison therefore provides weak evidence of a change in how reviewers combine their sub-scores and leaves Theorem~\ref{thm:rational}(b) untested.

\paragraph{Reviewer agreement and load.} The one-way intraclass correlations for soundness and contribution are 0.160 and 0.193 in 2024, and 0.166 and 0.186 in 2025. With these scores as operational measures, reviewers agree slightly less on soundness, the opposite of the auxiliary hypothesis P7; since soundness is a broad human assessment rather than a technical-validity test, this shows only that the Discovery Stack pattern cannot be assumed to generalise. Finally, submissions in the data rise from 7,404 to 19,814 and the withdrawn fraction from 22.3\% to 26.3\%. These figures describe potential review load; they do not identify the junk share, crowding out or changes in capacity.

\end{document}